\documentclass[english]{article}
\usepackage{lmodern}
\usepackage[T1]{fontenc}
\usepackage[latin9]{inputenc}
\usepackage{amsmath}
\usepackage{amssymb}
\usepackage{graphicx}
\usepackage{esint}
\usepackage{xcolor}
\usepackage{babel}
\usepackage{fullpage}
\usepackage{hyperref}
\usepackage{caption}
\usepackage{subcaption}
\usepackage{float}
\usepackage{cite}

\def\be{\begin{equation}}
\def\ee{\end{equation}}

\def\bea{\begin{eqnarray}}
\def\eea{\end{eqnarray}}

\numberwithin{equation}{section}

\begin{document}

\begin{titlepage}

\rightline{HIP-2014-01/TH}
\rightline{NORDITA-2014-8}
\rightline{OUTP-13-27P}
\rule{0pt}{1cm}


\centerline{
\Large \bf  
Holographic thermalization with Lifshitz scaling
}
\vspace{.3cm}
\centerline{
\Large \bf  
and hyperscaling violation
}

\vspace{1.3truecm}

 \centerline{\large
Piermarco Fonda$^{a,}$\footnote[1]{e-mail: piermarco.fonda@sissa.it},  
Lasse Franti$^{b,}$\footnote[2]{e-mail: lasse.franti@helsinki.fi},  
Ville Ker\"anen$^{d,}$\footnote[3]{e-mail: ville.keranen@physics.ox.ac.uk},  }
\vspace{.3cm}
 \centerline{\large
Esko Keski-Vakkuri$^{b,e,}$\footnote[4]{e-mail: esko.keski-vakkuri@helsinki.fi},  
Larus Thorlacius$^{c,f,}$\footnote[5]{e-mail: larus@nordita.org}
 and Erik Tonni$^{a,}$\footnote[6]{e-mail: erik.tonni@sissa.it}}   
    
\vspace{1cm}

\centerline{$^a${\it  SISSA and INFN, via Bonomea 265, 34136, Trieste, Italy}}
\vspace{.22cm}
\centerline{$^b${\it  Helsinki Institute of Physics and Department of Physics,}}
\centerline{{\it University of Helsinki, P.O.Box 64, FIN-00014, Finland}}
\vspace{.22cm}
\centerline{$^c${\it  Nordita, KTH Royal Institute of Technology and Stockholm University,}}
\centerline{{\it Roslagstullsbacken 23, SE-106 91 Stockholm, Sweden}}
\vspace{.22cm}
\centerline{$^d${\it  Oxford University, Rudolf Peierls Center for Theoretical Physics,}}
\centerline{{\it 1 Keble Road, OX1 3NP, United Kingdom}}
\vspace{.22cm}
\centerline{$^e${\it  Department of Physics and Astronomy, Uppsala University, SE-75108, Sweden}}
\vspace{.22cm}
\centerline{$^f${\it  University of Iceland, Science Institute, Dunhaga 3,
IS-107 Reykjavik, Iceland}}

\vspace{1.7truecm}

\centerline{\bf Abstract}
\vspace{.5truecm}

\noindent

A Vaidya type geometry describing gravitation collapse in asymptotically Lifshitz spacetime with
hyperscaling violation provides a simple holographic model for thermalization 
near a quantum critical point with non-trivial dynamic and hyperscaling violation exponents.
The allowed parameter regions are constrained by requiring that the matter energy momentum
tensor satisfies the null energy condition. We present a combination of analytic and numerical results
on the time evolution of holographic entanglement entropy in such backgrounds for different shaped boundary 
regions and study various scaling regimes, generalizing previous work by Liu and Suh. 

\end{titlepage}

\section{Introduction}

One of the interesting questions regarding quantum information is how fast quantum correlations can propagate in a physical system. In a groundbreaking study in 1972, Lieb and Robinson \cite{Lieb:1972wy} derived an upper bound for the speed of propagation of correlations in an interacting lattice system and in recent years there has been growing interest in this and related questions in connection with a number of new advances. The study of ultracold atom systems has developed to the level where experiments on the time evolution of quantum correlations are possible (see e.g. \cite{naturepaper}),  new techniques have been developed for the theoretical study of time evolution of observables in perturbed quantum lattices (see e.g. \cite{KGE}),analytical results have been obtained for the time evolution of observables after quenches in conformal field theory \cite{Calabrese:2005in,Calabrese:2006,Calabrese:2007rg} and entanglement entropy has been given a geometric interpretation \cite{Ryu:2006bv, Ryu:2006ef,  Hubeny:2007xt, Takayanagi:2012kg} in the context of the holographic duality of strongly interacting conformal field theory \cite{Maldacena:1997re}. The present paper follows up on this last direction.

In the context of holographic duality, different ways of introducing quenches in a conformal theory have been studied. One line of work focuses on constructing holographic
duals for quenches in strongly coupled theories \cite{Basu:2013soa,Buchel:2013gba,Hartman:2013qma, Buchel:2013lla,Buchel:2012gw},  in the spirit of similar work in weakly coupled quantum field theory  involving a sudden change in the parameters of the Hamiltonian \cite{Calabrese:2005in,Calabrese:2006,Calabrese:2007rg,Rigol:2007, Kollath:2006, Manmana:2007, Sotiriadis:2010si}. In another approach, the focus has instead been on perturbing the state of the system by turning on homogeneous sources for a short period of time. By a slight abuse of terminology, this process has also been called a ``quench'', although perhaps a ``homogenous explosion''  would be a closer term to describe the sudden change in the state of the boundary theory. There are two good reasons to study this model. One of them is that there is an elegant and tractable gravitational dual description of such a process in terms of the gravitational collapse of a thin shell of null matter to a black hole, the AdS-Vaidya geometry. The other good reason is
that the time evolution of quantum correlations manifested in the holographic entanglement entropy following such an explosion was found to behave in the same manner as in the 1+1 dimensional conformal field theory work \cite{Calabrese:2005in,Calabrese:2006,Calabrese:2007rg} -- in a relativistic case quantum correlations were found to propagate at the speed of light \cite{AbajoArrastia:2010yt,Albash:2010mv,Balasubramanian:2010ce,Balasubramanian:2011ur, Aparicio:2011zy, Balasubramanian:2011at, Allais:2011ys, Callan:2012ip, Hubeny:2013dea}. The interesting lesson there is that even a strongly coupled conformal theory with no quasiparticle excitations may behave as if the correlations were carried by free-streaming particles. 
The model also allows for an easy extrapolation of the results to higher dimensional field theory at strong coupling. In generic dimensions, it turns out that the time evolution of
holographic entanglement entropy has a more refined structure, characterized by different scaling regimes \cite{Liu:2013iza, Liu:2013qca}:  (I) a pre-local equilibrium power law growth in time, (II) a post-local equilibration linear growth in time, (III) a saturation regime. For entanglement surfaces of more general shape, one can also identify late-time memory loss, meaning
that near saturation the time-evolution becomes universal with no memory on the detailed shape of the surface.

Many condensed matter and ultracold atom systems feature more complicated critical behavior with anisotropic (Lifshitz) scaling \cite{Kachru:2008yh}, characterized by the dynamic critical exponent $\zeta > 1$, or hyperscaling violation characterized by a non-zero hyperscaling violation exponent $\theta$ \cite{Gouteraux:2011ce, Huijse:2011ef, Dong:2012se}. 
Hyperscaling violation leads to an effective dimension $d_{\theta}=d-\theta $. It was found that for a critical value $d_\theta =1$ the entanglement entropy exhibits a logarithmic
violation from the usual area law \cite{Ogawa:2011bz}, which is also generic for compressible states with hidden Fermi surfaces \cite{fermisurface}. 

By now there exist various holographic dual  models for critical points involving Lifschitz scaling and hyperscaling violation \cite{Gouteraux:2011ce, Huijse:2011ef, Dong:2012se,
Ogawa:2011bz,Charmousis:2010zz,Iizuka:2011hg,Singh:2010zs,Narayan:2012hk,Singh:2012un,Dey:2012tg,Dey:2012rs,
Charmousis:2012dw,Ammon:2012je,Bhattacharya:2012zu,Kundu:2012jn,Dey:2012fi,Shaghoulian:2011aa,Alishahiha:2012qu,Bueno:2012sd,Gath:2012pg,Gouteraux:2012yr}. 
In the light of the rich scaling structure in the time evolution of entanglement entropy, it is interesting to see how it carries over to systems with Lifshitz scaling and hyperscaling violation.
In \cite{Keranen:2011xs} a Lifshitz scaling generalization of the AdS-Vaidya geometry was constructed, and it was found that time evolution of entanglement entropy still contains a linear regime, where
entanglement  behaves as if it was carried by free streaming particles at finite velocity. This is non-trivial, since in the non-relativistic case $\zeta >1$ there is no obvious characteristic scale like the speed of light in relativistic theories. The authors of \cite{Liu:2013iza,Liu:2013qca}, on the other hand, considered a relativistic system with hyperscaling violation, and found that their previous analysis easily carries over to that case, with the spatial dimension $d$ replaced by the effective dimension $d_\theta$. In this paper we extend the analysis to systems that exhibit both Lifshitz scaling and hyperscaling violation. We do this by first constructing the extension of the Lifshitz-AdS-Vaidya geometry to the hyperscaling violating case, and then analyzing the time
evolution of the entanglement entropy for various boundary regions. 
We compute numerically the evolution of the holographic entanglement entropy for the strip and the sphere in backgrounds with non-trivial $\zeta$ and $\theta$. 
We then extract some analytic behavior in the thin shell limit for the temporal regimes (I), (II) and (III), generalizing the results of \cite{Liu:2013iza,Liu:2013qca} to the case of $\zeta \neq 1$ and $\theta \neq 0$.
In an appendix, we also consider briefly quench geometries where the critical exponents themselves are allowed to vary. This can be motivated from a quasiparticle picture and one could, for instance, consider a system where the dispersion relation is suddenly altered from $\omega \sim k^2 +\cdots $ to $\omega \sim k +\cdots$ or vice versa, by rapidly adjusting the chemical potential. We take some steps in this direction by considering holographic geometries where the dynamical critical exponent and the hyperscaling violation parameter are allowed to vary with time and show that such solutions can be supported by matter satisfying the null energy condition, at least in some simple cases. We leave a more detailed study for future work.

This paper is organized as follows. Hyperscaling violating Lifshitz-AdS-Vaidya solutions are introduced in Section~2 and parameter regions allowed by the null energy condition determined.
In Section~3 the holographic entanglement entropy for a strip and for a sphere is analyzed in static backgrounds and  Vaidya-type backgrounds are considered in Section~4. 
In Section~5 scaling regions in the time evolution of the entanglement entropy are studied for differently shaped surfaces. The details of some of the computations are presented in appendices along with a brief description of  holographic quench geometries where the hyperscaling violation parameter and the dynamical critical exponent are allowed to vary with time.

{\bf Note added.} As we were preparing this manuscript, \cite{Alishahiha:2014cwa} appeared with significant  overlap with some of our results. A preliminary check finds that where overlap exists, the results are compatible.

\section{Backgrounds with Lifshitz and hyperscaling exponents}
\label{sec backgrounds}

The starting point of our analysis is the following gravitational action 
\cite{Keranen:2011xs}
\be
\label{action}
S = \frac{1}{16 \pi G_N}
\int  \bigg(
 R  -\frac{1}{2} (\partial \phi)^2 -V(\phi)
 -\frac{1}{4} 
 \sum_{i=1}^{N_F} e^{\lambda_i \phi} F_i^2
 \bigg) \sqrt{-g} \, d^{d+2}x \,,
\ee
which describes the interaction between the metric $g_{\mu\nu}$, $N_F$ gauge fields and a dilaton $\phi$.
The simplest $d+2$ dimensional time independent background including the Lifshitz scaling $\zeta$ and the hyperscaling violation exponent $\theta$ is given by 
\cite{Gouteraux:2011ce, Huijse:2011ef, Dong:2012se}
\be
ds^{2}= z^{-2d_\theta/d}
(-z^{2-2\zeta}dt^{2}+dz^{2}+d\boldsymbol{x}^{2})\,,
\label{eq:purehs}
\ee
where $z>0$ is the holographic direction and the cartesian coordinates $\boldsymbol{x}$ parameterize $\mathbb{R}^d$ (we denote a vectorial quantity through a bold symbol). 
Hereafter the metric (\ref{eq:purehs}) will be referred as hvLif.
In (\ref{eq:purehs}) we have introduced the convenient combination
\be
d_\theta \equiv d-\theta\,.
\ee
When $\theta = 0$ and $\zeta=1$, (\ref{eq:purehs}) reduces to $AdS_{d+2}$ in Poincar\'e coordinates.

In the following, we will consider geometries that are asymptotic to the hyperscaling violating Lifshitz (hvLif) 
spacetime \eqref{eq:purehs}. In particular, static black hole solutions with Lifshitz scaling and hyperscaling violation have been studied in \cite{Dong:2012se, Alishahiha:2012qu, Bueno:2012sd}. The black hole metric is
\be
\label{eq:poincareHS}
ds^{2}=z^{-2d_\theta/d}
\left(-z^{2-2\zeta}F(z)dt^{2}+\frac{dz^{2}}{F(z)}+d\boldsymbol{x}^{2}\right) \,,
\ee
where the emblackening factor $F(z)$, which contains the mass $M$ of the black hole, is given by
\be 
\label{eq:emblackeningBH}
F(z)=1-Mz^{d_\theta+\zeta}\,.
\ee 
The position $z_h$ of the horizon is defined as $F(z_h) =0$ and the standard near horizon analysis of (\ref{eq:poincareHS}) provides the temperature of the black hole $T = z_h^{1-\zeta} |F'(z_h)|/(4\pi)$.
In order to have $F(z) \rightarrow 1$ when $z \rightarrow 0$, we need to require
\be
\label{eq:asympt}
d_\theta +\zeta \geqslant 0 \,.
\ee
The Einstein equations are $G_{\mu\nu} = T_{\mu\nu}$, where $G_{\mu\nu} $ is the Einstein tensor and $T_{\mu\nu} $ the energy-momentum tensor of the matter fields, {\it i.e.} the dilaton and gauge fields in \eqref{action}.
The Null Energy Condition (NEC) prescribes that $T_{\mu\nu} N^\mu N^\nu \geqslant 0$ for any null vector $N^\mu$.
On shell, the NEC becomes $G_{\mu\nu} N^\mu N^\nu \geqslant 0$ and, through an astute choice of $N^\mu$, one finds \cite{Dong:2012se}
\begin{eqnarray}
\label{eq:hsnec1}
d_\theta (\zeta-1-\theta/d) & \geqslant & 0\,,
 \\
 \label{eq:hsnec2}
(\zeta-1)(d_\theta+\zeta) & \geqslant & 0 \,.
\label{eq:hsnec}
\end{eqnarray}
\begin{figure}[t] 
\begin{center}
\vspace{-0.cm}
\hspace{-.5cm}
\includegraphics[width=.96\textwidth]{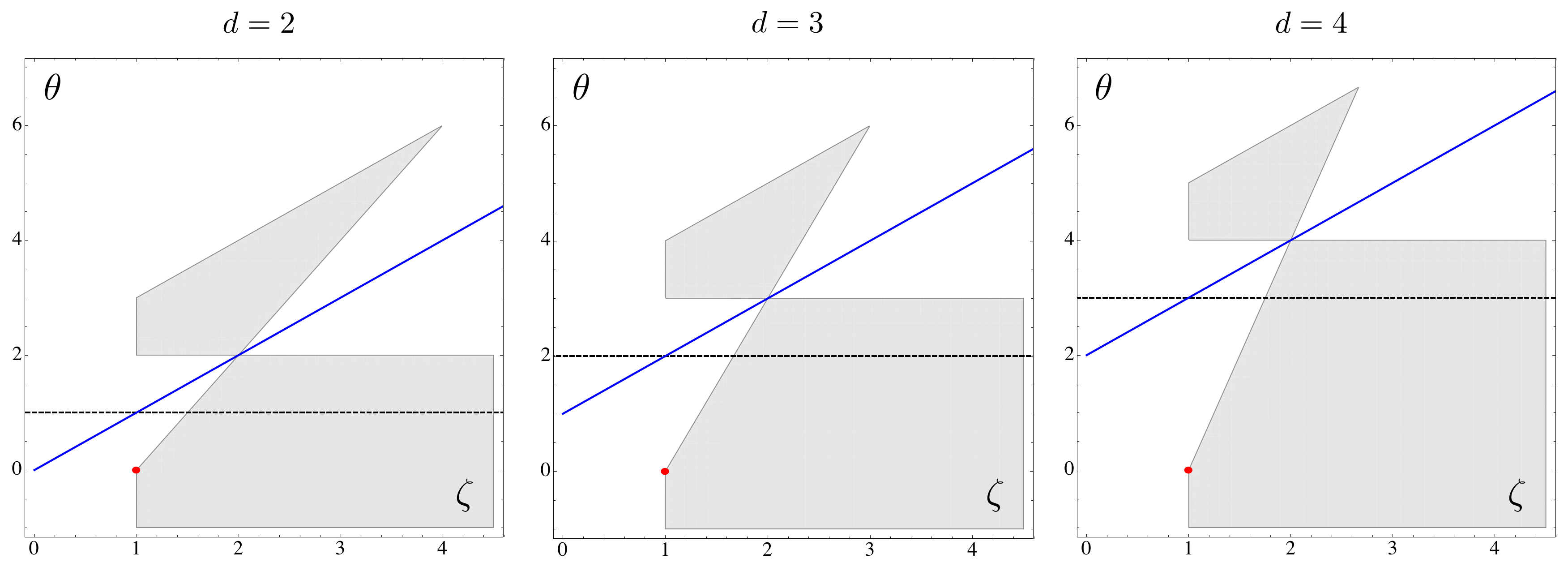}
\end{center}
\vspace{-.3cm}
\caption{\label{fig:necregions}
The grey area is the region of the $(\zeta,\theta)$ plane defined by (\ref{eq:hsnec1}) and (\ref{eq:hsnec2}), obtained from the Null Energy Condition, and also (\ref{eq:asympt}). The panels show $d=2,3,4$. The red dots denote $AdS_{d+2}$ and the horizontal dashed lines indicate the critical value $\theta =d-1$. The blue lines denote the upper bound defined by the condition (\ref{eq:lineargro3}). 
}
\end{figure}
In the critical case $\theta = d-1$, they reduce to $\zeta \geqslant 2- 1/d$.
In Fig. \ref{fig:necregions} we show the region identified by (\ref{eq:hsnec1}) and (\ref{eq:hsnec2}) in the $(\zeta, \theta)$ plane.

In order to construct an infalling shell solution, it is convenient to write the static metric (\ref{eq:poincareHS}) in an Eddington-Finkelstein-like coordinate system, by introducing a new time coordinate $v$ through the relation
\be
\label{v diff form}
dv=dt-\frac{dz}{z^{1-\zeta}F\left(z\right)}\,,
\ee
and rewriting (\ref{eq:poincareHS}) as
\be
\label{vaidya metric static}
ds^{2}= z^{-2d_\theta/d}
(-z^{2(1-\zeta)}F(z)dv^{2}-2z^{1-\zeta}\, dv\, dz+d\boldsymbol{x}^{2}) \,.
\ee
The dynamical background that we are going to consider is of Vaidya type \cite{Vaidya:1951zz, Bonnor:1970zz} and it is obtained by  promoting the mass $M$ in (\ref{vaidya metric static}) to a time dependent function $M(v)$, namely
\be
\label{vaidya metric}
ds^{2}= z^{-2d_\theta/d}
(-z^{2(1-\zeta)}F(v,z)dv^{2}-2z^{1-\zeta}\, dv\, dz+d\boldsymbol{x}^{2})\,,
\ee
where
\be
\label{Ffactor vaidya}
F(v,z)=1-M(v)z^{d_\theta+\zeta} \,.
\ee
The metric (\ref{vaidya metric}) with the emblackening factor (\ref{Ffactor vaidya}) is a solution of the equation of motion $G_{\mu\nu}=T_{\mu\nu}$, where the energy-momentum tensor is given by the one of the static case with $M$ replaced by $M(v)$, except for the component $T_{vv}$, which now contains the following additional term 
\be
\tilde{T}_{vv} = \frac{d_\theta}{2}\,z^{d_\theta} M'(v) \,.
\ee
Now consider the null vectors $N^\mu =(N^v, N^z, \boldsymbol{N}^{\boldsymbol{x}})$ given by
\be
\label{null vec vaidya}
N_{\textrm{I}}^{\mu} = \left(0,1,\mathbf{0}\right)\,,
\qquad
N_{\textrm{II}}^{\mu} = \left(-\frac{2z^{\zeta-1}}{F(v,z)},1,\mathbf{0}\right)\,,
\qquad
N_{\textrm{III}}^{\mu} =  \left(\pm\frac{z^{\zeta-1}}{\sqrt{F(v,z)}},0,\boldsymbol{n}_1\right)\,,
\ee
where $\boldsymbol{n}_1$ is a $d-1$ dimensional vector with unit norm.
The NEC for the vectors (\ref{null vec vaidya}) leads to the following inequalities
\begin{eqnarray}
\label{eq:necvaidya1}
d_\theta  ( \zeta-1-\theta/d) & \geqslant & 0\,,
\\
\label{eq:necvaidya2}
d_\theta\left[(\zeta-1-\theta/d)F^{2} - 2 z^{\zeta} F_{v}\right] & \geqslant & 0 \,,
\\
\label{eq:necvaidya3}
2(\zeta-1)(d_\theta+\zeta)F^2 
+[z F_{zz}-(d_\theta+3(\zeta-1))F_z ] z F - z^\zeta d_\theta F_v& \geqslant & 0\,,
\end{eqnarray}
where the notation $F_z \equiv \partial_z F$, $F_v \equiv \partial_v F$  and $F_{zz} \equiv \partial^2_z F$  has been adopted.
When $F(v,z)=1$ identically, (\ref{eq:necvaidya2}) and (\ref{eq:necvaidya3}) simplify to (\ref{eq:hsnec1}) and (\ref{eq:hsnec2}) respectively.
Plugging \ref{Ffactor vaidya} into (\ref{eq:necvaidya2}) and (\ref{eq:necvaidya3}), we get 
\bea
d_\theta\big[\,(\zeta-1-\theta/d) (1-M(v)z^{d_\theta+\zeta})^2 
+ 2 z^{d_\theta+2\zeta} M'(v)\big] & \geqslant & 0 \,,
\label{eq:necvaidya2a}\\
2(\zeta-1)(d_\theta+\zeta)(1-M(v)z^{d_\theta+\zeta})+z^{d_\theta+2\zeta} d_\theta M'(v)& \geqslant & 0 \,.
\label{eq:necvaidya3a}
\eea
In the special case of $\theta =0$ and $\zeta =1 $ we recover the condition $M'(v) \geqslant 0$, as expected.
Notice that the NEC for the AdS-Vaidya backgrounds modeling the formation of an asymptotically AdS charged black hole also leads to a non trivial constraint \cite{Caceres:2013dma}, similar to the ones in (\ref{eq:necvaidya2a}) and (\ref{eq:necvaidya3a}).
\\
In the following we will choose the following profile for $M(v)$ 
\be 
M(v)=\frac{M}{2}\big(1+\tanh(v/a) \big)\,,
\label{eq:kinkeq}
\ee 
which is always positive and increasing with $v$. It goes to $0$ when $v\rightarrow -\infty$ and to $M$ when $v \rightarrow +\infty$.
The parameter $a>0$ encodes the rapidity of the transition between the two regimes of $M(v)\sim 0$ and $M(v) \sim M$. In the limit $a \rightarrow 0$ the mass function becomes a step function $M(v) = M \theta(v)$. This is the thin shell regime and it applies to many of the calculations presented below. We have checked numerically that the profiles (\ref{eq:kinkeq}) that we employ satisfy the inequalities (\ref{eq:necvaidya2a}) and (\ref{eq:necvaidya3a}) for all $v$ and $z$.


\section{Holographic entanglement entropy for static backgrounds}

\subsection{Strip}
\label{sec strip static}

Let us briefly review the simple case when the region $A$ in the boundary theory is a thin long strip, which has two sizes $\ell \ll \ell_\perp$ \cite{Ryu:2006bv, Ryu:2006ef, Dong:2012se}. 
Denoting by $x$ the direction along the short length and by $y_i$ the remaining ones, the domain in the boundary is defined by $-\ell/2 \leqslant x \leqslant \ell/2$ and $0 \leqslant y_i \leqslant \ell_\perp$, for $i=1, \dots, d-1$. Since $\ell \ll \ell_\perp$, we can assume translation invariance along the $y_i$ directions and this implies that the minimal surface is completely specified by its profile $z=z(x)$, where $z(\pm \ell/2) = 0$. We can also assume that $z(x)$ is even. 
Computing from (\ref{eq:poincareHS}) the induced metric on such a surface, the area functional reads
\be
\label{area func}
\mathcal{A}[z(x)] = 2\ell_\perp^{d-1}\int_{0}^{\ell/2}
\frac{1}{z^{d_\theta}}\, \sqrt{1+\frac{z'^{2}}{F\left(z\right)}}\,dx \,.
\ee
Since the integrand does not depend on $x$ explicitly, the corresponding integral of motion is constant giving a first order equation for the profile 
\be 
z' =
- \, \sqrt{F(z) \big[(z_\ast/z)^{2 d_\theta}-1\big]} \,.
\label{eq:striphsdiff2}
\ee
Here we have introduced $z(0) \equiv  z_\ast$ and we have used that $z'(0) = 0$ and $z'(x) <0$. Plugging (\ref{eq:striphsdiff2}) into (\ref{area func}), it is straightforward to find that the area of the extremal surface is 
\begin{equation}
\mathcal{A} 
= 
2 \ell_\perp^{d-1}
z_\ast^{d_\theta}\int_{0}^{\ell/2-\eta} z(x)^{-2 d_\theta} dx
=
2 \ell_\perp^{d-1}
\int_{\epsilon}^{z_\ast} \frac{z_\ast^{d_\theta}}{
z^{d_\theta} \sqrt{F(z) \big[z_\ast^{2 d_\theta} - z^{2 d_\theta}  \big]}}\,  dz\,,
\label{eq:areastriphs}
\end{equation}
with $z(x)$ a solution of (\ref{eq:striphsdiff2}). 
A cutoff $z \geqslant \epsilon >0$ has been introduced to render the integral (\ref{eq:areastriphs})
finite, and a corresponding one along the $x$ direction
\be
\label{eta cutoff def}
z(\ell/2-\eta) = \epsilon \,.
\ee
The relation between $z_\ast$ and $\ell$ reads
\begin{equation}
\frac{\ell}{2}
=
\int^{z_\ast}_{0}
\frac{dz}{\sqrt{F(z) \big[(z_\ast/z)^{2 d_\theta}-1\big]}} \,.
\label{eq:lstripHS}
\end{equation}
The vacuum case of $F(z) =1$ can be solved analytically.
Indeed, one can then integrate (\ref{eq:striphsdiff2}), obtaining 
\be
x (z)=\frac{\ell}{2}-
\frac{z_\ast}{1+d_\theta}
\bigg(\frac{z}{z_\ast}\bigg)^{d_\theta+1}
\, _{2}F_{1}\bigg(
\frac{1}{2},\frac{1}{2}+\frac{1}{2 d_\theta };\frac{3}{2}+\frac{1}{2 d_\theta };
(z/z_\ast)^{2d_\theta}
\bigg)  \,,
\label{eq:analyticsolstrip}
\ee
where $_{2}F_{1}$ is the hypergeometric function.
Imposing $x(z_\ast) = 0$ in (\ref{eq:analyticsolstrip}) one finds
\be
\frac{\ell}{2} = 
\frac{\sqrt{\pi} \,\Gamma(\tfrac{1}{2}+\tfrac{1}{2d_\theta})}{\Gamma(\tfrac{1}{2d_\theta})}
\, z_\ast \,.
\ee
The area (\ref{eq:areastriphs}) with $F(z) =1$ is then \cite{Dong:2012se}
\be
\label{eq:areastrippurehs}
\mathcal{A} = 
\left\{
\begin{array}{ll}
\displaystyle 
\frac{2 \ell_\perp^{d-1}}{d_\theta-1}
\Bigg[
\frac{1}{\epsilon^{d_\theta-1}}
-
\frac{1}{\ell^{d_\theta-1}} 
\left(
\frac{\sqrt{\pi} \,\Gamma(\tfrac{1}{2}+\tfrac{1}{2d_\theta})}{\Gamma(\tfrac{1}{2d_\theta})}
\right)^{d_\theta}
\,\Bigg] + O\left(\epsilon^{1+d_\theta}\right)
\hspace{.6cm}&
d_\theta \neq 1 
\\
\rule{0pt}{.6cm}
\displaystyle 
2\ell_\perp^{d-1} \log(\ell/\epsilon) + O\left(\epsilon^{2}\right) 
& 
d_\theta = 1 
\end{array}
\right.
\ee
The critical value $d_\theta = 1 $ is characterized by this divergence, which is logarithmic instead of power-like.

\subsection{Sphere}
\label{sec sphere static}

If the perimeter between the two regions in the boundary theory is a $d-1$ dimensional sphere of radius $R$ it is convenient to adopt spherical coordinates in the bulk (we denote by $\rho$ the radial coordinate) for $\mathbb{R}^{d}$ in (\ref{eq:purehs}) and  (\ref{eq:poincareHS}), namely $d\boldsymbol{x}^2 = d\rho^2 +\rho^2 d\Omega^2_{d-1}$.
In this case, the problem reduces to computing $z=z(\rho)$.
The area functional reads
\be
\label{eq:spherarea}
\mathcal{A}[z(\rho)] =
 \frac{2\pi^{d/2}}{\Gamma(d/2)}
 \int_{0}^{R}
 \frac{\rho^{d-1}}{z^{d_\theta}}\sqrt{1+\frac{z'^{2}}{F(z)}}\, d\rho \,,
\ee
where the factor in front of the integral is the volume of the $d-1$ dimensional unit sphere. 
The key difference compared to the strip (see (\ref{area func})) is that now the integrand of (\ref{eq:spherarea}) depends explicitly on $\rho$ and one has to solve a second order ODE to find the $z(\rho)$ profile,
\be
\label{eq:sferatot}
z\big[ \rho F_z-2(d-1)z' \big] z'^{2}
-2F\big[ \rho\, z\, z''+(d-1)z\, z'+d_\theta \rho\, z'^{2}\big]
- 2 d_\theta \rho F^{2}
=0\,,
\ee
subject to the boundary conditions $z(R)=0$ and $z'(0)=0$.
For a trivial emblackening factor $F\left(z\right)=1$ the equation of motion (\ref{eq:sferatot}) simplifies to 
\be
\rho\, z\, z''+\big[d_\theta \rho+(d-1)z\, z'\big]\big(1+z'^{2}\big)=0 \,.
\label{eq:sferavac}
\ee
In the absence of hyperscaling violation ($\theta=0$) it is well known that $z(\rho)=\sqrt{R^2-\rho^2}$ describes an extremal surface for any dimension $d$ \cite{Ryu:2006ef}. Since the extremal surface is computed for $t=$ const., the Lifshitz exponent $\zeta$ does not enter in the computation but equation \eqref{eq:sferavac} does involve the hyperscaling exponent through the effective dimension $d_\theta$. The extremal surface cannot be found in closed form for general values of $d_\theta\neq 0$ but the leading behavior of the extremal surface area, including the UV divergent part, can be obtained from the small $z$ asymptotics when $\rho =R$ is approached from below. We find it convenient to rewrite \eqref{eq:sferavac} in terms of a dimensionless variables $z= R\,\tilde{z}(x)$, $\rho= R(1-x)$, 
\be
(1-x)\tilde{z}\ddot{\tilde{z}}+\big[d_\theta (1-x)-(d-1)\tilde{z}\dot{\tilde{z}}\big]\big(1+\dot{\tilde{z}}^{2}\big)=0 \,,
\label{eq:dimnless}
\ee
where $\dot{\tilde{z}}$ denotes $d\tilde z/dx$. \\
In the appendix  \S\ref{app sphere hvLif} we construct a sequence of parametric curves $\{x_i(s), \tilde{z}_i(s)\}$ for $i\in \mathbb{N}$ such that the asymptotic one $\{x_\infty(s), \tilde{z}_\infty(s)\}$ solves (\ref{eq:dimnless}). These curves are obtained in order to reproduce the behavior of the solution near the boundary (i.e. small $x$) in a better way as the index $i$ increases. 
Unfortunately, when $i$ is increasing, their analytic expressions become difficult to integrate to get the corresponding area.
Nevertheless, we can identify the following pattern. Given an integer $k_0 \geqslant 0$, which fixes the order in $\epsilon$ that we are going to consider, the procedure described in \S\ref{app sphere hvLif} leads to the following expansion for the area \eqref{eq:spherarea} 
\be
\label{Area dtheta neq odd}
\mathcal{A}[z(\rho)] \,=\,
\frac{2\pi^{d/2}R^{d-1}}{\Gamma(d/2)\,\epsilon^{d_\theta-1}} 
\left\{ \, \sum_{k=0}^{k_0}
\omega_k(d,d_\theta) 
\left(\frac{\epsilon}{R}\right)^{2k}
+O\big(\epsilon^{2(k_0+1)}\big)
\right\}\,,
\qquad  d_\theta \neq \{1,3, 5, \dots, 2k_0+1\} \,,
\ee
where
\be
\omega_k(d,d_\theta) \equiv 
\frac{\gamma_{2k}(d,d_\theta)}{\prod_{j=0}^k \big[d_\theta-(2j+1)\big]^{\alpha_{k,j}}}\,,
\qquad
\alpha_{k,j} \in \mathbb{N}\setminus\{0\}\,.
\ee
The coefficients $\gamma_{2k}(d,d_\theta)$ should be found by explicit integration.
For $k=0$, we get $\gamma_{0}(d,d_\theta)= 1/(d_\theta -1)$. 
The peculiar feature of  the values of $d_\theta$ excluded in (\ref{Area dtheta neq odd}) is the occurrence of a logarithmic divergence, namely, for $0\leqslant \tilde{k} \leqslant k_0$ we have
\be
\label{area sphere odd dtheta}
\mathcal{A}[z(\rho)] \,=\,
 \frac{2\pi^{d/2} R^{d-1}}{\Gamma(d/2)\, \epsilon^{2 \tilde{k} }} 
\left\{ \, \sum_{k=0}^{\tilde{k} -1} \omega_k(d,d_\theta)  \left(\frac{\epsilon}{R}\right)^{2k}
+ \beta_{2\tilde{k} }(d,d_\theta) \left(\frac{\epsilon}{R}\right)^{2\tilde{k} } \log(\epsilon/R) 
+ O\big(\epsilon^{2\tilde{k} }\big)
\right\}\,,
\quad
 d_\theta= 2\tilde{k} +1\,.
\ee
In \S\ref{sec parametric ref} the result for $i=2$ is discussed and it gives (see \S\ref{area parametric sec}) 
\be
\label{area sphere best}
\mathcal{A}[z(\rho)] \,=\,
\left\{ \begin{array}{ll}
\displaystyle
\frac{2\pi^{d/2}R^{d-1}}{\Gamma(d/2)\,\epsilon^{d_\theta-1}} 
\left[\,\frac{1}{d_\theta-1}-\frac{(d-1)^2(d_\theta-2)}{2(d_\theta-1)^2(d_\theta-3)} \frac{\epsilon^2}{R^2} + O(\epsilon^4)  \right] 
\hspace{1cm} & d_\theta\neq 1,3 \\
 \rule{0pt}{.8cm}
\displaystyle
-\frac{2\pi^{d/2}R^{d-1}}{\Gamma(d/2)} \,  \log(\epsilon/R) 
\left[\,1+
\frac{(d-1)^2}{4}\, \frac{\epsilon^2}{R^2} \log (\epsilon/R) + \dots 
\right]
 & d_\theta =1 \\
 \rule{0pt}{.8cm}
 \displaystyle
 \frac{2\pi^{d/2} R^{d-1}}{\Gamma(d/2)\, \epsilon^{2}} 
 \left[\,\frac{1}{2}-\frac{(d-1)(d-5)}{8}  \, \frac{\epsilon^2}{R^2} \log (\epsilon/R) + o(\epsilon^2) \right] 
\hspace{1cm} 
& d_\theta= 3 
\end{array}
\right. 
\ee
Notice that the first expression in (\ref{area sphere best}) for $\theta=0$ provides the expansion at this order of the hemisphere \cite{Ryu:2006ef}. \\
Comparing the result (\ref{area sphere best}) for the spherical region with the one in (\ref{eq:areastrippurehs}), which holds for a strip, it is straightforward to observe that, while for the sphere logarithmic divergences occur whenever $d_\theta$ is odd, for a strip this happens only when $d_\theta =1 $.
The logarithmic terms lead to an enhancement of the area for $d_\theta = 1$, but only contribute
at subleading order for higher odd integer $d_\theta$.


\section{Holographic entanglement entropy in Vaidya backgrounds}

\subsection{Strip}
\label{stripvaidyasec}

\begin{figure}[t] 
\begin{center}
\vspace{-0.cm}
\hspace{-.5cm}
\includegraphics[width=.98\textwidth]{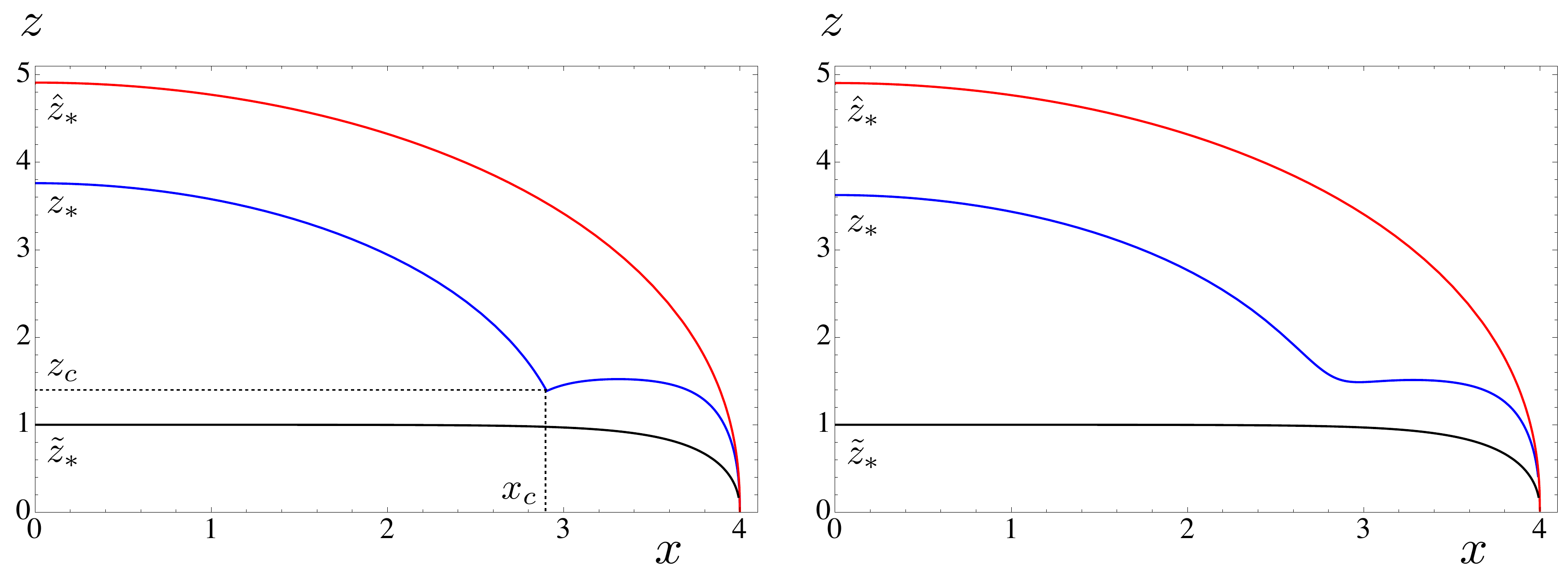}
\end{center}
\vspace{-.3cm}
\caption{\label{fig:profiles}
The profiles $z(x)$ of the extremal surfaces for a strip with $\ell=8$ for different boundary times: $t=0$ (hvLif regime, red curve), $t=3.6$ (intermediate regime, when the shell is crossed, blue curve) and $t=5$ (black hole regime, black curve). The final horizon is $z_h=1$. These plots have $d=2$, $\theta =2/3$ and $\zeta = 1.5$. The left panel shows the situation in the thin shell limit ($a=0.01$), while in the right panel $a=0.5$.
}
\end{figure}

\begin{figure}[t] 
\begin{center}
\vspace{-0.cm}
\hspace{-.5cm}
\includegraphics[width=.88\textwidth]{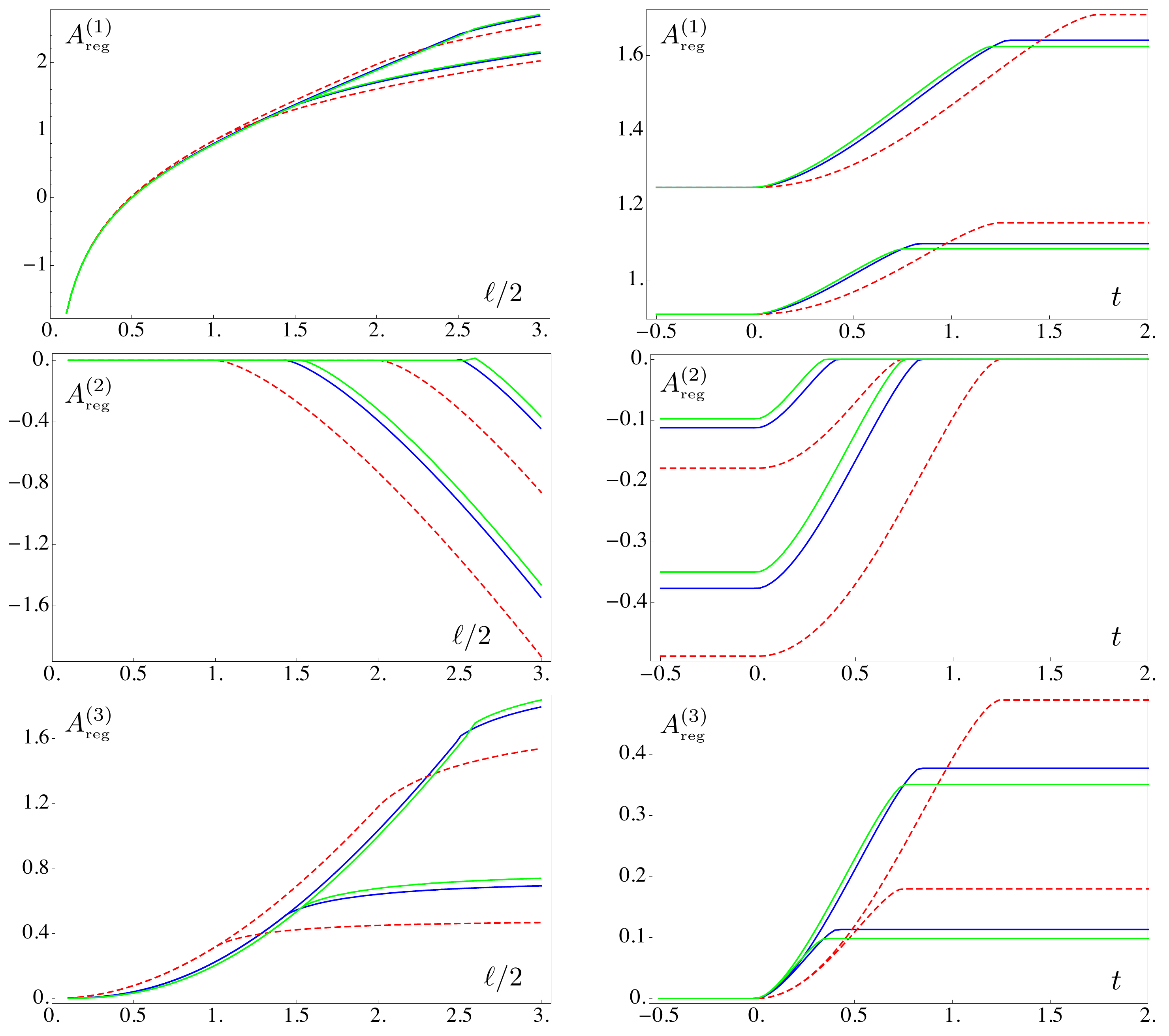}
\end{center}
\vspace{-.3cm}
\caption{\label{fig:A1A2A3}
Strip and $a=0.01$ (thin shell). 
Regularizations (\ref{eq:area1}), (\ref{eq:area2}) and (\ref{eq:area3}) of the area
for $d=1$ (dashed red), $d=2$ (blue) and $d=3$ (green) with $\theta=d-1$ and $\zeta=2-1/d$. Left panels: areas as functions of $\ell/2$ for fixed $t=1.5$ (bottom curves) and $t=2.5$ (upper curves). 
Right: area as functions of the boundary time $t$ with fixed $\ell=3$ and $\ell=5$. The latter ones are characterized by larger variations.
}
\end{figure}

In this section we consider the strip introduced in \S\ref{sec strip static} as the region in the boundary and compute holographically its entanglement entropy in the background given by the Vaidya metric (\ref{vaidya metric}), employing the prescription of 
\cite{Hubeny:2007xt}.
The problem is more complicated than in the static case considered in \S\ref{sec strip static} because the profile is now specified by two functions $z(x)$ and $v(x)$ which must satisfy $v(- \ell/2) =  v(\ell/2) =  t$ and $z(-\ell/2) = z(\ell/2) = 0$, with $t$ the time coordinate in the boundary. Since in our problem $v(x)$ and $z(x)$ are even, the area functional reads
\be
\label{area func vaidya}
\mathcal{A}[v(x), z(x)]
=2 \ell_\perp^{d-1}
\int_{0}^{\ell/2}
\frac{\sqrt{\mathcal{B}}}{z^{d_\theta}}
\,dx\,,
\qquad
\mathcal{B} \equiv 1- F(v,z) z^{2(1-\zeta)} v'^{2} -2z^{1-\zeta} z'v'\,,
\ee
and the boundary conditions for $v(x)$ and $z(x)$ are given by
\be
\label{bc vaidya}
z'(0)= v'(0) = 0\,,
\qquad
v(\ell/2) =  t\,,
\qquad
z(\ell/2) = 0 \,.
\ee
Since the integrand in (\ref{area func vaidya}) does not depend explicitly on $x$,
the corresponding integral of motion is constant, namely $z^{d_\theta} \sqrt{ \mathcal{B}}  = \textrm{const}$.
By recalling that $z(0) \equiv z_\ast$, this constancy condition can be written as
\be
\left(\frac{z_\ast}{z}\right)^{2d_\theta}
= \mathcal{B} \,.
\label{eq:striphamvaidya}
\ee
The equations of motion obtained extremizing the functional (\ref{area func vaidya}) are
\bea
\label{eom vaidya v2}
& &
\partial_{x} \big[ z^{1-\zeta} (z^{1-\zeta}F v' +z') \big]
\,=\,z^{2(1-\zeta)}F_v v'^{2}/2\,,
 \\
\label{eom vaidya v3}
& &
 \partial_{x}\big[ z^{1-\zeta} v' \big] 
 \,=\, 
 d_\theta \mathcal{B}/z
 + z^{2(1-\zeta)} F_z v'^2/2 
 + (1-\zeta) z^{-\zeta}(z'+z^{1-\zeta}F v') v' \,.
\eea
In Fig. \ref{fig:profiles} the typical profiles $z(x)$ obtained by solving these equations numerically are depicted. For $t\leqslant 0$ the extremal surface is entirely in the hvLif part of the geometry.
As time evolves and the black hole is forming, part of the surface enters into the shell and for large times, when the black hole is formed, the extremal surface stabilizes to its thermal result.
In the special case of $\theta =0$ and $\zeta = 1$, (\ref{eom vaidya v2}) and (\ref{eom vaidya v3}) simplify to 
\bea
& &
F_{v}v'^{2}=2\big[F v''+(F_{v}v'+F_{z}z')v'+z''\big]\,,
\\
& &
2zv'' = zF_{z} v'^{2}+2d(1-Fv'^{2}-2z'v') \,.
\eea
Once a solution of (\ref{eom vaidya v2}) and (\ref{eom vaidya v3}) satisfying the boundary conditions (\ref{bc vaidya}) has been found, the surface area is obtained by plugging the solution into (\ref{area func vaidya}). By employing (\ref{eq:striphamvaidya}), one finds that the area of the extremal surface reads 
\be
\label{area vaidya onshell}
\mathcal{A}
=2 \ell_\perp^{d-1} 
\int_{0}^{\ell/2}
\frac{z_\ast^{d_\theta}}{z^{2d_\theta}} 
\,dx \,.
\ee
The integral is divergent and we want to consider its finite part. 
As in the static case, one introduces a cutoff $\epsilon$ along the holographic direction and a corresponding one 
$\eta$ along the $x$ direction, as defined in (\ref{eta cutoff def}).
One way to obtain a finite quantity is to subtract the leading divergence, which, for the strip, is the only one (see (\ref{eq:areastrippurehs}) for the static case),
\bea
d_\theta \neq 1
& &
A^{(1)}_{\textrm{\tiny reg}} \equiv
\int_{0}^{\ell/2-\eta}
\frac{z_\ast^{d_\theta}}{z^{2d_\theta}}
\,dx
-\frac{1}{(d_\theta-1)\, \epsilon^{d_\theta-1}}\,,
\label{eq:area1}
\\
\rule{0pt}{.7cm}
d_\theta = 1
& & 
A^{(1)}_{\textrm{\tiny reg}} \equiv
\int_{0}^{\ell/2-\eta}
\frac{z_\ast^{d_\theta}}{z^{2d_\theta}}
\,dx
- \log(\ell/\epsilon)  
\nonumber \,.
\eea
Another way to get a finite result is by subtracting the area of the extremal surface at late time, 
after the black hole has formed
\be
A^{(2)}_{\textrm{\tiny reg}} \equiv
\int_{0}^{\ell/2-\eta}
\frac{z_\ast^{d_\theta}}{z^{2d_\theta}}
\,dx
-
\int_{0}^{\ell/2-\tilde{\eta}}
\frac{\tilde{z}_\ast^{d_\theta}}{\tilde{z}^{2d_\theta}}
\,dx\,,
\label{eq:area2}
\ee
or by subtracting the area of the extremal surface at early time, when the background is hvLif, namely
\be
A^{(3)}_{\textrm{\tiny reg}} \equiv
\int_{0}^{\ell/2-\eta}
\frac{z_\ast^{d_\theta}}{z^{2d_\theta}}
\,dx
-
\int_{0}^{\ell/2-\hat{\eta}}
\frac{\hat{z}_\ast^{d_\theta}}{\hat{z}^{2d_\theta}}
\,dx \,.
\label{eq:area3}
\ee
The quantities corresponding to the the black hole are tilded, while the ones associated to hvLif  are hatted. In particular, $\tilde{z}(\ell/2-\tilde{\eta}) = \epsilon$ and $\hat{z}(\ell/2-\hat{\eta}) = \epsilon$. In Fig. \ref{fig:A1A2A3} we compare the regularizations (\ref{eq:area1}), (\ref{eq:area2}) and (\ref{eq:area3}) as functions of $\ell$ and of the boundary time $t$ at the critical value $\theta =d-1$.

\subsubsection{Thin shell regime}
\label{ThinStrip}

Let us consider the limit $a \rightarrow 0$ in (\ref{eq:kinkeq}), which leads to a step function
\be
\label{M thin shell}
M(v) = M \theta(v)\,.
\ee
The holographic entanglement entropy in this background has been studied analytically for $\theta=0$, $\zeta =1$ and $d=1$ in \cite{Balasubramanian:2010ce, Balasubramanian:2011ur}. 
For more general values of $\theta$ and $\zeta$ the thin shell regime is obtained by solving the differential equations (\ref{eom vaidya v2}) and (\ref{eom vaidya v3}) in the vacuum (hvLif) for $v<0$ and in the background of a black hole of mass $M$ for $v>0$. The solutions are then matched across the shell. Thus, the metric is (\ref{vaidya metric}) with 
\be
\label{Fvz vaidya}
F(v,z) = \left\{ 
\begin{array}{lll}
1 & v< 0& \textrm{hvLif}\,,\\
F(z)  & v> 0& \textrm{black hole}\,,
\end{array}
\right.
\ee
where $F(z)$ is given by (\ref{eq:emblackeningBH}).
Recall that the symmetry of the problem allows us to work with $0 \leqslant x \leqslant \ell/2$. From Fig. \ref{fig:profiles} and by comparing Fig. \ref{fig:A1A2A3} with Fig. \ref{fig:A3dima05}, one can appreciate the difference between the thin shell regime and the one where $M(v)$ is not a step function. 
Denoting by $x_c$ the position where the two solutions match, we have 
\be 
\label{eq:cquantities}
v(x_c)=0 \,,  \qquad   z(x_c)\equiv z_c \,.
\ee 
Thus, when the extremal surface crosses the shell, the part having $0 \leqslant x < x_c$ is inside the shell (hvLif geometry) and the part with $x_c < x \leqslant  \ell/2$ is outside the shell (black hole geometry). \\
The matching conditions can be obtained in a straightforward way by integrating the differential equations (\ref{eom vaidya v2}) and (\ref{eom vaidya v3}) in a small interval which properly includes $x_c$ and then sending to zero the size of the interval. In this procedure, since both $v(x)$ and $z(x)$ are continuous functions with discontinuous derivatives, only a few terms contribute \cite{Hubeny:2006yu}. 
In particular,  $F_v = - M z^{d_\theta +\zeta} \delta(v)$ is the only term on the r.h.s.'s of (\ref{eom vaidya v2}) and (\ref{eom vaidya v3}) that provides a non vanishing contribution. Thus, considering (\ref{eom vaidya v3}) first, we find the following matching condition
\be 
v'_+=v'_-\equiv v'_c  \,,
\qquad 
\textrm{at $\;x=x_c$}\,.
\label{eq:vpm}
\ee 
Then, integrating across the shell (\ref{eom vaidya v2}) and employing (\ref{eq:vpm}) (we have also used that $\delta(v) =  \delta(x-x_c)/|v'_c|$, where $v'_c >0$, as discussed below), we find (notice that the term containing $v'$ on the l.h.s. provides a non vanishing contribution)
\be 
z_+' - z_-' =
\frac{z_c^{1-\zeta} v'_c}{2} \,\big( 1-F(z_c) \big) \,,
\qquad 
\textrm{at $\;x=x_c$} \,.
\label{eq:matching1}
\ee
\begin{figure}[t] 
\begin{center}
\vspace{-0.cm}
\hspace{-.5cm}
\includegraphics[width=.99\textwidth]{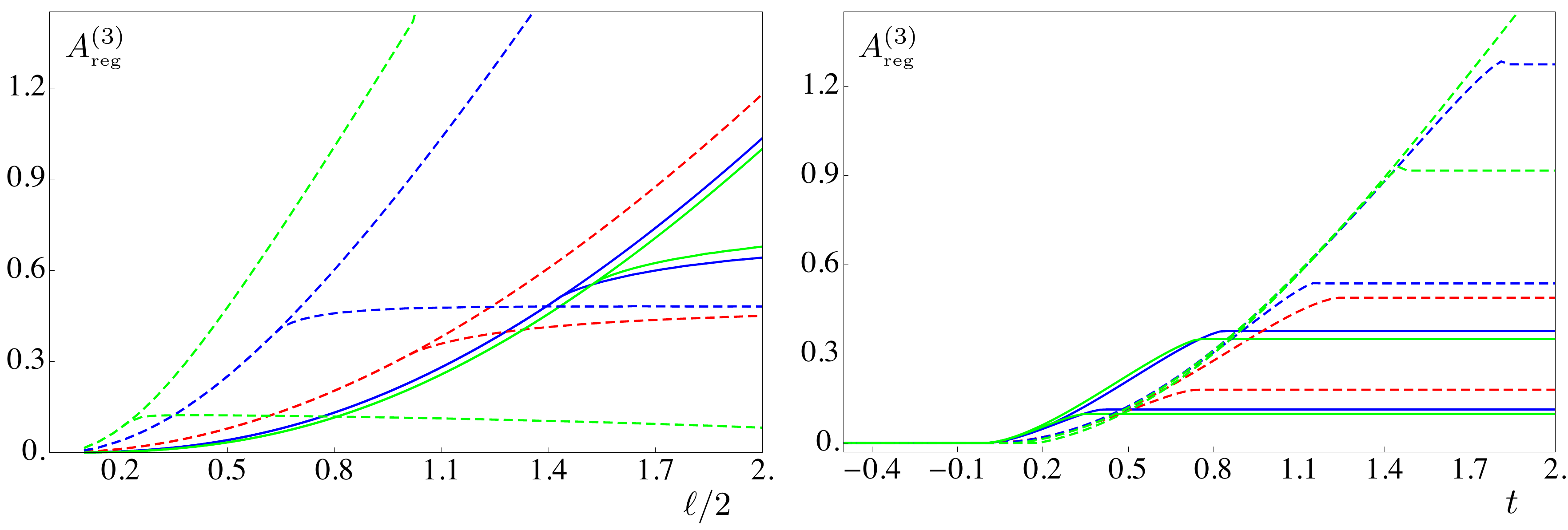}
\end{center}
\vspace{-.3cm}
\caption{\label{fig:A3dim}
Regularized area (\ref{eq:area3}) for the strip in the thin shell regime ($a=0.01$) for the critical value $\theta =d-1$ and $\zeta=2-1/d$ (continuous curves) compared with the corresponding cases without hyperscaling $\theta = 0$ (dashed curves). We plot $d=1$ (red), $d=2$ (blue) and $d=3$ (green).
Left panel: plots at fixed $t=1.5$ (bottom curves) and $t=2.5$ (upper curves). 
Right panel: plots at fixed $\ell=3, 5$ (larger strips have larger variations for $A^{(3)}_{\textrm{\tiny reg}}$). Strips with smaller $\ell$ thermalize earlier.
}
\end{figure}

\begin{figure}[t] 
\begin{center}
\vspace{-0.cm}
\hspace{-.5cm}
\includegraphics[width=.98\textwidth]{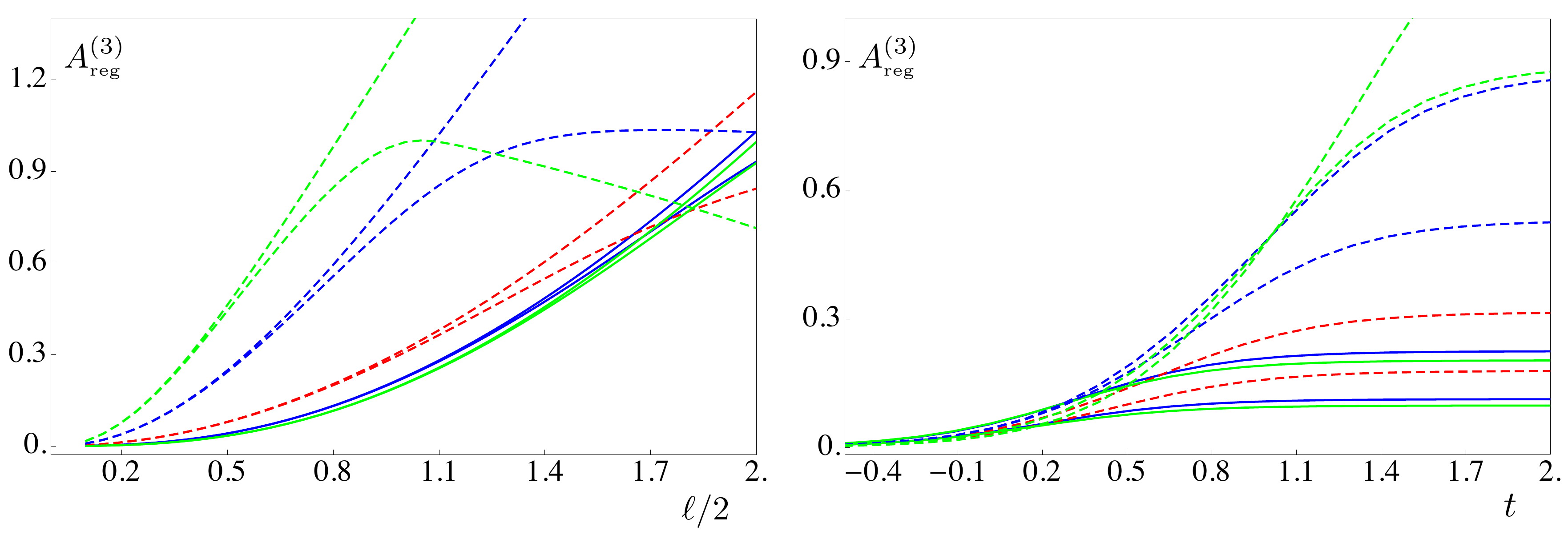}
\end{center}
\vspace{-.3cm}
\caption{\label{fig:A3dima05}
Regularized area (\ref{eq:area3}) for the strip with $a=0.5$.
These plots should be compared with Fig. \ref{fig:A3dim}, because the parameters $d$, $\theta$ and $\zeta$ and the color code are the same.
}
\end{figure}
\begin{figure}[t] 
\begin{center}
\vspace{-0.cm}
\hspace{-.5cm}
\includegraphics[width=.98\textwidth]{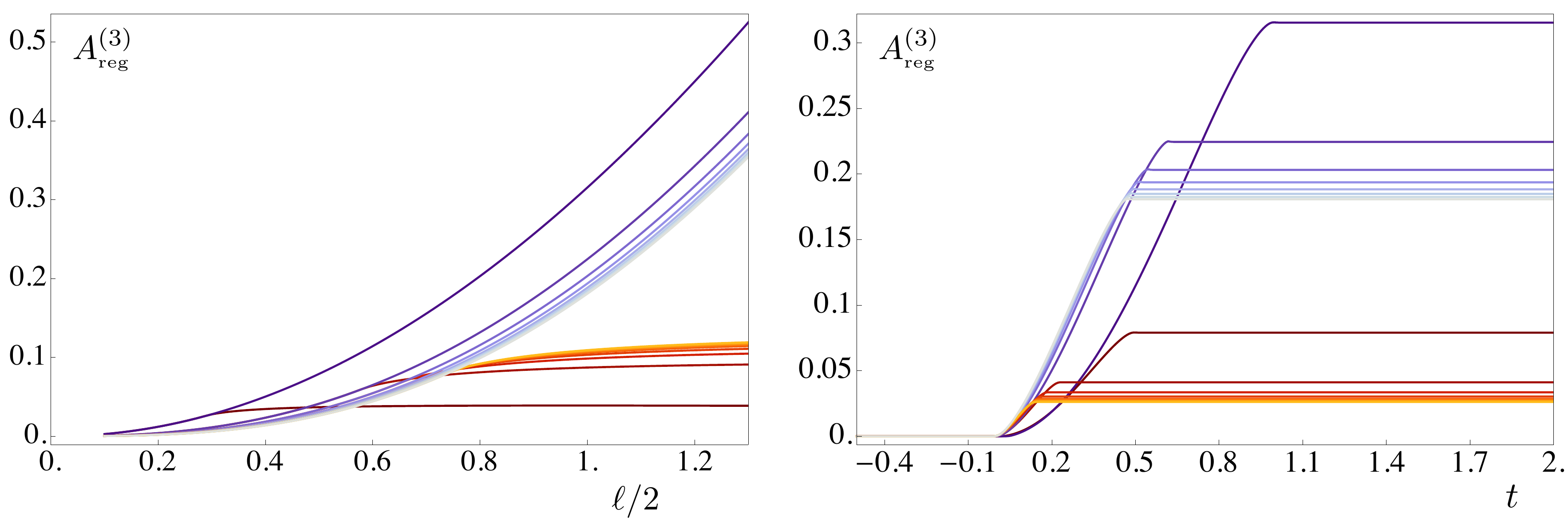}
\end{center}
\vspace{-.3cm}
\caption{\label{fig:A3differentdims}
Regularized area (\ref{eq:area3}) for the strip in the thin shell regime ($a=0.01$.) with $\theta=d-1$ and $\zeta = 2-1/d$ for various dimensions $d=1, 2, 3, \dots, 8$. 
The darkest curve within each group has $d=1$ and the brightest one has $d=8$.
Left panel: the red curves have $t=0.15$ and the blue ones have $t=0.7$.
Right panel: the red curves have $\ell=1$ and the blue ones have $\ell=2$.
}
\end{figure}

Since $F_v$ vanishes for $v \neq 0$, the differential equation (\ref{eom vaidya v2}) tells us that 
\be
\label{E def}
z^{1-\zeta} \left(v'z^{1-\zeta}F+z'\right) = \textrm{const}
\equiv 
\left\{ \begin{array}{lll}
E_{-} &   0 \leqslant x < x_c & \textrm{hvLif}\,,\\
\rule{0pt}{.4cm}
E_{+} &   x_c < x \leqslant \ell/2 & \textrm{black hole}\,.
\end{array}
\right.
\ee
Let us consider the hvLif part ($v<0$) first, where $F = 1$. 
Since $v'(0) = 0$ and $z'(0)=0$, (\ref{E def}) tells us that $E_{-} = 0$. Thus, (\ref{E def}) implies that 
\be
\label{vprime ads}
v' = - \,z^{\zeta -1} z'\,,
\qquad
0 \leqslant x < x_c \,.
\ee
Plugging this result into (\ref{eq:striphamvaidya}) with $F=1$, it reduces to the square of (\ref{eq:striphsdiff2}) with $F=1$, as expected.
Taking the limit $x \rightarrow x_c^{-}$ of (\ref{vprime ads}), one finds a relation between the constant value $v'_c$ defined in (\ref{eq:vpm}) and $z'_-$, i.e.
\be
\label{vprimec}
v'_c = - \,z_c^{\zeta -1} z'_{-} > 0\,,
\ee
where we have used that $z'_- < 0$.
Integrating (\ref{vprime ads}) from $x=0$ to $x=x_c$, we obtain that
\be
z_c^\zeta= z_\ast^\zeta + \zeta v_\ast\,.
\label{eq:zetac}
\ee
Now we can consider the region outside the shell ($v>0$), where the geometry is given by the black hole.
From (\ref{E def}) with $F=F(z)$ given in (\ref{eq:emblackeningBH}) we have that
\be
\label{vprime bh region}
v' = \frac{1}{z^{1-\zeta}F(z)} \, \bigg( \frac{E_+}{z^{1-\zeta}} - z'\bigg)\,,
\qquad
x_c < x \leqslant \ell/2 \,.
\ee
Then, plugging this result into (\ref{eq:striphamvaidya}), one gets
\be
\label{zprime bh Eplus}
z'^2 = 
F(z)\bigg[
\bigg( \frac{z_\ast}{z}\bigg)^{2d_\theta}-1\bigg]
+\frac{E^2_+}{z^{2(1-\zeta)}} \,,
\qquad
x_c < x \leqslant \ell/2\,.
\ee
We remark that (\ref{zprime bh Eplus}) becomes (\ref{eq:striphsdiff2}) when $E_+ = 0$.
The constant $E_+$ can be related to $z'_-$ by taking the difference between the equations in (\ref{E def}) across the shell. By employing (\ref{eq:vpm}), the result reads
\be
E_+ - E_- = 
z_c^{1-\zeta} \big[ 
z'_+ - z'_-
+ z_c^{1-\zeta} v'_c  \big(F(z_c) -1\big) 
\big]\,.
\ee
Then, with $E_- = 0$, the matching conditions (\ref{eq:matching1}) and (\ref{vprimec}) lead to
\be 
\label{Eplus soln}
E_+= \frac{z_c^{1-\zeta} }{2} \big(1-F(z_c)\big) z_-'  \,,
\ee
where $E_+ < 0$ because of (\ref{vprime ads}). Moreover, from  (\ref{eq:striphamvaidya}), one finds that 
\be
\mathcal{B}_+=\mathcal{B}_-= \bigg(\frac{z_\ast}{z_c}\bigg)^{2d_\theta}\,,
\qquad 
\textrm{at $\;x=x_c$} \,.
\ee
Finally, the size $\ell$ can be expressed in terms of the profile function $z(x)$ (we recall that $z' < 0$) by summing the contribution inside the shell (from (\ref{zprime bh Eplus})  with $F(z)=1$) and the one outside the shell (from (\ref{zprime bh Eplus}))
\be 
\label{elle shell tot}
\frac{\ell}{2} = 
\int_{z_c}^{z_\ast}  
z^{d_\theta} \big( z_\ast^{2d_\theta} - z^{2d_\theta} \big)^{-1/2} dz
+
\int_0^{z_c}
\bigg\{ 
F(z)\bigg[
\bigg( \frac{z_\ast}{z}\bigg)^{2d_\theta}-1\bigg]
+\frac{E^2_+}{z^{2(1-\zeta)}}
\bigg\}^{-1/2} dz\,.
\ee
Notice that we cannot use (\ref{zprime bh Eplus}) for the part outside the shell because $E_+ \neq 0$.
Similarly, we can find the boundary time $t$ by considering first (\ref{bc vaidya}) and (\ref{eq:cquantities}), and then employing (\ref{vprime bh region}).
We find 
\be
\label{t shell tot}
t = \int_0^t dv = \int_{x_c}^{\ell/2} v' dx 
= \int_0^{z_c} 
\frac{z^{\zeta-1}}{F(z)} 
\Bigg[  1+ E_+ z^{\zeta -1 } 
\bigg\{ 
F(z)\bigg[
\bigg( \frac{z_\ast}{z}\bigg)^{2d_\theta}-1\bigg]
+\frac{E^2_+}{z^{2(1-\zeta)}}
\bigg\}^{-1/2}
\, \Bigg] dz\,,
\ee
where in the last step (\ref{vprime bh region}) and (\ref{zprime bh Eplus}) have been used
(we recall that $z' < 0$).\\
The area of the extremal surface (\ref{area vaidya onshell}) is obtained by summing the contributions inside and outside the shell in a similar manner. The result is
\be
\label{A shell tot}
\mathcal{A} = 2\ell_\perp^{d-1} 
z_\ast^{d_\theta}
\Bigg(
\int_{z_c}^{z_\ast}  
z^{-d_\theta} \big( z_\ast^{2d_\theta} - z^{2d_\theta} \big)^{-1/2}dz
+
\int_\epsilon^{z_c}
z^{-2d_\theta}  \bigg\{ 
F(z)\bigg[
\bigg( \frac{z_\ast}{z}\bigg)^{2d_\theta}-1\bigg]
+\frac{E^2_+}{z^{2(1-\zeta)}}
\bigg\}^{-1/2}dz
\Bigg)\,,
\ee
where the cutoff $\epsilon$ must be introduced to regularize the divergent integral, as already discussed. 
In Fig. \ref{fig:A3differentdims} we show $A^{(3)}_{\textrm{\tiny reg}}$ for various dimensions. It seems that a limiting curve is approached as $d$ increases.

It is straightforward to generalize the above analysis to the case of $n$ dimensional surfaces extended in the bulk which share the boundary with an $n$ dimensional spatial surface in the boundary, i.e. surfaces with higher codimension than the extremal surface occurring for the holographic entanglement entropy.
For a strip whose sides have length 
$\ell$ in one direction and $\ell_\perp$ in the remaining $n-1$ ones, the area functional to be extremized reads 
\be
\label{area func vaidya n-dim}
\mathcal{A}[v(x), z(x)]
=2 \ell_\perp^{n-1}
\int_{0}^{\ell/2}
\frac{\sqrt{\mathcal{B}}}{z^{n d_\theta/d}}
\,dx\,,
\ee
where $\mathcal{B}$ has been defined in (\ref{area func vaidya}). This functional reduces to the one in (\ref{area func vaidya}) for the holographic entanglement entropy when $n=d$.
The extrema of the functional (\ref{area func vaidya n-dim}) with $n=2$ are employed to study the holographic counterpart of the spacelike Wilson loop, while the $n=1$ case describes the holographic two point function. \\
The equations of motion of (\ref{area func vaidya n-dim}) are simply given by (\ref{eom vaidya v2}) and (\ref{eom vaidya v3}) where the $d_\theta$ in the r.h.s. of  (\ref{eom vaidya v3}) is replaced by $n d_\theta/d$, while $F(v,z)$ is kept equal to (\ref{Ffactor vaidya}).
Similarly, we can adapt all the formulas within \S\ref{stripvaidyasec} to the case $n \neq d$ by replacing $d_\theta$ by $n d_\theta/d$ whenever it does not occur through $F(v,z)$ or $F(z)$, which remain equal to (\ref{Ffactor vaidya}) and (\ref{eq:emblackeningBH}) respectively.


\subsection{Sphere }
\label{sec sphere hee}

Let us consider a circle of radius $R$ in the boundary of the asymptotically hvLif spacetime. As discussed in \S\ref{sec sphere static} for the static case, it is more convenient to adopt spherical coordinates in the Vaidya metric (\ref{vaidya metric}) for $\mathbb{R}^d$. The area functional is given by
\be
\label{area func vaidya sphere}
\mathcal{A}[v(\rho), z(\rho)]
=
 \frac{2\pi^{d/2}}{\Gamma(d/2)}
 \int_{0}^{R}
 \frac{\rho^{d-1}}{z^{d_\theta}}\sqrt{\mathcal{B}}\, d\rho\,,
 \qquad
 \mathcal{B} \equiv 1- F(v,z) z^{2(1-\zeta)} v'^{2} -2z^{1-\zeta} z'v'\,,
\ee
where now the prime denotes the derivative w.r.t. $\rho$. An important difference compared to the strip, as already emphasized for the static case, is that the Lagrangian of (\ref{area func vaidya sphere}) depends explicitly on $\rho$. This implies that we cannot find an integral of motion which allows to get a first order differential equation to describe the extremal surface.
Thus, we have to deal with the equations of motion, which read
\bea
\label{eomsp vaidya v2}
& &
\frac{z^{d_\theta} \sqrt{\mathcal{B}}}{\rho^{d-1}}
\,\partial_{\rho} 
\bigg[\,
  \frac{\rho^{d-1} z^{1-\zeta-d_\theta}}{\sqrt{\mathcal{B}}}
  \, (v'z^{1-\zeta}F+z' )
\bigg] =
\frac{z^{2(1-\zeta)} }{2} \, F_v  v'^{2}\,,
 \\
\label{eomsp vaidya v3}
& &
\frac{z^{d_\theta} \sqrt{\mathcal{B}}}{\rho^{d-1}}
\,\partial_{\rho} 
\bigg[\,
\frac{\rho^{d-1} z^{2(1-\zeta) - d_\theta}}{\sqrt{\mathcal{B}}} \, v'
\bigg] =
 \frac{d_\theta}{z} \, \mathcal{B} 
 +\frac{z^{2(1-\zeta)} }{2} \, F_z v'^2 
 + \frac{1-\zeta}{z^{\zeta}} (z'+z^{1-\zeta}F v') v' \,.
\eea
These equations have to be supplemented by the following boundary conditions
\be
v(R) = t \,,\qquad v'(0) = 0\,,
\hspace{1cm} \textrm{and} \hspace{1cm} 
z(R) = 0 \,, \qquad z'(0) = 0 \,.
\ee
We are again mainly interested in the limiting case of a thin shell (\ref{M thin shell}).


\begin{figure}[t] 
\begin{center}
\vspace{-0.cm}
\hspace{-.5cm}
\includegraphics[width=.98\textwidth]{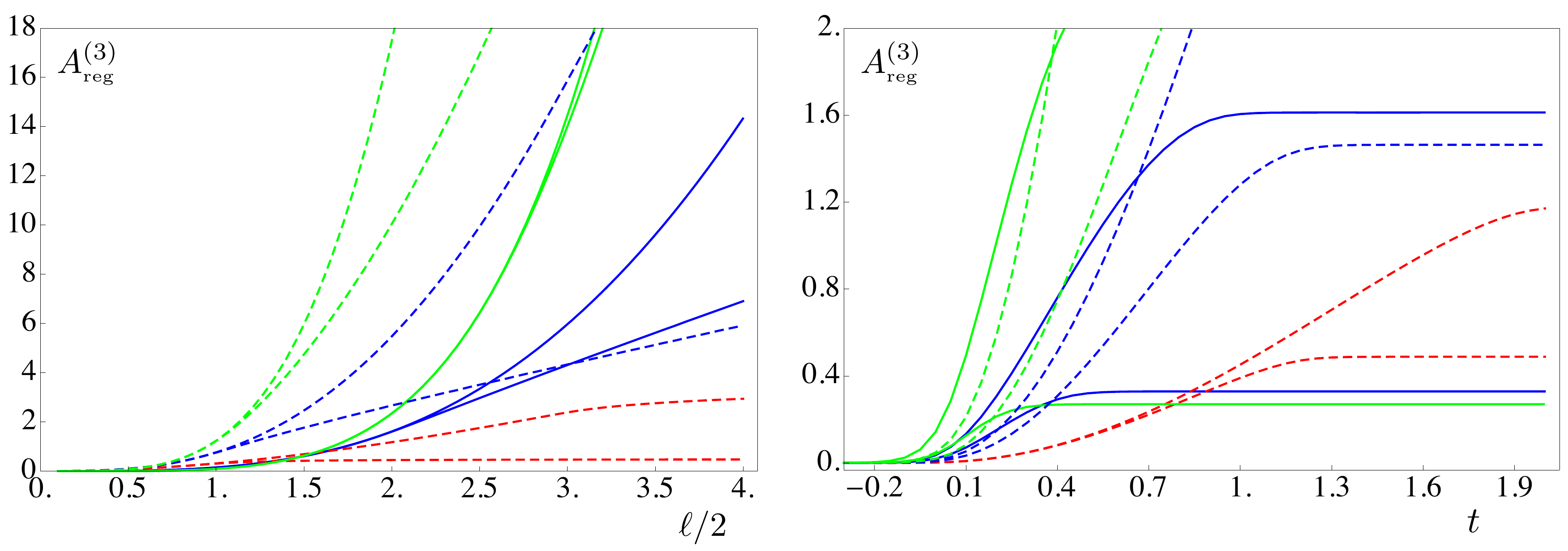}
\end{center}
\vspace{-.3cm}
\caption{\label{fig:A3dima001sphere}
Holographic entanglement entropy for the sphere in the thin shell regime with $a=0.01$ (see \S\ref{sec sphere hee}).
The parameters $d$, $\theta$ and $\zeta$ are the same of Fig. \ref{fig:A3dim} (same color coding). Left panel: fixed $t=1.5$ (lower curve) and $t=3$ (upper curve). Right panel: fixed $R=2$ and $R=4$ (larger spheres thermalize later).
}
\end{figure}

\subsubsection{Thin shell regime}

Considering the thin shell regime, defined by (\ref{M thin shell}), we can adopt  to the sphere some of the observations made in \S\ref{ThinStrip} for the strip. 
Again, there is a value $\rho_c$ such that for $0 \leqslant \rho < \rho_c$ the extremal surface is inside the shell (hvLif geometry), while for $\rho_c < \rho \leqslant R$ it is outside the shell (black hole geometry). \\
The matching conditions can be found by integrating (\ref{eomsp vaidya v2}) and (\ref{eomsp vaidya v3}) across the shell, as was done in \S\ref{ThinStrip} for the strip. 
Introducing
\be
\label{primed bar def}
\check{v}' \equiv \frac{v'}{\sqrt{\mathcal{B}}} \,,
\qquad
\check{z}' \equiv \frac{z'}{\sqrt{\mathcal{B}}} \,,
\ee 
we can use (\ref{eomsp vaidya v3}), whose r.h.s. does not contain $F_v$, to obtain
\be 
\check{v}'_+ = \check{v}'_-\,,
\qquad 
\textrm{at $\;\rho=\rho_c$}\,,
\label{eq:vpm sphere}
\ee 
while from (\ref{eomsp vaidya v2}) and employing (\ref{eq:vpm sphere}) as well, we get
\be
\label{eq:matching1sphere}
\check{z}_+' - \check{z}_-' =
\frac{z_c^{1-\zeta} \, \check{v}'_c}{2} \,\big( 1-F(z_c) \big) \,,
\qquad 
\textrm{at $\;x=x_c$} \,.
\ee
Considering (\ref{eomsp vaidya v2}), since $F_v = 0$ for $v\neq 0$, we have
\be
\label{E def sphere}
  \frac{\rho^{d-1} z^{1-\zeta-d_\theta}}{\sqrt{\mathcal{B}}}
  \, (v'z^{1-\zeta}F+z' )
= \textrm{const}
\equiv 
\left\{ \begin{array}{lll}
E_{-} &   0 \leqslant \rho < \rho_c & \textrm{hvLif}\,,\\
\rule{0pt}{.4cm}
E_{+} &   \rho_c < \rho \leqslant R & \textrm{black hole}\,,
\end{array}
\right.
\ee
where $E_- = 0$ because $v'(0)=0$ and $z'(0)=0$.
By using (\ref{primed bar def}), one can write
\bea
\label{eq:oneoverbplus} 
1/\mathcal{B}_+ &=& 
1+\check{v}'_+ z_c^{(1-\zeta)}\big[z_c^{(1-\zeta)} \check{v}'_+F(z_c)+2 \check{z}'_+\big] \,, \\
\label{eq:oneoverbminus} 
1/\mathcal{B}_- &=& 
1+\check{v}'_- z_c^{(1-\zeta)}(z_c^{(1-\zeta)} \check{v}'_-+2 \check{z}'_-)\,.
\eea
Taking the difference of these expressions and using (\ref{eq:vpm sphere}) and (\ref{eq:matching1sphere}), one finds
\be 
\mathcal{B}_+=\mathcal{B}_-\,.
\ee
By using (\ref{eq:vpm sphere}), (\ref{eq:matching1sphere}) and (\ref{E def sphere}), we get
\be 
\label{Eplus sphere}
E_+=\frac{\rho_c^{d-1} z_c^{2(1-\zeta)-d_\theta}}{2\sqrt{\mathcal{B}_+}}(F(z_c)-1)v'_c \,.
\ee
Then, from (\ref{E def sphere}) in the black hole region, one obtains
\be 
\label{eq:vprimesphere}
v'=\frac{z^{\zeta-1}}{F(z)}
\left(\frac{A E_+\sqrt{1+z'^2/F(z)}}{\sqrt{1+A^2 E^2/F(z)}} -z' \right) \,,
\qquad 
A\equiv\frac{z^{d_\theta+\zeta-1}}{\rho^{d-1}}\,.
\ee
Plugging this expression into (\ref{eomsp vaidya v3}) leads to
\bea
\label{eq:spherethineqz}
& & \hspace{-3cm}
2 d_\theta \rho F^{2}
+z\big[ \rho F_z-2(d-1)z' \big] z'^{2}
-2F\big[ \rho\, z\, z''+(d-1)z\, z'+d_\theta \rho\, z'^{2}\big]
\\
& & \hspace{2cm}
+ \,E_+^2 A^2 \rho \big[ z (F_z+2z'')-2(\zeta-1)(F+z'^2) \big]
=0 \,,
\nonumber 
\eea
which reduces to  (\ref{eq:sferatot}) when $E_+ = 0$, as expected.
The boundary time $t$ is obtained by integrating (\ref{eq:vprimesphere}) outside the shell 
$\rho_c \leqslant \rho < R$ (see e.g. (\ref{t shell tot}) for the strip)
\be
\label{bdy time sphere}
t=\int_{\rho_c}^{R} 
\frac{z^{\zeta-1}}{F(z)}
\left(\frac{A E_+\sqrt{1+z'^2/F(z)}}{\sqrt{1+A^2 E_+^2/F(z)}} -z' \right) 
d\rho\,.
\ee
Notice that we cannot provide a similar expression for $R$, like we did for the strip in (\ref{elle shell tot}).
Finally, the area of the extremal surface at time $t$ is the sum of two contributions, one inside (finite) and one outside (infinite) the shell, and is given by
\be
\label{area total sphere}
\mathcal{A} = \frac{2\pi^{d/2}}{\Gamma(d/2)}
\left(
\int_{0}^{\rho_c} \frac{\rho^{d-1}\, \sqrt{1+z'^2}}{z^{d_\theta}}\, d\rho
+
\int_{\rho_c}^{R}d\rho 
\frac{\rho^{d-1}\, \sqrt{1+ z'^2/F(z)}}{z^{d_\theta}\,\sqrt{1+ A^2 E_+^2/F(z)}} 
\right) \,.
\ee
Numerical results for the regularized extremal area $A^{(3)}_{\textrm{\tiny reg}}$ for a sphere (defined via an appropriate adaptation of (\ref{eq:area3})) in the thin shell regime are shown in Fig. \ref{fig:A3dima001sphere}.


\begin{figure}[t] 
\begin{center}
\vspace{-0.cm}
\hspace{-.5cm}
\includegraphics[width=1\textwidth]{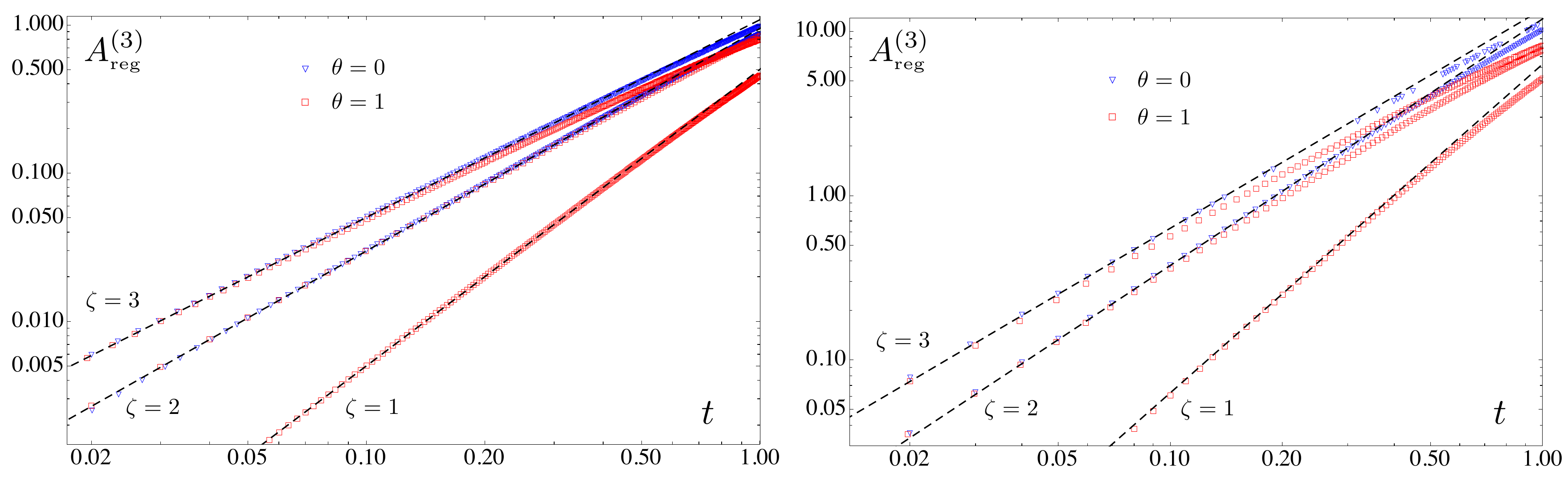}
\end{center}
\vspace{-.3cm}
\caption{\label{fig:initialgrowth}
Initial growth of the holographic entanglement entropy for $d=2$ (see \S\ref{sec initial growth}). 
The points come from the numerical solution of (\ref{eom vaidya v2})-(\ref{eom vaidya v3}) for the strip (left) and (\ref{eomsp vaidya v2})-(\ref{eomsp vaidya v3}) for the sphere (right) in the thin shell regime. The black dashed lines are obtained through the formula (\ref{eq:quadgrowth}), which is independent of $\theta$ and of the shape of the region in the boundary.
Left panel: strip with $\ell=4$. Right panel: sphere with $R=4$.
}
\end{figure}

\section{Regimes in the growth of the holographic entanglement entropy}

In this section we extend the analysis performed in \cite{Liu:2013iza,Liu:2013qca} to $\theta \neq 0$ and $\zeta \neq 1$. For $t<0$ we have $A_{\textrm{\tiny reg}}^{(3)}=0$ because the background is hvLif. 
When $t>0$, it is possible to identify three regimes: an initial one, when the growth is characterized by a power law, an intermediate regime where the growth is linear and a final regime, when $A_{\textrm{\tiny reg}}^{(3)}(t)$ saturates to the thermal value.
We report our results for the different regimes in the main text while the details of the computation are described in Appendix \S\ref{app computational details}.

\begin{figure}[t] 
\begin{center}
\vspace{-0.cm}
\hspace{-.5cm}
\includegraphics[width=.7\textwidth]{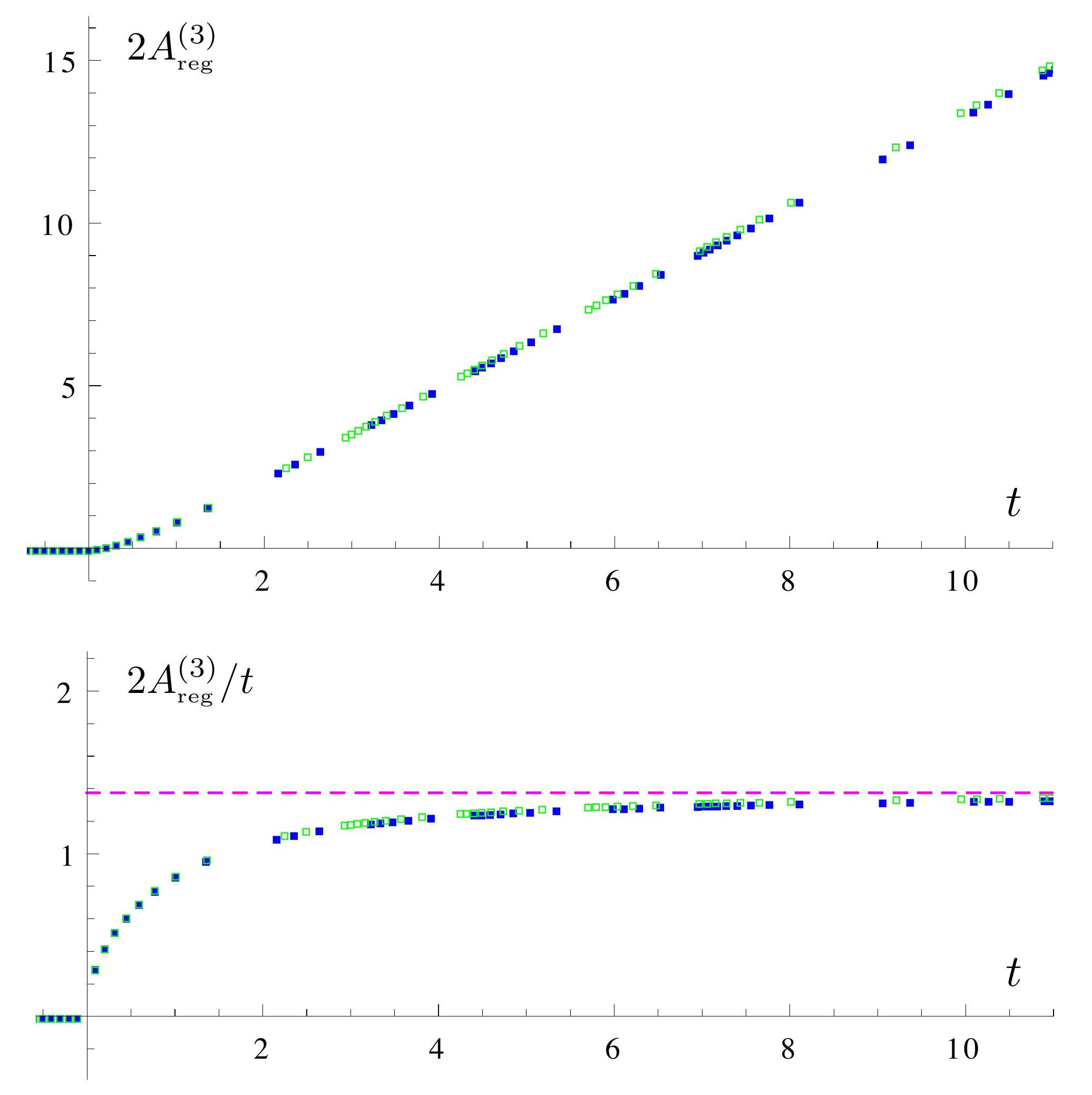}
\end{center}
\vspace{-.3cm}
\caption{\label{fig:lineargrowth}
Typical example of linear growth for the holographic entanglement entropy in the thin shell regime. Here $d=2$, $z_h=1$, $\theta =1$ and $\zeta=2$ for two large strips: $\ell \sim 16$ (green squares) and $\ell \sim 20$ (blue squares). In the bottom panel, the dashed line is obtained through (\ref{linear growth formula}) and (\ref{vE def}). 
}
\end{figure}

\subsection{Initial growth}
\label{sec initial growth}

The initial regime is characterized by times that are short compared to the horizon scale
\be
0 <  t \ll z_h \,.
\ee
In Appendix \S\ref{app sec initial}, following \cite{Liu:2013qca}, we expand $A^{(3)}_{\textrm{\tiny reg}}$ around $t=0$ and consider the first non trivial order for an $n$ dimensional spatial region whose boundary $\Sigma$ has a generic shape. Given the metric (\ref{vaidya metric}) with (\ref{Fvz vaidya}), the final result for this regime is (see (\ref{eq:variationq})) 
\be 
\label{eq:A2oftn}
\mathcal{A}^{(3)}_{\textrm{\tiny reg}}(t)
=  \frac{M A_\Sigma (\zeta t)^{[d_\theta(1-n/d)+\zeta+1]/\zeta}}{2[d_\theta(1-n/d)+\zeta+1]} \,,
\ee
where $A_\Sigma$ is the area of $\Sigma$. Notice that for the holographic entanglement entropy $n=d$, for the holographic counterpart of the Wilson loop $n= 2$ and for the holographic two point function $n= 1$. Explicitly, for the holographic entanglement entropy, (\ref{eq:A2oftn}) becomes 
\be 
\label{eq:quadgrowth}
\mathcal{A}^{(3)}_{\textrm{\tiny reg}}(t)
= \frac{M A_\Sigma \,\zeta^{1+1/\zeta} }{2(\zeta +1)}  
\;  t^{1+1/\zeta} \,,
\ee
which is independent of $d$ and $\theta$.
This generalizes the result of \cite{Liu:2013qca} (see \cite{Hubeny:2013hz} for $d=1$).
In Fig. \ref{fig:initialgrowth} we show some numerical checks of (\ref{eq:quadgrowth}) both for the strip and for the sphere.

\subsection{Linear growth}

\begin{figure}[t] 
\begin{center}
\vspace{-0.cm}
\hspace{-.5cm}
\includegraphics[width=.7\textwidth]{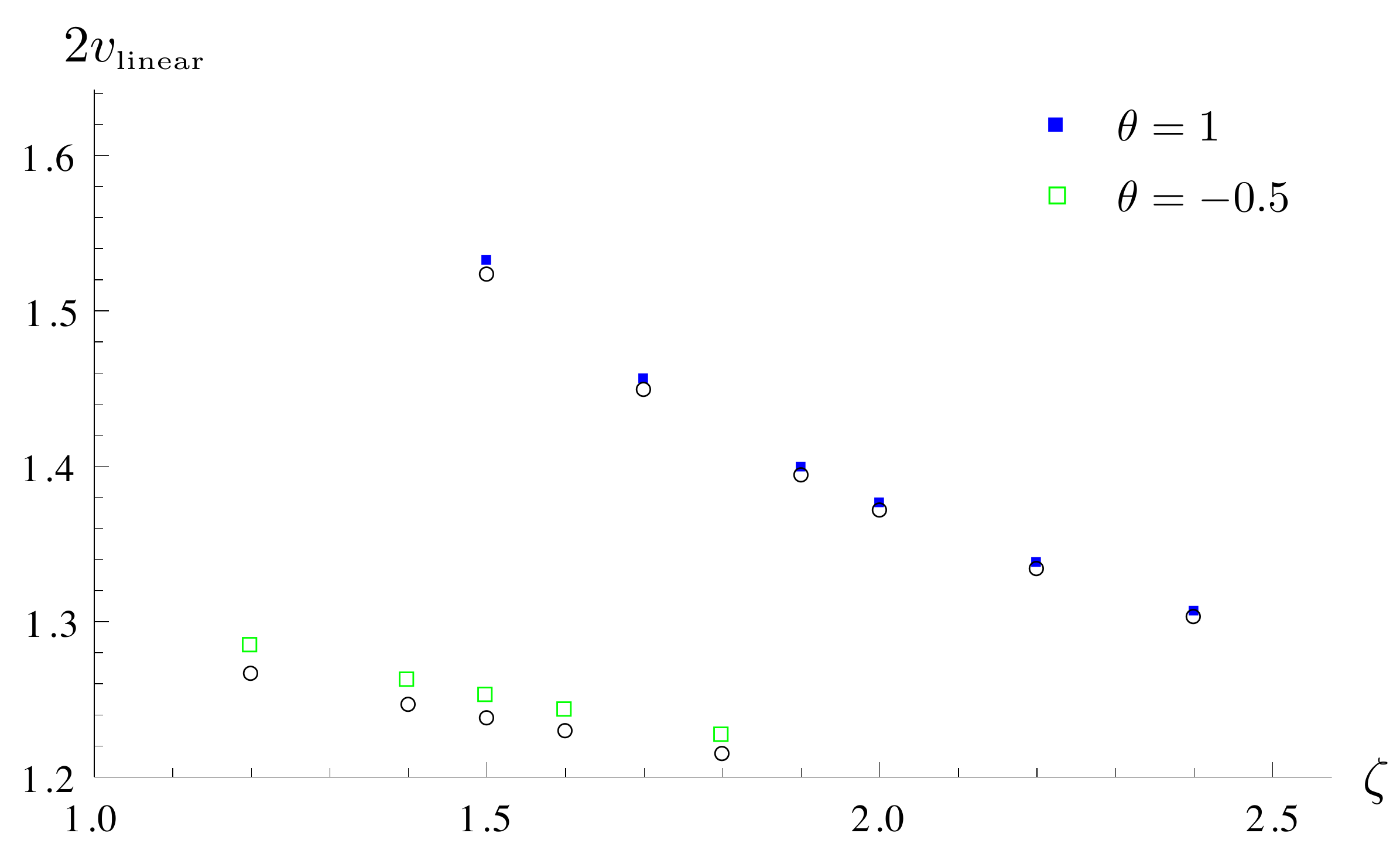}
\end{center}
\vspace{-.3cm}
\caption{\label{fig:vlinear}
Linear regime for the strip: the colored squares are values of the slope (see (\ref{vlinear def})) found from the numerical data as in the bottom panel of Fig. \ref{fig:lineargrowth}.  
The black empty circles denote the corresponding results of $v_E$ from (\ref{vE def}).
In this plot $z_h=1$.
}
\end{figure}

When $z_\ast$ is large enough, the holographic entanglement entropy displays a linear growth in time. The computational details for the strip are explained in Appendix \S\ref{linearstrip}. 
The result for (\ref{Fvz vaidya}) is that, in the regime given by
\be 
z_h\ll t \ll \ell\,,
\ee
and if the following condition is satisfied
\be 
d_\theta \geqslant 2- \zeta\,,
\label{eq:lineargro3}
\ee
we find a linear growth in time for the holographic entanglement entropy, namely
\be
\label{vlinear def}
\mathcal{A}^{(3)}_{\textrm{\tiny reg}}(t) \equiv 
2\ell_\perp^{d-1} v_{\textrm{\tiny linear}} \, t \,.
\ee
The method of \cite{Liu:2013qca} for the thin shell regime, extended to $\theta \neq 0$ and $\zeta \neq 1$, tells us that
\be
\label{linear growth formula}
\mathcal{A}^{(3)}_{\textrm{\tiny reg}}(t)=
2\ell_\perp^{d-1} A^{(3)}_{\textrm{\tiny reg}}(t)\,,
\qquad
A^{(3)}_{\textrm{\tiny reg}}(t)
=\frac{\sqrt{-F(z_m)}}{z_m^{d_\theta+\zeta-1}}\,t
\,\equiv \frac{v_E}{z_h^{d_\theta+\zeta-1}}\, t \,,
\ee
where, for $F(z)$ given by (\ref{eq:emblackeningBH}), $v_E$ reads
\be
\label{vE def}
v_E
=
\frac{(\eta-1)^{\frac{\eta-1}{2}}}{\eta^{\frac{\eta}{2}}}\,,
\qquad
\eta=\frac{2(d_\theta+\zeta-1)}{d_\theta +\zeta} \,.
\ee
It can be easily seen that $v_E = 1$ when $\eta =1$ and $v_E \rightarrow 0$ as $\eta \rightarrow +\infty$ monotonically. Notice that the linear regime depends only on the combination $d_\theta + \zeta $.
In Fig. \ref{fig:lineargrowth}, where the points are computed using the numerical solutions of (\ref{eom vaidya v2}) and (\ref{eom vaidya v3}), we see a typical linear behavior in time for two strips with large $\ell$. The agreement between the slope of the numerical data and the value computed from (\ref{vE def}) is quite good.
In Fig. \ref{fig:vlinear} we compare the slopes of the numerical curves with the values obtained from (\ref{vE def}) for other values of $\theta$ and $\zeta$. We consider the linear growth regime for more generic backgrounds in Appendix \ref{app generic metrics} In order to get a better understanding of the origin of the $\zeta$ dependence in (\ref{linear growth formula}).

\subsection{Saturation}

\begin{figure}[t] 
\begin{center}
\vspace{-0.cm}
\hspace{-.5cm}
\includegraphics[width=.7\textwidth]{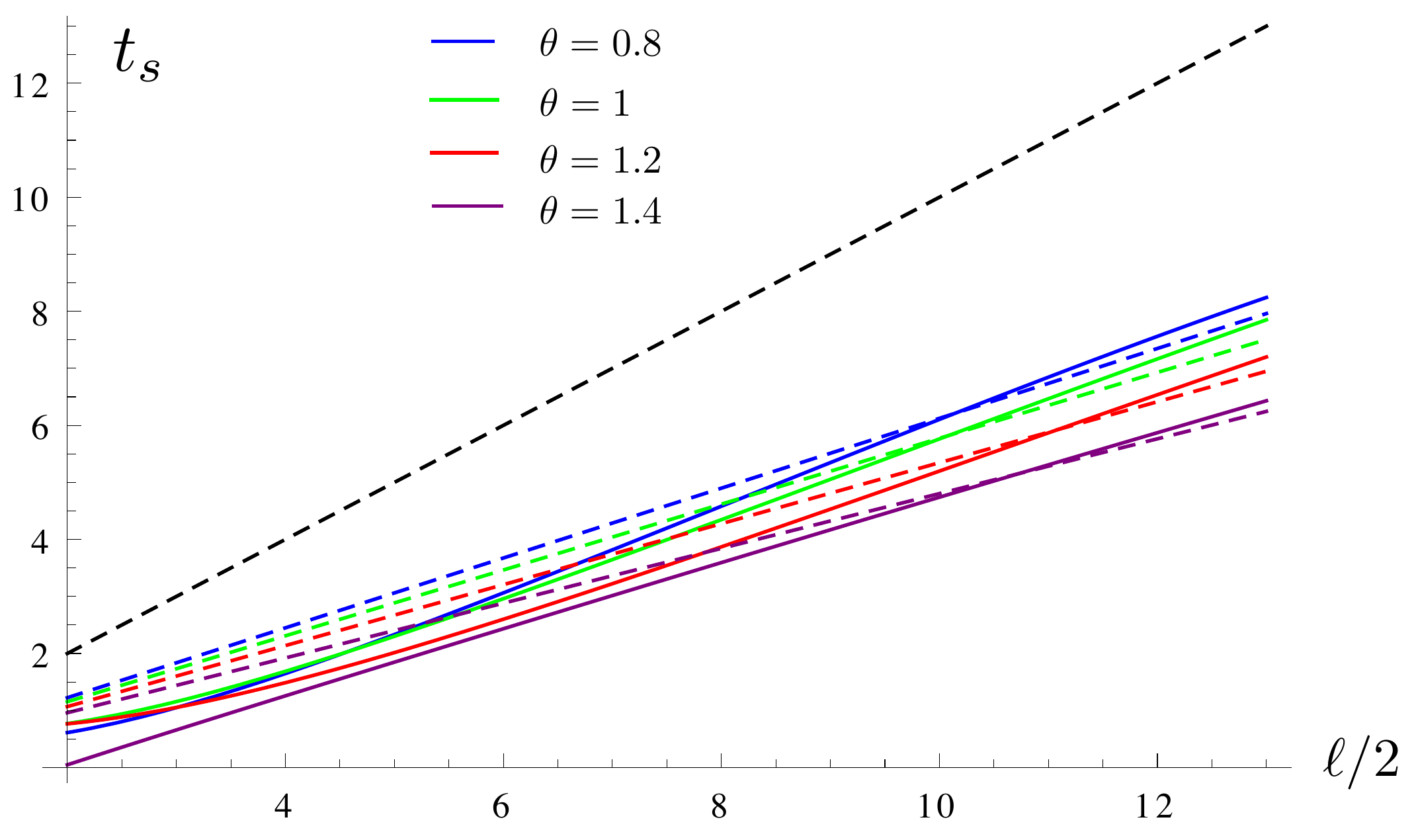}
\end{center}
\vspace{-.3cm}
\caption{\label{fig:tsgeod}
Saturation time as a function of the transverse length scale $\ell$ for geodesic correlators. The dashed black line is a reference line with slope equal to 1, while the colored ones are obtained through (\ref{ts strip text n}) with $n=1$, $\zeta =2$ and the corresponding values of $\theta$ indicated in the legend. The agreement improves for large $\ell$.}
\end{figure}

\begin{figure}[t] 
\begin{center}
\vspace{.5cm}
\hspace{-.5cm}
\includegraphics[width=.81\textwidth]{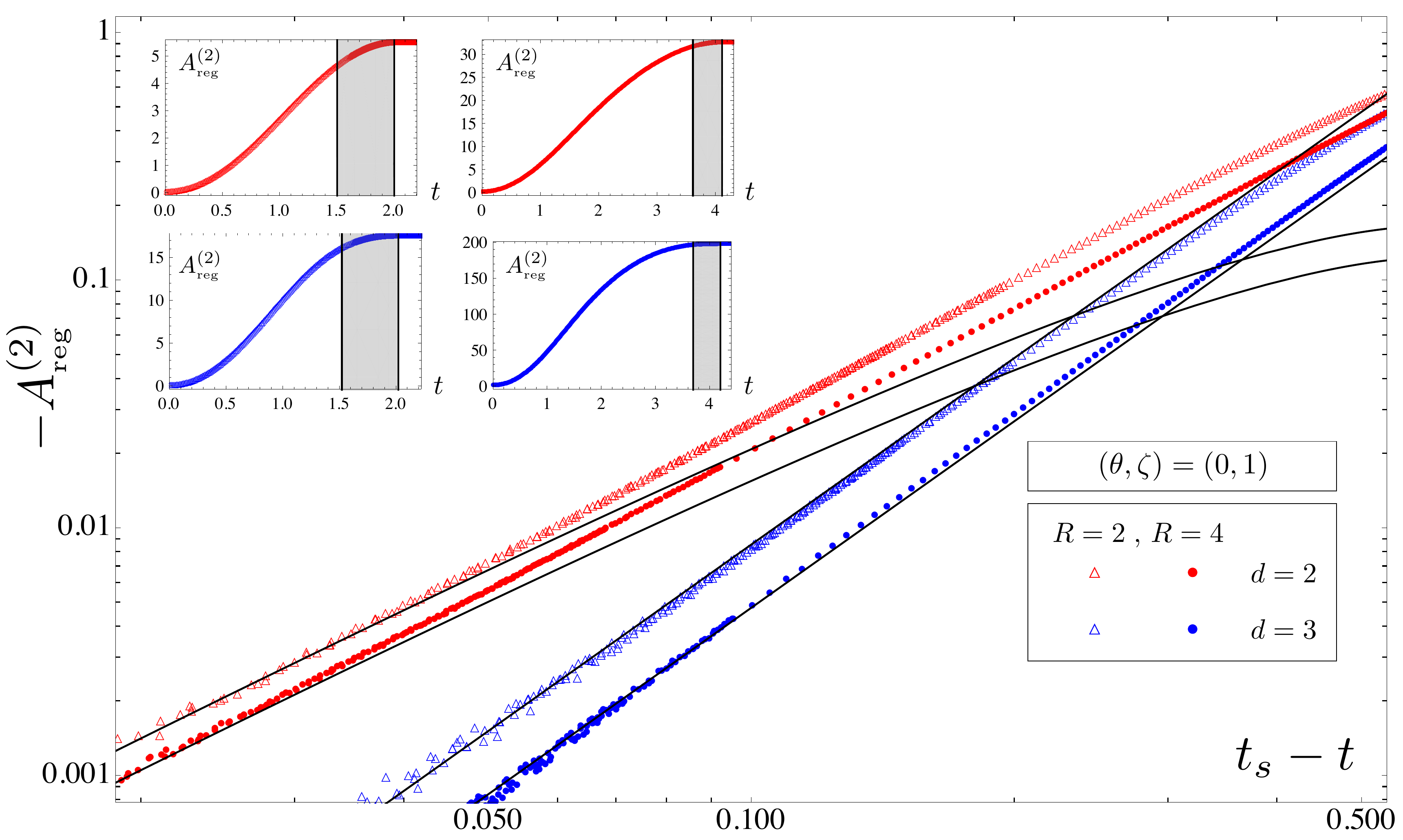}
\end{center}
\vspace{-.3cm}
\caption{
\label{fig:sat1}
Saturation regime for the holographic entanglement entropy in the thin shell limit ($a=10^{-4}$) for a spherical region of radius $R$ in the boundary. In this plot $\theta=0$ and $\zeta=1$, which is the situation considered in \cite{Liu:2013iza, Liu:2013qca}. The continuos black curves are obtained through (\ref{A2reg saturation sphere}) with the corresponding values of $d$.
The inset shows the entire sets of data describing the complete evolution of the four cases considered (the plots are shown in the same positions of the corresponding points in the legend). The gray regions have $\Delta t=0.5$ and show the parts of the curves which have been reported in the main plot.
}
\end{figure}

\begin{figure}[t]
\begin{center}
\vspace{0.cm}
\includegraphics[width=.81\textwidth]{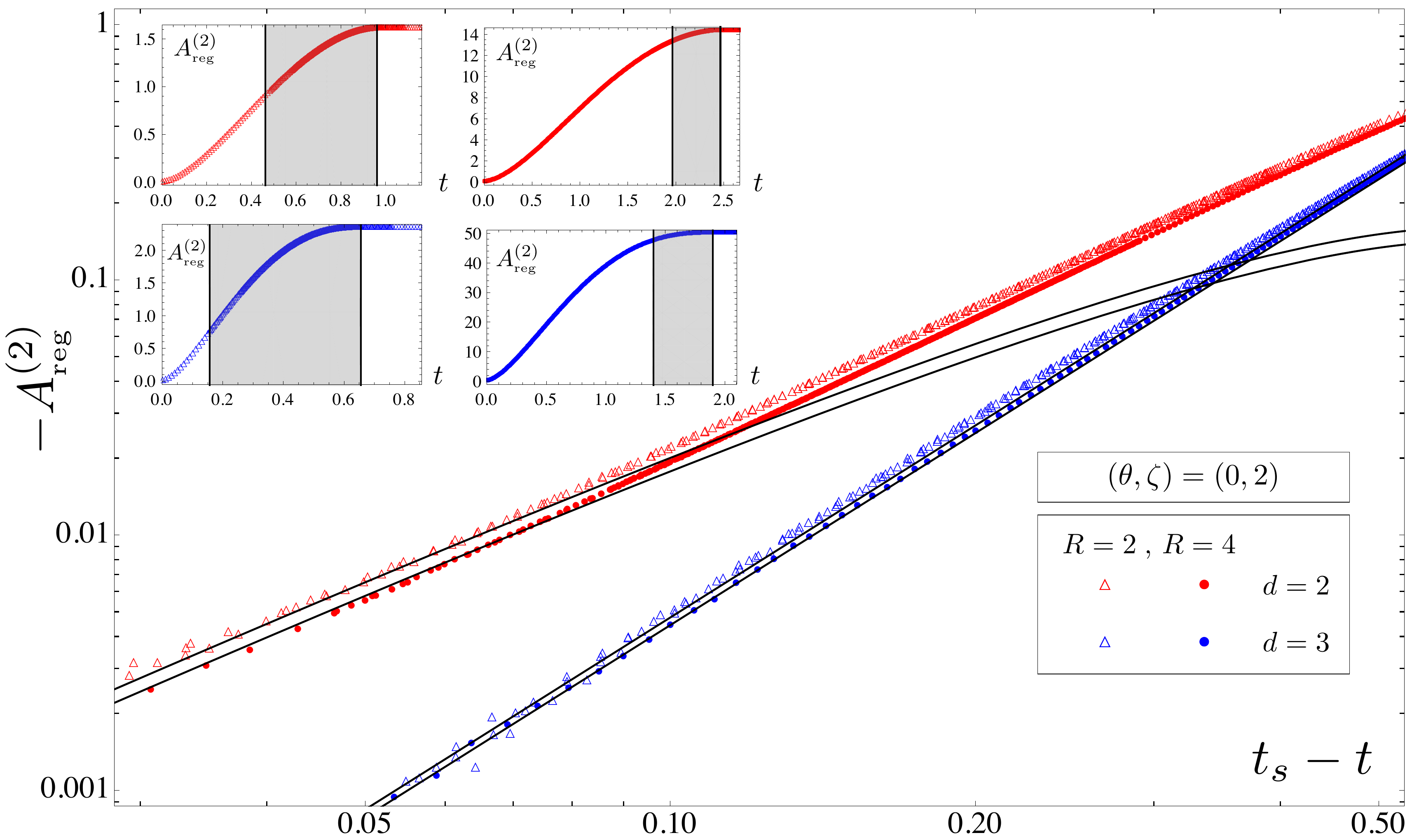}\\
\vspace{.6cm}
\hspace{.0cm}
\includegraphics[width=.81\textwidth]{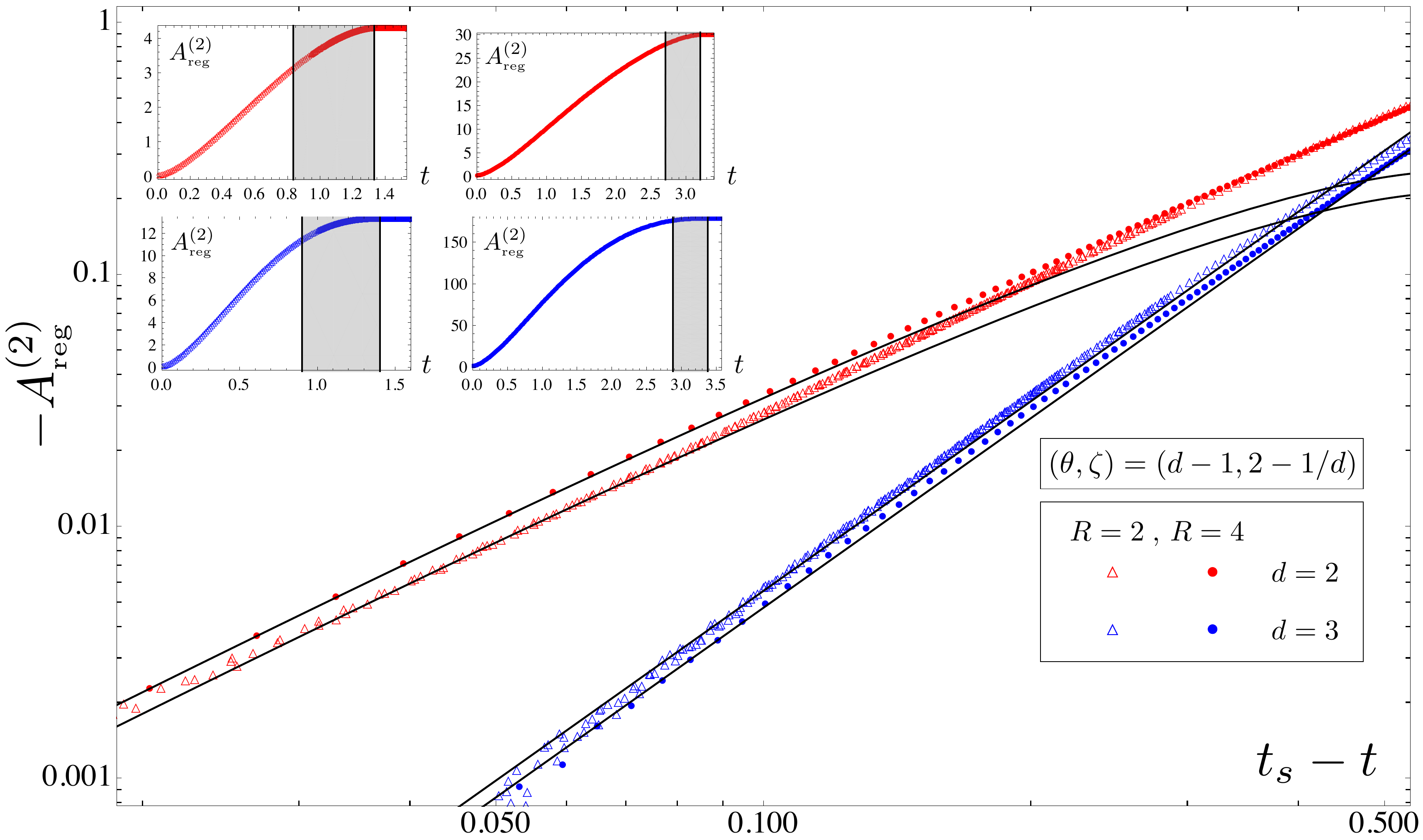}
\end{center}
\vspace{-.0cm}
\caption{
\label{fig:sat23}
Saturation regime for the holographic entanglement entropy in the thin shell limit ($a=10^{-4}$) for a spherical region of radius $R$ in the boundary. 
Here $(\theta,\zeta)=(0,2)$ (top panel) and $(\theta,\zeta)=(d-1,2-1/d)$ (bottom panel). The plots are constructed as in Fig. \ref{fig:sat1}.
The agreement with the continuos black curves from (\ref{A2reg saturation sphere}) indicates that  the saturation regime is independent of  $(\theta,\zeta)$.
}
\end{figure}

We define the saturation time $t_s$ as the boundary time such that, for $t>t_s$, the extremal surface probes only the black hole part of the geometry. It is possible to estimate $t_s$ as a function of $z_\ast$ for sufficiently large regions with generic shapes. 
The relevant computations for this regime are explained in Appendix \S\ref{app sec saturation}.
To leading order, $t_s$ is given by
\be
t_s=-\frac{z_h^{\zeta-1}}{F'_h}\log (z_h-z_\ast)\,,
\ee
where $F'_h \equiv -\partial_z  F(z) |_{z=z_h}$. 
Since the relation between $z_\ast$ and the characteristic length of the boundary region depends on its shape, we have to consider the strip and the sphere separately.
For a strip, if  $\partial_t A_{\textrm{\tiny reg}}(t)$ is continuous at $t=t_s$,  we find the following linear relation
\be
\label{ts strip text}
t_s= z_h^{\zeta-1}\sqrt{\frac{d_\theta}{ 2z_h F'_h}}\;\ell+\dots\,,
\ee
where the dots denote subleading orders at large $\ell$. 
Notice that (\ref{ts strip text}) can be generalized to $n$ dimensional spatial surfaces in the boundary according to the observation made in the end of \S\ref{ThinStrip}, namely $d_\theta$ should be replaced by $n d_\theta/d$ while $F(z)$ kept equal to (\ref{eq:emblackeningBH}). This gives
\be
\label{ts strip text n}
t_s = z_h^{\zeta-1}\sqrt{\frac{n d_\theta}{ 2d z_h F'_h}}\;\ell+\dots\;.
\ee
It can also be shown that, whenever $\partial_t \mathcal{A}_{\textrm{\tiny reg}}(t)$ is  continuous at saturation, we have
\be
\mathcal{A}_{\textrm{\tiny reg}}^{(2)}(t)\propto (t-t_s)^2 + o((t-t_s)^2) \,,
\ee 
for a strip for any values of $\zeta$ and $\theta$ (see Appendix \ref{app sat hee}).\\
The saturation time has also been evaluated numerically for the geodesic correlator, with the following procedure from \cite{Keranen:2011xs}. The action
for the geodesics has solutions with turning points either inside or outside the horizon.
We first choose turning points $z_*$  inside the horizon, generate the corresponding geodesic and find the coordinates of the endpoints at the boundary and the length of the geodesic. The results are regulated by subtracting the vacuum value. For  sufficiently large $\ell$, at early times the bulk geodesics will all have turning points inside the horizon, and also pass through the infalling shell extending into the part of the spacetime with vacuum geometry. In this case the corresponding
observable will not be thermal. At later times the turning point will be outside the horizon and the observable takes a thermal value. The conversion between these two types of behavior is 
sharp and defines the saturation time. Following \cite{Keranen:2011xs}, the saturation times can be calculated by fitting surfaces to the data of the above solution. The
intersection of the surfaces then defines the curve for the saturation time as a function of the transverse length scale. In Fig. \ref{fig:tsgeod} the numerical results for the saturation time of the geodesics are compared with the corresponding results from (\ref{ts strip text n}). Notice that the agreement improves for large $\ell$, as expected.\\
When the boundary region is a sphere and in the regime of large $R$, the transition to the saturated value is always smooth. In Appendix \S\ref{app sat time}, we show that
\be 
t_s =z_h^{\zeta-1} \sqrt{\frac{2 d_\theta}{ z_h F'_h}}\,R
-z_h^{\zeta-1}\frac{(d-1)}{ F^{(1)}(z_h)}  \log R+\dots\;.
\ee
Moreover, by extending the analysis of \cite{Liu:2013qca} to backgrounds with non trivial $\zeta$ and $\theta$, in \S\ref{app sat hee sphere} we find that
\be
\label{A2reg saturation sphere}
\mathcal{A}_{\textrm{\tiny reg}}^{(2)} \,\propto\, 
\left\{\begin{array}{ll} 
\displaystyle
-(t_s-t)^2 \log(t_s-t)  \hspace{.7cm}&d=2  \\ 
\rule{0pt}{.5cm}
\displaystyle
(t_s-t)^{1+d/2}
&d> 2 \end{array}\right.
\ee
telling us that the saturation regime is independent of $\zeta$ and $\theta$.
In Figs. \ref{fig:sat1} and \ref{fig:sat23} we show the saturation regime for the holographic entanglement entropy in the thin shell limit ($a=10^{-4}$) for the two cases of $R=2$ and $R=4$
with the dimensionality given by either $d=2$ or $d=3$.
The agreement between the numerical data and the expression (\ref{A2reg saturation sphere}) is quite good.

\subsection*{Acknowledgments}

We thank Alice Bernamonti and Rob Myers for useful discussions.
EKV, LT and ET are grateful to the Isaac Newton Institute for Mathematical Sciences for 
hospitality during the program on the {\it Mathematics and Physics of the Holographic Principle},
September 16 - October 11, 2013, 
where part of this work was carried out.
ET also acknowledges the Perimeter Institute for hospitality.
The research of VK was supported by the European Research Council
under the European Union's Seventh Framework Programme (ERC Grant
agreement 307955).
The work of LT is supported in part by grants from the Icelandic Research Fund and the
University of Iceland Research Fund. 
EKV is in part supported by the Academy of Finland grant no 1268023. 
LF is in part supported by the Finnish Academy of Science and Letters.


\appendix

\section{Spherical region for hvLif}
\label{app sphere hvLif}

In this appendix we construct a sequence of curves $\{x_i(s), z_i(s)\}$ for $i\in \mathbb{N}$ defined in a parametric way, whose asymptotic one $\{x_\infty(s), z_\infty(s)\}$ is the solution of (\ref{eq:sferavac}).\\
The extremal surface ending on the sphere of radius $R$ and extended in the $t=\textrm{const}$ section of the $d+2$ dimensional spacetime hvLif  obeys (\ref{eq:sferavac})
with the boundary conditions $z'(0)=0$ and $z(R)=0$. 
We recall that for hvLif without black holes the Lifshitz exponent $\zeta$ does not enter in the equation.
The equation (\ref{eq:sferavac}) can be rewritten as
\be 
\label{eq:sferavac v1 theta}
d\left(\frac{z'}{\rho}+\frac{1}{z}\right)+\left(\frac{z''}{1+z'^2}-\frac{z'}{\rho}\right)=\frac{\theta}{z}\,.
\ee
We find it convenient to introduce
\be
\label{ztilde def}
\tilde{z}(x)  \equiv \frac{z(\rho(x))}{R}\,,
\hspace{.7cm}
x \equiv 1-\frac{\rho}{R} \in [0,1]
\hspace{.2cm}
\qquad \Longrightarrow \qquad
\hspace{.2cm}
z'(\rho) = -\dot{\tilde{z}}(x)\,,
\hspace{.7cm}
z''(\rho) = \frac{\ddot{\tilde{z}}(x)}{R}\,.
\ee
By employing (\ref{ztilde def}), (\ref{eq:sferavac v1 theta}) becomes (\ref{eq:dimnless}), which can be written as follows
\be 
d\left(\frac{1}{\tilde{z}}
-\frac{\dot{\tilde{z}}}{1-x}\right)
+\frac{\ddot{\tilde{z}}}{1+\dot{\tilde{z}}^2}
+\frac{\dot{\tilde{z}}}{1-x}
=
\frac{\theta}{\tilde{z}} \,,
\qquad \hspace{1cm}\tilde{z}(0)=0\,,
\qquad \dot{\tilde{z}}(1)=0\,.
\label{eq:sferavacx}
\ee
The well known hemispherical solution for $\theta=0$ becomes
\be
\label{eq:sferadseries}
\tilde{z}(x)\big|_{\theta=0}
=\sqrt{x(2-x)}=\sqrt{2x}\left(\sum_{n=0}^{+\infty}\frac{\Gamma(n-3/2)}{\Gamma(-3/2)n!}x^n\right)\,,
\ee
which evidences that $\tilde{z}(x)=\sqrt{2x}$ when $x \rightarrow 0$. 
Also for $\theta \neq 0$ we have $\tilde{z}\simeq0$ near the boundary $x\simeq 0$ and here we are interested in the way it vanishes.
First, from (\ref{eq:sferavacx}) one observes that, when $d-\theta\neq0$ (the case $d-\theta=0$ is not allowed by NEC), the solution must have a divergent $\tilde{z}'(0)$.
Introducing the following ansatz for the solution close to the boundary 
\be
\label{ansatz tildez}
\tilde{z}(x) = c_0 x^\alpha\,,
\qquad
0<\alpha<1\,,
\qquad
x \sim 0\,,
\ee
and plugging it into (\ref{eq:sferavac v1 theta}), the first order for $x \rightarrow 0$ provides the following equation
\be
\label{eq alpha and c0}
\left( d-\theta +1 - \frac{1}{\alpha} \, \right) x^{-\alpha} 
+ c_0^2 \alpha (1-d) x^{\alpha-1}  = 0\,.
\ee
We can recognize three cases:
\begin{enumerate}
\item $d=1$. In this case  we find
\be
\tilde{z}(x)\simeq c_0 x^{\frac{1}{2-\theta}}\,,
\label{eq:d1spherebdy} 
\ee
where the condition $0<\alpha<1$ becomes $\theta<1$.
In particular, for $\theta=0$ we recover the expected $\sqrt{x}$ behaviour, although the overall constant is not fixed. Since for $d=1$ the calculations from the strip hold, we have that (see (\ref{eq:analyticsolstrip}))
\be
\label{x ztilde}
x(\tilde{z})=\frac{\tilde{z}_\ast}{2-\theta}
\bigg(\frac{\tilde{z}}{\tilde{z}_\ast}\bigg)^{2-\theta}
\, _{2}F_{1}\bigg(
\frac{1}{2},\frac{1}{2}+\frac{1}{2 (1-\theta)};\frac{3}{2}+\frac{1}{2(1-\theta)};
(\tilde{z}/\tilde{z}_\ast)^{2(1-\theta)}
\bigg)  \,,
\ee
where the constant $\tilde{z}_\ast$ reads 
\be
\tilde{z}_\ast=\frac{\Gamma(1/(2-2\theta))}{\sqrt{\pi}\,\Gamma(1/2+1/(2-2\theta))}\,.
\ee
Since the hypergeometric function in (\ref{x ztilde}) goes to $1$ at the boundary, from (\ref{eq:d1spherebdy}) we can write
\be
c_0=\left(\tilde{z}_\ast^{1-\theta} (2-\theta)\right)^\frac{1}{2-\theta}\,,
\ee
which simplifies to $c_0 = \sqrt{2}$ when $\theta=0$ because $\tilde{z}_\ast|_{\theta=0} = 1$.
\item $d\neq1$ and $d_\theta\neq1$.
In this regime one finds that
\be
\tilde{z}(x)= \sqrt{\frac{d_\theta-1}{d-1}\, 2x} 
\left[1- \frac{1}{4}\left(1+\frac{d-1}{d_\theta-1}-	\frac{d-3}{d_\theta-3}\right)x +O(x^2)\right]\,,
\label{eq:limitzx}
\ee
again, notice how when $d_\theta=d$ we recover the AdS solution but now with even the correct value of the coefficient, $c_0=\sqrt{2}$. We included also the $c_1$ correction to show the emergence of poles in the coefficient for any odd integer value of $d_\theta$. It is possible to compute the expansion up to arbitrary order, but it appears the terms in the series cannot be written in any compact or recursive form. 
\item $d\neq1$ and $d_\theta=1$. In this case (\ref{eq alpha and c0}) becomes
\be
 x^{-\alpha}  \frac{2\alpha-1}{\alpha} + x^{\alpha-1} c_0^2 \alpha (1-d)=0 \,,
 \ee
which gives $\alpha=1/2$ and $c_0=0$. This tells us that the ansatz (\ref{ansatz tildez}) is meaningless in this case.
\end{enumerate}

\subsection{A parametric reformulation}
\label{sec parametric ref}

\begin{figure}
\center
\includegraphics[width=0.6\textwidth]{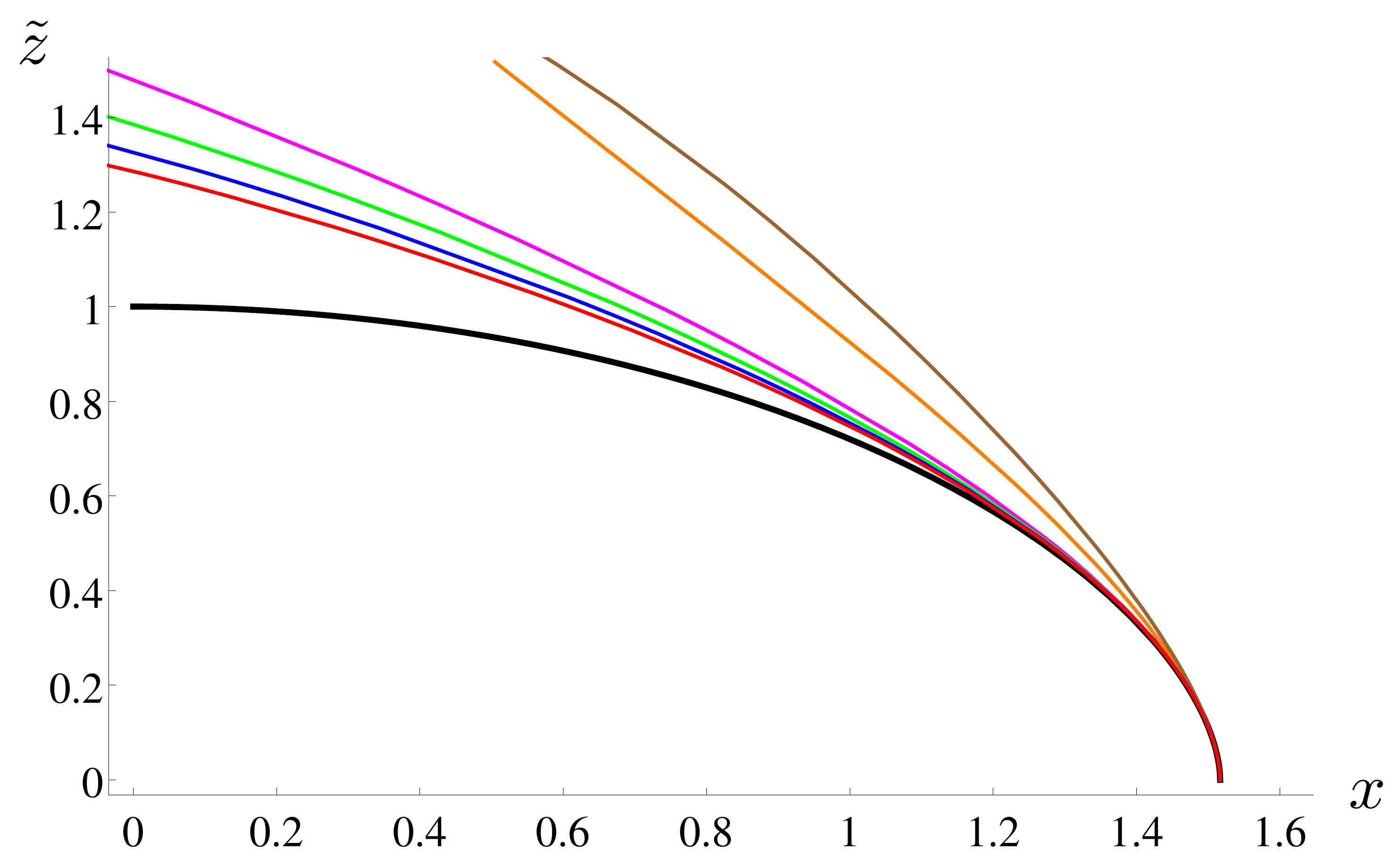}
\caption{
The black curve is the numerical solution $\tilde{z}=\tilde{z}(x)$ of (\ref{eq:sferavacx}) for $d=2$ and $d_\theta=1$. The remaining curves are $(x_i(s),\tilde{z_i}(s))$ for $i=1,2,3,4,5,6$ (respectively orange, brown, magenta, green, blue and red), constructed in \S\ref{sec parametric ref}.}
\label{fig:comparison}
\end{figure}

In order to improve this analysis and understand better the last case, following \cite{Keranen:2013vla} (where the $d=2$ case has been studied) we introduce
\be
s \equiv \frac{1}{\tilde{z}^{d_\theta}\sqrt{1+\dot{\tilde{z}}^2}}\,.
\label{eq:defs}
\ee
This allows to write the term containing $\ddot{\tilde{z}}$ in (\ref{eq:sferavacx}) as follows
\be
\frac{\ddot{\tilde{z}}}{1+\dot{\tilde{z}}^2}
=
-\frac{d_\theta}{\tilde{z}} - \left(s \,\frac{d \tilde{z}}{ds} \right)^{-1}\,.
\ee
Thus, the equation (\ref{eq:sferavacx}) can be written as
\be
\frac{d-1}{x-1} \, \frac{\frac{d \tilde{z}}{ds}}{\frac{dx}{ds}}  
-\left(s \,\frac{d \tilde{z}}{ds} \right)^{-1} \,=\, 0\,.
\label{eq:parameq0}
\ee
From (\ref{eq:defs}) it is straightforward to write that
\be
\label{dxds v1}
\frac{dx}{ds} \,=\,
\frac{s\tilde{z}^{d_\theta}}{\sqrt{1-s^2 \tilde{z}^{2d_\theta}}}\,\frac{d\tilde{z}}{ds}\,.
\ee
Then, by isolating $x$ in (\ref{eq:parameq0})  and employing (\ref{dxds v1}), the differential equation (\ref{eq:parameq0}) becomes 
\be
\label{eq:parameq}
x  \,=\, 1+\frac{(d-1)\sqrt{1-s^2 \tilde{z}^{2d_\theta}}}{\tilde{z}^{d_\theta}}\, 
\frac{d \tilde{z}}{ds}\,.
\ee
We find it convenient to rewrite (\ref{eq:parameq}) and (\ref{dxds v1}) respectively as follows
\be
\label{eq:paramequa}
\left\{ \begin{array}{ll}
\displaystyle
\frac{d}{d s}\tilde{z}(s)^{-(d_\theta-1)} 
\,=\,
\frac{(d_\theta-1)[1-x(s)]}{(d-1)\sqrt{1-s^2 \tilde{z}(s)^{2d_\theta}}}
\hspace{.3cm}& d_\theta\neq 1
\\
\rule{0pt}{.7cm}
\displaystyle
\frac{d}{d s}\log \tilde{z}(s) 
\,=\,
- \frac{1-x(s)}{(d-1)\sqrt{1-s^2 \tilde{z}(s)^{2}}}
\hspace{.3cm}& d_\theta = 1
\end{array}\right.\,,
\hspace{.5cm}
\qquad
\frac{d}{d s}x(s)\,=\,
- \frac{[1-x(s)] \,s \tilde{z}(s)^{2d_\theta} }{(d-1)[1-s^2 \tilde{z}(s)^{2d_\theta}]}\,. 
\hspace{.5cm}
\ee
Integrating these equations, one finds
\be
\label{integraleq z}
\tilde{z}(s) \,=\,
\left\{ 
\begin{array}{cc}
\displaystyle
\left(\frac{d_\theta-1}{d-1}
\sum_{n=0}^{+\infty} 
\frac{\Gamma(n+1/2)}{\sqrt{\pi} n!}
\int^{s}_{s_{\textrm{\tiny{min}}}}
[1-x(r)] r^{2n} \tilde{z}(r)^{2d_\theta n} dr\right)^{-\frac{1}{d_\theta}-1}
&\hspace{.7cm}  d_\theta\neq 1
\\
\rule{0pt}{.8cm}
\displaystyle
\exp\left(- \frac{1}{d-1}
\sum_{n=0}^{+\infty}
\frac{\Gamma(n+1/2)}{\sqrt{\pi} n!}
\int^{s}_{s_{\textrm{\tiny{min}}}}
[1-x(r)] r^{2n} \tilde{z}(r)^{2n} dr\right) 
& \hspace{.7cm}  d_\theta=1
\end{array}
\right. 
\ee
and
\be
\label{integraleq x}
x(s)\,=\, 
-\frac{1}{d-1}\sum_{n=0}^{+\infty} \int^{s}_{s_{\textrm{\tiny{min}}}} 
[1-x(r)]r^{1+2n} \tilde{z}(r)^{2d_\theta(1+n)} dr\,,
\ee
where the expansion of $(1-w)^{-\alpha}$ for $w\rightarrow 0$ has been used. This can be done because (\ref{eq:defs}) implies that $s \tilde{z}^{d_\theta}$ is infinitesimal when $s$ is large.
Moreover, $s_{\textrm{\tiny{min}}}$ is the value of $s$ at which the tip of the minimal surface is reached and it can be found from (\ref{eq:defs})
\be
 s_{\textrm{\tiny{min}}} =\tilde{z}_\ast^{-d_\theta}\,.
\ee
It is evident that (\ref{integraleq z}) and (\ref{integraleq x}) is only a formal solution and it does not even allow to plot the solution numerically. Nevertheless, this form allows us to construct  the solution $\{ \tilde{z}(s), x(s) \}$ recursively through an inductive procedure.\\
Since large $s$ corresponds to the boundary, we have that  $x(s) = o(1)$ for large $s$. This allows us to observe that the leading order of the integrals in (\ref{integraleq z}) and (\ref{integraleq x}) can be obtained by neglecting $x(r)$ within the square brackets occurring in the integrands.
We find it  convenient to define the first pair of functions in the inductive process through the boundary conditions $x(s)\rightarrow 0$ and  $\tilde{z}(s) \rightarrow 0$ for $s\rightarrow \infty$, namely
\be
\label{0step}
\tilde{z}_0(s)=0 \,,
\qquad
x_0(s)= 0 \,.
\ee
Then for $i>0$ we define
\be
\label{integraleq z i}
\tilde{z}_{i+1}(s) \,=\,
\left\{ 
\begin{array}{cc}
\displaystyle
\left(\frac{d_\theta-1}{d-1}
\sum_{n=0}^{i} 
\frac{\Gamma(n+1/2)}{\sqrt{\pi} n!}
\int^{s}_{s_{\textrm{\tiny{min}}}}
[\tilde{z}_{i-n+1}(r)^{2d_\theta n}-x_{i-n}(r)\tilde{z}_{i-n}(r)^{2d_\theta n}] r^{2n}  dr\right)^{-\frac{1}{d_\theta-1}}
&\hspace{.4cm}  d_\theta\neq 1
\\
\rule{0pt}{.8cm}
\displaystyle
\exp\left(- \frac{1}{d-1}
\sum_{n=0}^{i}
\frac{\Gamma(n+1/2)}{\sqrt{\pi} n!}
\int^{s}_{s_{\textrm{\tiny{min}}}}
[\tilde{z}_{i-n+1}(r)^{2n} -x_{i-n}(r)\tilde{z}_{i-n}(r)^{2n} ] r^{2n} dr\right) 
& \hspace{.4cm}  d_\theta=1
\end{array}
\right. 
\ee
and
\be
\label{integraleq x i}
x_{i+1}(s)\,=\, 
-\frac{1}{d-1}
\sum_{n=0}^{i} 
\int^{s}_{s_{\textrm{\tiny{min}}}} 
[\tilde{z}_{i-n+1}(r)^{2d_\theta(1+n)}-x_{i-n}(r)\tilde{z}_{i-n}(r)^{2d_\theta(1+n)}]r^{1+2n}  dr\,.
\ee
Given the pairs $\{\tilde{z}_j(s), x_j(s)\}$ for $j\leqslant i$, this equation give $\{\tilde{z}_{i+1}(s), x_{i+1}(s)\}$. Notice that $x_{i+1}$ depends on $\tilde{z}_{i+1}$ through the $n=0$ term and this means that one has to solve (\ref{integraleq z i}) first and then (\ref{integraleq x i}).
This procedure defines a sequence of pairs $\{\tilde{z}_i(s), x_i(s)\}$ for $i\in \mathbb{N}$ and the exact solution of (\ref{eq:paramequa}) is the asymptotic one $\{\tilde{z}_\infty(s), x_\infty(s)\}$  for $i \rightarrow +\infty$. 
The pair $\{\tilde{z}_i(s), x_i(s)\}$ for some finite $i$ gives a better approximation of the asymptotic solution the higher $i$ is, starting from the regime of large $s$.\\
Given (\ref{0step}), for $i=1$ we find
\be
\label{eq:z1x1}
\tilde{z}_1(s)\,=\,
\left\{ \begin{array}{l}
\displaystyle
\left(\frac{d-1}{d_\theta-1}\right)^{\frac{1}{d_\theta-1}} (s-s_\textrm{\tiny min})^{-\frac{1}{d_\theta-1}} 
\\
\rule{0pt}{.5cm}
\displaystyle
e^{-\frac{s-s_\textrm{\tiny min}}{d-1}} 
\end{array}
\right. \,,
\qquad 
x_1(s) = \left\{ \begin{array}{ll}
\displaystyle
\frac{1}{2}
\left(
\frac{d_\theta-1}{d-1} 
\right)^{\frac{d_\theta+1}{d_\theta-1}}
(s-s_\textrm{\tiny min})^{-\frac{2}{d_\theta-1}}
& \hspace{.4cm} d_\theta\neq 1
\\ 
\rule{0pt}{.7cm}
\displaystyle
\frac{2s + d-1}{4} \, e^{-2\frac{s-s_\textrm{\tiny min}}{d-1}} 
&  \hspace{.4cm}  d_\theta =1 
\end{array}
\right. 
\ee
From (\ref{eq:z1x1}) for $d_\theta\neq 1$ and large $s$, we can write 
\be
s= \left(
\frac{d-1}{d_\theta-1} 
\right)^{\frac{1+d_\theta}{2}} (2x_1)^{-\frac{d_\theta-1}{2}} \,.
\ee 
Plugging this back into the corresponding $\tilde{z}_1$ in (\ref{eq:z1x1}), we get the first term of (\ref{eq:limitzx}) and  the first term of (\ref{eq:sferadseries}) when $\theta=0$, as expected.
By employing (\ref{0step}) and (\ref{eq:z1x1}), for $i=2$ in the regime of large $s$ we find
\be
\label{eq:z2}
\tilde{z}_2(s) = 
\left\{ \begin{array}{ll}
\displaystyle
\left(\frac{d-1}{d_\theta-1}\right)^{\frac{1}{d_\theta-1}} s^{-\frac{1}{d_\theta-1}} 
\left[
1
-
\left(
\frac{d-1}{d_\theta-1}
\right)^{\frac{d_\theta+1}{d_\theta-1}}
\frac{\theta\, s^{-\frac{2}{d_\theta-1}}}{2(d_\theta-3)}
\right]^{-\frac{1}{d_\theta-1}} 
& d_\theta\neq 1,3
\\
\rule{0pt}{.7cm}
\displaystyle
e^{-\frac{1}{d-1}s}+\frac{(d-1)(d-3) + 2 (d-2) s + 2 s^2}{8}
\, e^{-\frac{3}{d-1}s}
& d_\theta =1 
\\  
\rule{0pt}{.8cm}
\displaystyle
\left[
\frac{2s}{d-1}
+\frac{(d-1)(d-3)}{8}
\log s 
\right]^{-\frac{1}{2}}
& d_\theta=3 
\end{array}
\right. 
\ee
The expression for $x_2$ is quite complicated even at large $s$ and we do not find it useful to write it here.
We have neglected $s_\textrm{\tiny min}$ because $s$ is large, but it must be taken into account properly to obtain the plot in Fig. \ref{fig:comparison}.
Higher orders are rather complicated as well and therefore we do not write them.
Repeating the procedure we can find the various curves in Fig. \ref{fig:comparison}, from which it is evident that the exact solution of (\ref{eq:dimnless}) is better approximated as $i$ increases. 

\subsubsection{Area}
\label{area parametric sec}

The area functional is given by (\ref{eq:spherarea}) $F(z) =1$), namely
\be
\label{area func xz}
\mathcal{A}  \,=\,
\frac{2\pi^{d/2}R^{\theta}}{\Gamma(d/2)} \int_{0}^1 \frac{(1-x)^{d-1}}{\tilde{z}^{d_\theta}}\sqrt{1+\dot{z}^{2}}\,dx
\,=\,
\frac{2\pi^{d/2}R^{\theta}}{(d-1)\Gamma(d/2)} 
\int_{s_{\textrm{\tiny min}}}^{+\infty} \frac{(1-x)^{d}}{1-s^2\tilde{z}^{2 d_\theta}}\,ds\,.
\ee
Since the integral is divergent, we must introduce the UV cutoff $\tilde{\epsilon}=\epsilon/R$ in the $\tilde{z}$ variable.
It corresponds to a large value $s_{\textrm{\tiny max}}$ such that $z(s_{\textrm{\tiny max}})=\epsilon$. By employing the expressions  $\{\tilde{z}_i(s), x_i(s)\}$ discussed above in (\ref{area func xz}), one gets a corresponding area $\mathcal{A}_i$. Thus, we have
\be
\label{area func xz i-th}
\mathcal{A} \, = \lim_{i \rightarrow \infty}\mathcal{A}_i  \,,
\hspace{1cm}
\qquad
\mathcal{A}_i 
\,\equiv\,
\frac{2\pi^{d/2}R^{\theta}}{(d-1)\Gamma(d/2)} 
\int_{s_{\textrm{\tiny min}}}^{s_{\textrm{\tiny max}}}
\frac{(1-x_{i-1})^{d}}{1-s^2\tilde{z}_{i-1}^{2 d_\theta}}\,ds\,.
\ee
A crucial point consists in finding $s_{\textrm{\tiny max}}(\epsilon)$, but the relation $\tilde{z}_i(s_{\textrm{\tiny max}})=\tilde{\epsilon}$
is typically transcendental and therefore it cannot be inverted. 
Introducing $s_{\textrm{\tiny max},i}$ as the solution of $\tilde{z}_i(s_{\textrm{\tiny max},i})=\tilde{\epsilon}$, we have that for $i=1$ the inversion can be performed, giving
\be
\label{smax i=1}
s_{\textrm{\tiny max},1} =
\left\{ \begin{array}{ll}
\displaystyle
\frac{d-1}{(d_\theta-1)\tilde{\epsilon}^{d_\theta-1}}
\hspace{.5cm} & d_\theta\neq 1 \\
 \rule{0pt}{.5cm}
\displaystyle
-(d-1)\log \tilde{\epsilon}
 & d_\theta =1 
\end{array}
\right. 
\ee
which gives
\be
\label{area xz i=1}
\mathcal{A}_1 \,=\,
\left\{ \begin{array}{ll}
\displaystyle
\frac{2\pi^{d/2}}{(d_\theta-1)\Gamma(d/2)} \, \frac{R^{d-1}}{\epsilon^{d_\theta-1}} 
\hspace{1cm} & d_\theta\neq 1 \\
 \rule{0pt}{.8cm}
\displaystyle
-\frac{2\pi^{d/2}}{\Gamma(d/2)} \, R^{d-1} \log(\epsilon/R)
 & d_\theta =1 
\end{array}
\right. 
\ee
For $i=2$, it is clear from (\ref{eq:z2}) that $\tilde{z}_2(s_{\textrm{\tiny max},2})=\tilde{\epsilon}$ cannot be inverted.
Nevertheless, we can find the first terms of the expansion of $s_{\textrm{\tiny max}}(\tilde{\epsilon})$ for $\tilde{\epsilon} \rightarrow 0$ as follows.
The relation $\tilde{z}_2(s_{\textrm{\tiny max},2})=\tilde{\epsilon}$  can be written as
\be 
\tilde{\epsilon}=f_1(s)+f_2(s) \,, \label{zss1}
\ee
where both $f_1$ and $f_2$ vanish for $s \rightarrow \infty$, while $f_1/f_2 \rightarrow 0$.
Assuming that $f_1$ is invertible, we have that
\be
 \label{zss3}
s 
\,=\, f_1^{-1}(\tilde{\epsilon}-f_2(s))
\,=\, f_1^{-1}(\tilde{\epsilon}) 
- [ \partial_{\tilde{\epsilon}} f^{-1}_1(\tilde{\epsilon}) ] f_2(s) + O\big(f_2(s)^2\big) 
\,=\, f_1^{-1}(\tilde{\epsilon}) 
- [ \partial_{\tilde{\epsilon}} f^{-1}_1(\tilde{\epsilon}) ]
f_2(f_1^{-1}(\tilde{\epsilon}) ) +\dots
\ee
where in the second step we have employed that $f_2/\tilde{\epsilon} = (f_2/f_1)/(1+f_2/f_1) \rightarrow 0$ when $s \rightarrow \infty$, while in the last one the first order of the expansion has been used. The dots denote higher orders that we are neglecting.
Thus, for $i=2$ we find
\be
s_{\textrm{\tiny max},2}\,=\,
\left\{ \begin{array}{ll}
\displaystyle
\frac{d-1}{(d_\theta-1)\tilde{\epsilon}^{d_\theta-1}}
\left[\,1
-\frac{(d-1) \theta}{2(d_\theta-1)(d_\theta-3)}\, \tilde{\epsilon}^2 + \dots 
\right]
\hspace{.5cm} & d_\theta\neq 1,3 \\
 \rule{0pt}{.7cm}
\displaystyle
-(d-1)\log \tilde{\epsilon}
\left[\,1-
\frac{(d-1)^2}{4}\, \tilde{\epsilon}^2 \log \tilde{\epsilon} + \dots 
\right]
 & d_\theta =1 \\ 
 \rule{0pt}{.8cm}
 \displaystyle
\frac{d-1}{2\tilde{\epsilon}^{2}}-\frac{(d-1)^2(d-3)}{8}\log \tilde{\epsilon} + \dots 
& d_\theta=3 
\end{array}
\right. 
\ee
As for the integral (\ref{area func xz i-th}) with $i=2$, we find
\be
\label{area xz i=2}
\mathcal{A}_2 \,=\,
\left\{ \begin{array}{ll}
\displaystyle
\frac{2\pi^{d/2}R^{d-1}}{\Gamma(d/2)\,\epsilon^{d_\theta-1}} 
\left[\,\frac{1}{d_\theta-1}-\frac{(d-1)^2(d_\theta-2)}{2(d_\theta-1)^2(d_\theta-3)} \frac{\epsilon^2}{R^2} + O(\epsilon^4)  \right] 
\hspace{1cm} & d_\theta\neq 1,3 \\
 \rule{0pt}{.8cm}
\displaystyle
-\frac{2\pi^{d/2}R^{d-1}}{\Gamma(d/2)} \,  \log(\epsilon/R) 
\left[\,1+
\frac{(d-1)^2}{4}\, \frac{\epsilon^2}{R^2} \log (\epsilon/R) + \dots 
\right]
 & d_\theta =1 \\
 \rule{0pt}{.8cm}
 \displaystyle
 \frac{2\pi^{d/2} R^{d-1}}{\Gamma(d/2)\, \epsilon^{2}} 
 \left[\,\frac{1}{2}-\frac{(d-1)(d-5)}{8}  \, \frac{\epsilon^2}{R^2} \log (\epsilon/R) + o\big(\epsilon^2\big) \right] 
\hspace{1cm} 
& d_\theta= 3 
\end{array}
\right. 
\ee
As a check of this formula, notice that the first expression for $\theta=0$ provides the expansion at this order of the hemisphere \cite{Ryu:2006ef}. Moreover, we have also checked that the first expression in (\ref{area xz i=2}) can be found by plugging (\ref{eq:limitzx}) into (\ref{area func xz}), properly regulated through the introduction of $x_{\textrm{\tiny min}}>0$ such that $x_{\textrm{\tiny min}}=x(s_{\textrm{\tiny max}})$.

\section{Computational details for the entanglement growth}
\label{app computational details}

\subsection{Initial growth: generic shape}
\label{app sec initial}

Let us consider a $n$ dimensional region embedded into $\mathbb{R}^d$, which is the spatial part of the boundary (i.e. $z=0$) of the Vaidya background (\ref{vaidya metric}). The boundary of such region will be denoted by $\Sigma$ and it has a generic shape. The submanifold $\Sigma$ is $n-1$ dimensional and therefore it can be parameterized through a $n-1$ dimensional vector of intrinsic coordinates $\xi^\alpha$. Thus, being $x_a$ the cartesian coordinates of $\mathbb{R}^d$, the submanifold $\Sigma$ is specified by 
\be
x_a (\xi^\alpha)\,, \qquad 
a\in \{1,\dots,d\} \,,\qquad 
\alpha\in \{1,\dots,n-1\} \,.
\ee
The surface $\Gamma_\Sigma$ we are looking for is also $n$ dimensional and it extends into the bulk, arriving to the boundary along $\Sigma$, i.e. $\partial \Gamma_\Sigma = \Sigma$ at certain time $t$.
It is described by the functions
\be 
\label{GammaSigma coords}
v(\xi^\alpha,z)\,, \qquad X_a(\xi^\alpha,z) \,,
\ee  
satisfying the following boundary conditions
\be
v(\xi^\alpha,0)=t \,,  \qquad X_a(\xi^\alpha,0) =  x_a (\xi^\alpha) \,.
\ee
We remark that for the holographic entanglement entropy $n=d$, for the holographic counterpart of the Wilson loop $n= 2$ and for the holographic two point function $n= 1$ ($\Gamma_\Sigma$ is a  geodesic and $\Sigma$ is made by two points spacelike separated). \\
The area $A_\Sigma$ of $\Gamma_\Sigma$ is given by
\be 
\mathcal{A}_{\Gamma_\Sigma} =\int_0^{z_\ast} dz\int d\xi^\alpha 
\, \frac{\sqrt{\det{\gamma}}}{z^{n d_\theta/d}}\,,
\ee
where $z^{-2d_\theta/d}\gamma_{ab}$ is the induced metric on $\Gamma_\Sigma$ and $\det{\gamma}$ denotes the determinant of $\gamma_{ab}$.
Differentiating (\ref{GammaSigma coords}) and plugging the results into (\ref{vaidya metric}), we find that
\bea
\label{gamma eq1}
& & 
\gamma_{zz} \,=\,  
-z^{2(1-\zeta)} F v_z^2-2 z^{1-\zeta}v_z 
+\boldsymbol{X}_z \cdot \boldsymbol{X}_z\,,
\\
\label{gamma eq2}
& &
\gamma_{\alpha z} \,=\,  
 -z^{2(1-\zeta)} F v_\alpha v_z  - z^{1-\zeta}v_\alpha  
+\boldsymbol{X}_\alpha \cdot \boldsymbol{X}_z\,,
\\
\label{gamma eq3}
& &
\gamma_{\alpha\beta} \,=\,  
 -z^{2(1-\zeta)} F v_\alpha v_\beta
+\boldsymbol{X}_\alpha \cdot \boldsymbol{X}_\beta\,,
\eea
where $\boldsymbol{X}$ denotes the vector whose components are $X_a$, the dots stand for the scalar product and the subindices indicate the corresponding partial derivatives.\\
Here we consider the analogue of $A^{(3)}_{\textrm{\tiny reg}}$ defined in (\ref{eq:area3}), namely the area of $\Gamma_\Sigma$ regularized through the area of $\widehat{\Gamma}_\Sigma$ computed in hvLif, when $F=1$. Given that the hatted quantities refer to hvLif, it reads
\be
\label{A3diff}
\mathcal{A}^{(3)}_{\textrm{\tiny reg}}(t)
=
\int \bigg[ 
\int_0^{z_\ast} \frac{\sqrt{\det{\gamma}}}{z^{nd_\theta/d}} \,dz
- \int_0^{\hat{z}_\ast} 
\frac{\sqrt{\det{\hat{\gamma}}}}{z^{nd_\theta/d}} \,dz 
\bigg] d^{n-1}\xi \,.
\ee
The initial regime is characterized by $0 < t \ll z_h$ and we want to compute 
$\mathcal{A}^{(3)}_{\textrm{\tiny reg}}(t)$ for small $t$.
Keeping the first order in (\ref{A3diff}) and repeating the same arguments discussed in \cite{Liu:2013qca}, we find 
\be
\label{eq:variationq}
\mathcal{A}^{(3)}_{\textrm{\tiny reg}}(t)=
\int \Bigg[ 
 \int_0^{\hat{z}_\ast}
 \frac{\partial_F \big(\sqrt{\det{\gamma}}\,\big)\big|_{F=1}}{z^{nd_\theta/d}} \, \delta F \,dz
+ \frac{\sqrt{\det{\hat{\gamma}}}}{z_\ast^{nd_\theta/d}} \, \delta z_\ast 
+ \sum_{A=0, a}\frac{\partial}{\partial X_{A,z}}
\bigg(   \frac{\sqrt{\det{\hat{\gamma}}}}{z^{nd_\theta/d}}  \bigg) \delta X_A
\bigg|_0^{\hat{z}_\ast}
\Bigg] d^{n-1}\xi \,,
\ee
where $X_0 \equiv v$, $X_{A,z} \equiv \partial_z X_A$ and only the first term within the square brackets provides a non-vanishing contribution. In order to find it, we employ the well known formula for the variation of the determinant
\be 
\label{det formula}
\partial_F\big(  \sqrt{\det\gamma} \, \big) 
= \frac{\sqrt{\det\gamma}}{2}\,
\textrm{Tr} \big( \gamma^{-1} \partial_F \gamma \big)\,.
\ee
From (\ref{gamma eq1}), (\ref{gamma eq2}) and (\ref{gamma eq3}), we get respectively
\be
\label{partialF gammas}
\partial_F (\gamma_{zz}) \big|_{F=1} 
=-\frac{v_z^2}{z^{2(\zeta-1)} }  \,,
\qquad
\partial_F (\gamma_{\alpha z}) \big|_{F=1} 
=-\frac{v_\alpha v_z}{z^{2(1-\zeta)}}  \,,
\qquad
\partial_F (\gamma_{\alpha \beta}) \big|_{F=1} 
=-\frac{v_\alpha v_\beta}{z^{2(1-\zeta)}}   \,.
\ee
Now, from (\ref{v diff form}) with $F=1$ we find that $\hat{v} = t - z^\zeta/\zeta$.
Since $t$ is a constant in terms of $\xi^\alpha$, in (\ref{partialF gammas}) we have that $v_\alpha = o(t)$ and $v_z = -z^{\zeta-1} +o(t)$. 
Plugging these behaviors into (\ref{partialF gammas}), only the first expression is non vanishing and equal to $-1$. Then, by using that $X_a(\xi^\alpha,z)  = x_a(\xi^\alpha) + o(z)$, where $o(z)$ vanishes fast enough when $z \rightarrow 0$, we have
\be
\label{gamma comp bdy}
\gamma_{\alpha\beta} = h_{\alpha\beta} +o(z)\,,
\qquad
\gamma_{\alpha z} = o(z)\,,
\qquad
\gamma_{zz} = 1+o(z)\,,
\ee
where $h_{\alpha\beta} \equiv \partial_\alpha x_a \partial_\beta x_a$ is the induced metric on $\Sigma$. Notice that (\ref{gamma comp bdy}) tells us that the contribution of the term $\textrm{Tr} ( \gamma^{-1} \partial_F \gamma)$ to $\partial_F ( \sqrt{\det \gamma}\,)|_{F=1}$ is equal to $-1$.
Collecting these observations, we find
\be  
\partial_F \big( \sqrt{\det \gamma}\, \big)\big|_{F=1}
=-\frac{\sqrt{\det h}}{2} \,.
\ee
Finally, since in our case $\delta F =F(z)-1= - M z^{d_\theta+\zeta}$ is non vanishing only for $0< z< z_c$, the first term in (\ref{eq:variationq}) becomes
\be
\label{eq:variationq}
\mathcal{A}^{(3)}_{\textrm{\tiny reg}}(t)
= \frac{M A_\Sigma}{2} \int_0^{z_c}  z^{ d_\theta(1-n/d)+\zeta} dz
= \frac{M A_\Sigma \,z_c^{d_\theta(1-n/d)+\zeta+1}}{2[d_\theta(1-n/d)+\zeta+1]}
=  \frac{M A_\Sigma (\zeta t)^{[d_\theta(1-n/d)+\zeta+1]/\zeta}}{2[d_\theta(1-n/d)+\zeta+1]}\,. 
\ee
In the last step we have used that $z_c=(\zeta t)^{1/\zeta}$ to the first order, which is obtained from $\hat{v} = t-z^\zeta/\zeta$ and the condition $v=0$ at the shell.

\subsection{Linear growth}
\label{linearstrip}

In order to study this regime, we consider the strip (see \S\ref{stripvaidyasec}).
Following \cite{Liu:2013qca}, let us start from (\ref{zprime bh Eplus}) for the black hole regime.
By employing (\ref{Eplus soln}) and (\ref{eq:striphsdiff2}) with $F(z) =1$, we can write it as follows
\be
\label{zprime bh Eplus linear}
z'^2 = 
F(z)\bigg[
\bigg( \frac{z_\ast}{z}\bigg)^{2d_\theta}-1\bigg]
+
g(z)\bigg[
\bigg( \frac{z_\ast}{z_c}\bigg)^{2d_\theta}-1\bigg]
\equiv 
H(z)\,,
\qquad
x_c < x \leqslant \ell/2\,,
\ee
where 
\be
g(z)\equiv\frac{(F(z_c)-1)^2}{4}
\bigg(\frac{z_c}{z}\bigg)^{2(1-\zeta)} \,.
\ee
Notice that the dependence on $z$ of $g(z)$ disappears when $\zeta =1$.
Assuming that $H(z)$ has a minimum at  $z=z_m$ with $z_m < z_\ast$, its defining equation $\partial_z H(z)|_{z_m} = 0$ gives
\be	
z_\ast^{2d_\theta}
=\frac{z_mF'(z_m)+2(\zeta-1)g(z_m)}{z_mF'(z_m) -2d_\theta F(z_m)
+2(\zeta-1) (z_m/z_c)^{2d_\theta}g(z_m)}
\, z_m^{2d_\theta} \,.
\label{eq:zmstrip}
\ee
Assuming also that at $z=z_m$, it is possible to find $z_c = z_c^\ast$ such that $H(z_m)=0$ (thus $z_c^\ast = z_c^\ast(z_m)$). Then, $z_c^\ast$ is given by
\be
\frac{2d_\theta F(z_m)\big[ F(z_m)+g(z_m)|_{z_c = z_c^\ast}\big]
+\big[(z_m/z_c^\ast)^{2d_\theta}-1\big]
\big[2(1-\zeta)F(z_m)+z_mF'(z_m) \big]g(z_m)|_{z_c = z_c^\ast}}{
z_m F'(z_m)  -2d_\theta (z_m)F(z_m)+2(\zeta-1) (z_m/z_c^\ast)^{2d_\theta}
g(z_m)|_{z_c = z_c^\ast}}
=0 \,.
\label{eq:zcaststrip}
\ee
When $F(z)$ is given by (\ref{eq:emblackeningBH}), (\ref{eq:zmstrip}) and (\ref{eq:zcaststrip}) become respectively
\be	
z_\ast^{2d_\theta}=
\frac{(d_\theta+\zeta)(z_m/z_h)^{d_\theta+\zeta}
+(1-\zeta)(z_c/z_h)^{2(d_\theta+\zeta)}(z_c/z_m)^{2(1-\zeta)}}{
4d_\theta-2(d_\theta-\zeta)(z_m/z_h)^{d_\theta+\zeta}+(1-\zeta)(z_m/z_h)^{2(d_\theta+\zeta)}(z_c/z_m)^{2(1-\zeta)}} 
\, z_m^{2d_\theta}\,,
\ee
and 
\bea
2 d_\theta \big[1-(z_m/z_h)^{d_\theta+\zeta}\big]^2
\,= & & \\
& & 
\hspace{-4cm}
=\,-\frac{(z_c^\ast)^{2(d_\theta+\zeta)}}{(z_c^\ast/z_m)^{2(1-\zeta)}}
\left\{
1-\bigg(  \frac{z_m}{z_h} \bigg)^{d_\theta+\zeta}
+\left[ \bigg(  \frac{z_m}{z_c^\ast} \bigg)^{2d_\theta}  -1 \right]
\left[
2(1-\zeta)-(d_\theta+2-\zeta) \bigg(  \frac{z_m}{z_h} \bigg)^{d_\theta+\zeta} 
\right]
\right\}\,.
\nonumber
\label{eq:zcstargeneral}
\eea
At this point, let us consider the limit $z_\ast\to\infty$ with both $z_m$ and $z_c^\ast$ kept fixed.
For the moment we just assume to be in a regime where this is allowed.
The equations (\ref{eq:zmstrip}) and (\ref{eq:zcaststrip})  become respectively
\be
\label{eqs zast large}
z_mF'(z_m) -2d_\theta F(z_m)=2(1-\zeta)
\bigg(\frac{z_m}{z_c}\bigg)^{2d_\theta} g(z_m)\,,
\hspace{1.5cm}
F(z_m)=-\left(\frac{z_m}{z_c^\ast}\right)^{2d_\theta} g(z_m)\big|_{z_c^\ast} \,.
\ee
Plugging the second equation in (\ref{eqs zast large}) into the first one, one finds
\be 
z_m F'(z_m)+2(1-\zeta-d_\theta) F(z_m)=0\,,
\qquad
\textrm{at \; $z_c = z_c^\ast$}\,,
\label{eq:largezs3}
\ee
which can be written also in the following form
\be
\partial_{z_m}
\bigg( \frac{F(z_m)}{z_m^{2(d_\theta+\zeta-1)}}\bigg) = 0\,.
\ee
For $F(z)$ given by (\ref{eq:emblackeningBH}) this equation tells us that
\be 
\frac{z_m}{z_h}=
\left(\frac{2(d_\theta+\zeta-1)}{d_\theta+\zeta-2} \right)^{\frac{1}{d_\theta+\zeta}} 
=
\left(\frac{\eta}{\eta-1} \right)^{\tfrac{1}{2}(2- \eta)}  \,,
\hspace{1.3cm}
\eta \equiv \frac{2(d_\theta+\zeta-1)}{d_\theta+\zeta} \,.
\label{eq:etacond}
\ee
Notice that in this expression, the dimensionality, the Lifshitz and the hyperscaling exponents occur only through the combination $d_\theta + \zeta$. 
In order to have a positive expression within the brackets of the first equation in (\ref{eq:etacond}), we need to require $\eta > 1$, i.e. 
\be 
d_\theta +\zeta > 2 \,.
\label{eq:etacondition}
\ee
Plugging (\ref{eq:etacond}) into the second equation of (\ref{eqs zast large}) computed for (\ref{eq:emblackeningBH}), we find that
\be 
\label{zcast def}
\frac{z_c^\ast}{z_h} 
=
\frac{2(\eta-1)^{\frac{1}{2}(\eta-1)}}{\eta^{\frac{1}{2}\eta}} \,.
\ee
\begin{figure}[t] 
\begin{center}
\vspace{-0.cm}
\hspace{-.5cm}
\includegraphics[width=1\textwidth]{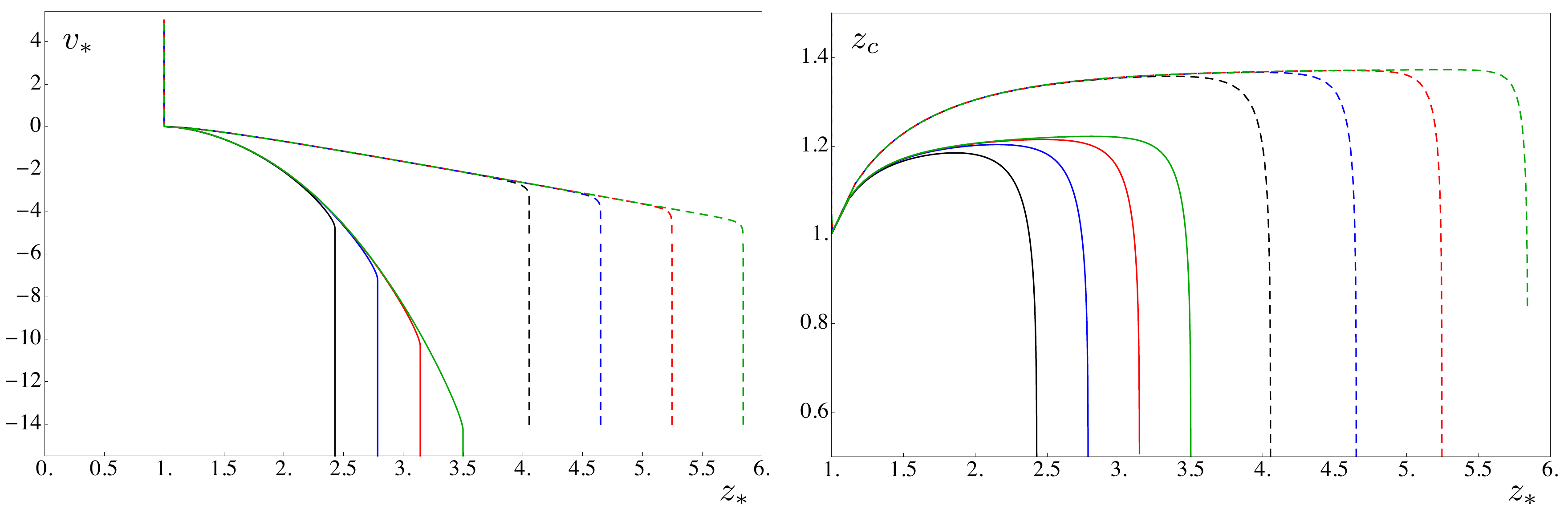}
\end{center}
\vspace{-.3cm}
\caption{\label{fig:criticalcurve}
Thin shell regime: $v_\ast$ and $z_c$ for the strip in terms of $z_\ast$ at constant size $\ell$. Here $d=2$ and The horizon is $z_h = 1$.
Dashed curves correspond to $\theta=0$ and $\zeta =1$, while continuous curves have $\theta=1$ and $\zeta=3$. Different colors denote different strips:
$\ell = 4$ (black), $\ell = 5$ (blue), $\ell = 6$ (red) and $\ell = 7$ (green).
}
\end{figure}
It is useful to plot curves $C_\ell$ with constant $\ell$ in the plane $(z_\ast, z_c)$ or $(v_\ast, z_c)$ as done in Fig. \ref{fig:criticalcurve}. 
As $t$ evolves, $z_\ast$ decreases along each curve. After some time (which changes with $\ell$), all the curves lie on a limiting one $C^\ast$. For any fixed $\ell$, it will be shown that $A_{\textrm{\tiny reg}}(t)$ is linear when the curve $C_\ell$ coincides with $C^\ast$.
From Fig. \ref{fig:criticalcurve} it is clear that, as $\ell$ increases, also the linear regime increases.
Thus, now we are considering 
\be
z_\ast\to\infty\,,
\qquad
\eta>1 \,,
\qquad
z_c= (1-\varepsilon) z_c^\ast\,,
\label{eq:limitlineartrue}
\ee
where $0<\varepsilon \ll 1$. When $z_\ast$ is large, for $F(z)$ given by (\ref{eq:emblackeningBH}), from (\ref{eq:etacond}) and (\ref{zcast def}) we have that
\be
\frac{z_m}{z_c^\ast} = \frac{\eta}{2\sqrt{\eta -1}} > 1\,.
\ee
This tells us that the solutions $z(x)$ are not injective for  $0 \leqslant x \leqslant \ell/2$, which implies that we cannot employ (\ref{elle shell tot}), (\ref{t shell tot}) and (\ref{A shell tot}) because they have been derived assuming that $z(x)$ is invertible. In this case we have to use the following ones (see \cite{Liu:2013qca} for a detailed discussion)
\bea 
 \label{eq:stripmultl} 
 \frac{\ell}{2} &=& 
 \int_{z_c}^{z_\ast}\frac{z^{d_\theta}}{\sqrt{z_\ast^{2d_\theta} -z^{2d_\theta} }}\,dz
 +\int_{z_c}^{z_m}\frac{dz}{\sqrt{H(z)}} 
 +\int_{0}^{z_m}\frac{dz}{\sqrt{H(z)}} \,,
\\
\label{eq:stripmultt}
\rule{0pt}{.7cm}
t &=& 
\int_{z_c}^{z_m}
\frac{1}{z^{1-\zeta}F(z)}  \bigg(\frac{E_+}{z^{1-\zeta}\sqrt{H(z)}}+1\bigg)dz
+\int_{0}^{z_m}
\frac{dz}{z^{1-\zeta}F(z)} \bigg(\frac{E_+}{z^{1-\zeta}\sqrt{H(z)}}+1\bigg) dz\,, 
\\
\rule{0pt}{.7cm}
 \label{eq:stripmulta} 
 \mathcal{A} &=& 
 2\ell_\perp^{d-1} z_\ast^{d_\theta}
 \left(
 \int_{z_c}^{z_\ast}  \frac{dz}{z^{d_\theta} \sqrt{z_\ast^{2d_\theta} - z^{2d_\theta} }}
 \int_{z_c}^{z_m} \frac{dz}{z^{2d_\theta}\sqrt{H(z)}} 
 +\int_{0}^{z_m} \frac{dz}{z^{2d_\theta}\sqrt{H(z)}}\right)\,.
 \eea
 Comparing these equations with (\ref{elle shell tot}), (\ref{t shell tot}) and (\ref{A shell tot}), notice that only the part outside the shell is different.\\
Since the point $z=z_m$ and $z_c=z_c^\ast$ is a quadratic zero of $H(z)$, it provides 
a leading contribution to the integrals in (\ref{eq:stripmultl}), (\ref{eq:stripmultt}) and (\ref{eq:stripmulta}). Thus, expanding $H(z)$ around $z=z_m$, we find
\be 
H(z)=H_2 (z-z_m)^2+ b \varepsilon \,.
\ee
By employing that for a smooth function $f(z)$ we have
\be 
\int \frac{f(z)}{\sqrt{H_2 (z-z_m)^2+b\epsilon}}\,dz
=
\frac{f(z_m)}{\sqrt{H_2}} 
\textrm{arcsinh} \big(H_2(z-z_m)/(b\varepsilon)\big)
+\dots
=
-\frac{f(z_m)}{\sqrt{H_2}} \log \varepsilon + \dots\,,
\ee
we conclude that
\bea
\ell /2&=& 
\frac{\sqrt{\pi}\, \Gamma\big(1/(2 d_\theta)+1/2\big)}{ \Gamma\big(1/(2 d_\theta)\big)} \,z_\ast
-\frac{\log \varepsilon}{\sqrt{H_2}}\,,
\label{eq:linearl}
\\
\rule{0pt}{.6cm}
t&=& 
-\frac{E_+(z_c^\ast)}{z_m^{2(1-\zeta)}F(z_m) \sqrt{H_2}}
\,\log \varepsilon
= -\frac{z_\ast^{d_\theta}}{z_m^{d_\theta+1-\zeta}\sqrt{-H_2F(z_m)}}
\,\log \varepsilon\,,
\label{eq:lineart}
\\
\rule{0pt}{.6cm}
\mathcal{A}^{(3)}_{\textrm{\tiny reg}}
&=&
-2\ell_\perp^{d-1} \frac{z_\ast^{d_\theta}}{z_m^{2d_\theta} \sqrt{H_2}}
\,\log \varepsilon\,,
\label{eq:lineara3}
\eea
where in the second equality of (\ref{eq:lineart}) we used the second equation of (\ref{eqs zast large}). 
Combining (\ref{eq:lineart}) and (\ref{eq:lineara3}), we also obtain that
\be
\mathcal{A}^{(3)}_{\textrm{\tiny reg}}
=2\ell_\perp^{d-1}\frac{\sqrt{-F(z_m)}}{z_m^{d_\theta+\zeta-1}}\,t
\,\equiv 
2\ell_\perp^{d-1} \frac{v_E}{z_h^{d_\theta+\zeta-1}}\, t \,.
\label{eq:finalareaapp}
\ee
For a $F(z)$ given by (\ref{eq:emblackeningBH}), the linear growth velocity reads
\be
v_E
=
 \bigg( \frac{z_h}{z_m} \bigg)^{d_\theta +\zeta -1}  \sqrt{-F(z_m)} 
=\frac{(\eta-1)^{\frac{\eta-1}{2}}}{\eta^{\frac{\eta}{2}}}\,,
\ee
where $\eta$ has been defined in (\ref{eq:etacond}).

\subsection{Saturation}
\label{app sec saturation}

\subsubsection{Large regions in static backgrounds}

In order to understand the regime of saturation, when the holographic entanglement entropy approaches the thermal value, let us consider the static case when the size of the boundary region is large with respect to $z_h$. 
In this case a large part of the extremal surface is very close to the horizon.
\\
Starting with the strip, when $\ell \gg z_h$, we have that
(we recall that tilded values of $z$ refer to the static black hole case, following the notation introduced in \S\ref{stripvaidyasec})
\be
\label{strip eps1}
\tilde{z}_\ast = (1-\varepsilon) z_h\,,
\ee
where $\varepsilon$ is a positive infinitesimal parameter. Expanding (\ref{eq:lstripHS}), we find
\be
\label{eq:StripBHlarge}
\frac{\ell}{2} 
=-\frac{z_h \log\varepsilon }{\sqrt{2d_\theta z_h F'_h}} +\dots \,,
\qquad
F'_h \equiv - \partial_z F(z) \big|_{z=z_h}\,.
\ee 
In a similar way, plugging (\ref{strip eps1}) into (\ref{area func}) and keeping the first divergent term as $\varepsilon  \rightarrow 0$, we get
\be 
\mathcal{A} =
-\frac{\sqrt{2}\, \ell_\perp^{d-1} \log\varepsilon}{z_h^{d_\theta-1}\sqrt{d_\theta F'_h}}+\dots =\frac{\ell_\perp^{d-1} \ell}{z_h^{d_\theta}} +\dots\;.
\ee
For a sphere, the analysis is slightly more complicated because we have to expand the differential equation for the minimal surface \cite{Liu:2013una}. Setting
\be
z(\rho)=z_h-\varepsilon \, a(\rho) +O(\varepsilon^2)\,,
\label{eq:spherperturb}
\ee
and expanding (\ref{eq:sferatot}), the first order reads
\be
2 z_h \big[(d-1)a'+\rho a''\big] a - z_h\rho a'^2 - 2d_\theta F'_h a^2  = 0 \,.
\ee
This equation cannot be solved exactly, but, at large $\rho$, we can find that the solution behaves as  
\be 
a(\rho)= C\,\frac{e^{\rho \sqrt{2 d_\theta F'_h/z_h}}}{\rho^{d-1}} + \dots \,,
\ee
where $C$ is an arbitrary constant. 
Keeping only the first order in $\varepsilon$ in (\ref{eq:spherperturb})  and imposing $z(R) =0$, one finds $a(R) = z_h /\varepsilon$, whose logarithm gives
\be
-\log \varepsilon = R \sqrt{-2 d_\theta F'(z_h)/z_h} - (d-1) \log R + \dots \,.
\label{eq:SphereBHlarge}
\ee
As for the area, plugging (\ref{eq:spherperturb}) into (\ref{eq:spherarea}) and keeping the first divergent term as $\varepsilon  \rightarrow 0$, (\ref{eq:SphereBHlarge}) allows us to conclude that
\be 
\mathcal{A}=  \frac{2\pi^{d/2} R^d}{d \Gamma(d/2) z_h^{d_\theta} } + \dots\;.
\ee

\subsubsection{Saturation time}
\label{app sat time}

In the thin shell regime and whenever the saturation to the thermal value of the holographic entanglement entropy is smooth (the derivative does not jump), we can define the saturation time $t_s$ as the time such that $\tilde{v}_\ast=0$.
For $t>t_s$, the extremal surface is entirely within the black hole region. 
Thus, the equation for $t_s$ reads
\be
\label{t_s def}
0=\tilde{v}_\ast(t_s)=t_s-\int_{0}^{\tilde{z}_\ast} \frac{dz}{z^{1-\zeta}F(z)} \,. 
\ee
For $F(z)$ given by (\ref{eq:emblackeningBH}) the integral can be solved explicitly, finding
\be
t_s=
\frac{(\tilde{z}_\ast)^\zeta}{\zeta} 
\;_{2}F_{1}
\Big(1, \zeta/(d_\theta+\zeta);1+\zeta/(d_\theta+\zeta);
(\tilde{z}_\ast / z_h)^{d_\theta+\zeta}\Big) \,.
\label{eq:ts}
\ee
For very large regions, $\tilde{z}_\ast = z_h (1-\varepsilon)$ and therefore (\ref{eq:ts}) expanded to the first order in $\varepsilon$ gives
\be
t_s=-\frac{z_h^{\zeta-1} \log \varepsilon}{F'_h}
=-\frac{z_h^{\zeta} \log \varepsilon}{d_\theta+\zeta}\,,
\ee
where in the second step we have employed (\ref{eq:emblackeningBH}).
If the region on the boundary is a strip, we can use (\ref{eq:StripBHlarge}) to obtain
\be
\label{ts strip}
t_s= 
z_h^{\zeta-1}\sqrt{\frac{2 d_\theta}{ z_h F'_h }}\;\frac{\ell}{2}
+ \dots 
= z_h^{\frac{\zeta-d_\theta}{2}-1}\sqrt{\frac{d_\theta }{2(d_\theta +\zeta)}}\; \ell
+ \dots\;.
\ee
For a sphere, (\ref{eq:SphereBHlarge}) gives us
\be 
t_s =z_h^{\zeta-1} \sqrt{\frac{2 d_\theta}{ z_h F'_h}}\;R
-\frac{(d-1) z_h^{\zeta-1}}{ F'_h}  \log R+\dots\;.
\ee

\subsubsection{Saturation of the holographic entanglement entropy: strip}
\label{app sat hee}

In this section we try to estimate $A^{(2)}_{\textrm{\tiny reg}}(t)$ as a function of $t-t_s$, being $t_s$ the saturation time computed above. As the holographic entanglement entropy approaches  its thermal value, the extremal surface is almost entirely within the black hole region. 
This means that the point $z_c$, where the extremal surface crosses the shell, is very close to $z_\ast$. \\
Let us consider the strip first and introduce a positive infinitesimal parameter $\varepsilon$ as follows
\be
z_c=z_\ast \left(1-\frac{\varepsilon^2}{2d_\theta} \right) \,.
\ee
Plugging this expansion into (\ref{Eplus soln}), at first order we get
\be 
E_+=
\frac{z_c^{1-\zeta}(F(z_c)-1)}{2}\sqrt{\left(\frac{z_\ast}{z_c}\right)^{2d_\theta}-1}
=\frac{z_\ast^{1-\zeta}(F(z_\ast)-1)}{2} \,\varepsilon +O(\varepsilon ^2) \,.
\ee
Since we are approaching the extremal surface corresponding to the one of the static black hole, $z_\ast$ is close to its thermal value $\tilde{z}_\ast$, namely we are allowed to introduce another positive infinitesimal parameter $\delta$ as
\be
z_\ast=\tilde{z}_\ast \left(1-\frac{\delta}{2d_\theta} \right)\,.
\ee
We want to estimate $t-t_s$ in terms of the infinitesimal parameters $\varepsilon$ and $\delta$. Using (\ref{t shell tot}) and (\ref{eq:ts}), we find that
\bea
t-t_s &=&
 \int_{0}^{z_c}\frac{z^{\zeta-1}}{F(z)} 
 \left(\frac{E_+ }{z^{1-\zeta}\sqrt{H(z)}}+1\right)-
 \int_{0}^{\tilde{z}_\ast}\frac{z^{\zeta-1}}{F(z)}\,dz \,,
 \\
 \rule{0pt}{.6cm}
&=&\int_{\tilde{z}_\ast}^{z_c}\frac{z^{\zeta-1}}{F(z)}\,dz 
+\int_0^{z_\ast}  \frac{E_+z^{2(\zeta-1)}}{F(z)\sqrt{H(z)}}\,dz
-\int^{z_\ast}_{z_c}  \frac{E_+z^{2(\zeta-1)}}{F(z)\sqrt{H(z)}}\,dz\,,
\\
\label{t minus ts}
&=&-\frac{\tilde{z}_\ast^\zeta}{2d_\theta F(\tilde{z}_\ast)}\;\delta
+\frac{\tilde{z}_\ast^{1-\zeta}(F(\tilde{z}_\ast)-1)Q_1(\tilde{z}_\ast)}{2}\;\varepsilon +\dots\,,
\eea
where $H(z)$ is defined as the r.h.s. of (\ref{zprime bh Eplus}) (see also (\ref{zprime bh Eplus linear})), $Q_1(z_\ast)$ is defined as follows 
\be
Q_1(z_\ast)\equiv
\int_0^{z_\ast} \frac{z^{2(\zeta-1)}}{F(z)\sqrt{F(z)\big[ (z_\ast/z)^{2d_\theta}-1\big]}}\,dz\,,
\ee
and the dots denote higher orders in $\varepsilon$ and $\delta$.
Following \cite{Liu:2013qca}, one can find a relation between $\delta$ or $\varepsilon$ from the expansion of (\ref{elle shell tot}). The presence of $\zeta$ does not modify the result, which reads
\be
\delta = \frac{1-F(\tilde{z}_\ast)}{F(\tilde{z}_\ast)Q_2'(\tilde{z}_\ast)}\,\varepsilon
+O(\varepsilon^2)\,,
\ee 
where (see \cite{Liu:2013qca} for further details)
\be
Q_2(z_\ast)\equiv
\int_0^{z_\ast} \frac{dz}{\sqrt{F(z)\big[ (z_\ast/z)^{2d_\theta}-1\big]}}\,.
\ee
Thus, plugging this result into (\ref{t minus ts}), one finds
\be
t-t_s \propto \varepsilon + O(\varepsilon^2) \,,
\ee 
where the coefficient in front of $\varepsilon$ depends on $\zeta$ and $\theta$, as can be clearly seen from (\ref{t minus ts}), but the power of $\varepsilon$ does not. 
Given this result, one can repeat precisely the computation of \cite{Liu:2013qca} and show that in this regime
\be
A_{\textrm{\tiny reg}}^{(2)}(t) \propto \varepsilon^2 + O(\varepsilon^3) \,,
\ee 
i.e.
\be
A_{\textrm{\tiny reg}}^{(2)}(t) \propto (t-t_s)^2 + O\big((t-t_s)^3\big)\,.
\ee 
Notice that the exponent is independent of $\theta$ and $\zeta$.

\subsubsection{Saturation of the holographic entanglement entropy: sphere}
\label{app sat hee sphere}

Given a black hole in the hvLif spacetime, the corresponding area functional is (\ref{eq:spherarea}), whose extremization gives (\ref{eq:sferatot}).
Since (\ref{eq:sferatot}) is invariant under the change $\rho \to -\rho$, its solution $z(\rho)$ is an even function. In particular, its Taylor series expansion contains only positive even powers of  $\rho$.
Introducing $z(0)=\tilde{z}_\ast$, the expansion of $z(\rho)$ for  $\rho \sim 0$ gives
\be 
z(\rho)=\tilde{z}_\ast - \frac{d_\theta}{2d\tilde{z}_\ast}F(\tilde{z}_\ast) \rho^2 + O(\rho^4)\,. \label{eq:bhexp}
\ee
For the Vaidya spacetime in the thin shell regime, the equation for $z(\rho)$ for $0<\rho_c < \rho < R$ is (\ref{eq:spherethineqz}), where $E_+$ has been defined in (\ref{Eplus sphere}).
We recall that the quantities associated to the hvLif vacuum part can be obtained by sustituting $E_+$ with $E_-=0$ in all the corresponding expressions for the black hole part.
The relation defining $v$ in the black hole part of the metric is (\ref{eq:vprimesphere}).
The total area is (\ref{area total sphere}),
while the boundary time $t$ is obtained by integrating $v'$ (see (\ref{eq:vprimesphere})) outside the shell, i.e. (\ref{bdy time sphere}).\\

\noindent
The assumption in the following is that we are at a boundary time such that the minimal surface lies almost entirely outside the shell and has almost reached its static configuration, that is
\be 
\label{eq:latetimez}
z(\rho)=z_0(\rho)+\delta z_1(\rho)+O(\delta^2) \,,
\ee
where  $\delta$ is supposed small and $z_0$ is solution of (\ref{eq:sferatot}) which is just (\ref{eq:spherethineqz}) with $E_+=0$. The boundary conditions are such that $z_0(R)=z_1(R)=0$.
Expanding (\ref{eq:spherethineqz}) to the first order in $\delta$, we find the following differential equation for $z_1$
\be
z_1''+ P(\rho) z_1' + Q(\rho) z_1 = S(\rho)\,,
\qquad
z'_1(0)=z_1(R)=0\,,
\ee 
where 
\bea
\label{Prho def}
P(\rho)&=&
\frac{d-1}{\rho}+
\left( \frac{2d_\theta}{z_0} + \frac{3(d-1)z_0'-\rho F'(z_0)}{\rho F(z_0)} \right) z_0'\,,
\\ 
\label{Qrho def}
Q(\rho)&=&
\frac{d_\theta}{z_0}\left(F'(z_0)-\frac{F(z_0)+z_0'^2}{z_0}\right)-\frac{1}{2F(z_0)}\left(z_0'^2 F''(z_0)-\frac{F'(z_0)z_0'^2\big[ \rho F'(z_0)-2(d-1)z_0' \big]}{\rho F(z_0)}\right)\,,
\hspace{.6cm}
\\
\label{Srho def}
S(\rho)&=&
\frac{E_+^2}{\delta}
\left(1+ \frac{z_0'^2}{F(z_0)}\right)
\left(\frac{(d_\theta+\zeta-1)\rho}{z_0}
+\frac{2(d-1)z_0'-\rho F'(z_0)}{2F(z_0)}\right)
\rho^{2(1-d)} z_0^{2(d_\theta+\zeta-1)} \,.
\eea
Notice that $S(\rho)$ depends on $E_+^2/\delta$. Indeed, since $E_+\to 0$ when $\delta \to 0$, we could have  $E_+^2/\delta = O(1)$ as $\delta \to 0$. 
In the following the correct relation between $E_+$ and $\delta$ will be obtained and $E_+ /\delta =O(1)$ (see (\ref{eq:epluss}) and (\ref{eq:deltad})). \\
It is useful to remind that, given a second order linear differential equation
\be 
\label{second order diff eq}
f''(x)+A(x) f'(x)+ B(x) f(x)= C(x)\,,
\ee
a solution can be written in terms of the solutions $f_{j}(x)$ ($j=1,2$) of the corresponding homogenous differential equation (i.e. (\ref{second order diff eq}) with $C=0$). It reads
\be 
f_{\textrm{\tiny inh}}(x)= f_1(x)\int^{x}_{x_0}\frac{f_2(y)C(y)}{f_1(y)f_2'(y)-f_2(y)f_1'(y)}\,dy
-f_2(x)\int^{x}_{x_0} \frac{f_1(y)C(y)}{f_1(y)f_2'(y)-f_2(y)f_1'(y)} \,dy\,,
\label{eq:inhomo}
\ee
where $x_0$ is arbitrary and  $f_{\textrm{\tiny inh}}(x_0)=0$ is trivially satisfied. 
Then, since (\ref{second order diff eq}) is linear, its most general solution  is $f_{\textrm{\tiny inh}}+A f_1+B f_2$.

\paragraph{Expansion for $\rho\simeq0$.}
In this regime we can expand $z_0(\rho)$ as in (\ref{eq:bhexp}). 
Then, (\ref{Prho def}), (\ref{Qrho def}) and (\ref{Srho def}) become respectively
\bea
P(\rho) 
&=&\frac{d-1}{\rho}
+\frac{d_\theta }{d}\, \frac{(d-3)F(\tilde{z}_\ast)+\tilde{z}_\ast F'(\tilde{z}_\ast)}{\tilde{z}_\ast^2}\,\rho+O(\rho^3)\,,
\\
Q(\rho) 
&=&d_\theta\frac{\tilde{z}_\ast F'(\tilde{z}_\ast)-F(\tilde{z}_\ast)}{\tilde{z}_\ast^2}+O(\rho^2)\,,
\\
S(\rho) 
&=& \frac{E_+^2}{\delta}\, 
\frac{\big[2(\theta-\zeta d) F(\tilde{z}_\ast)+ d \tilde{z}_\ast F'(\tilde{z}_\ast)\big]
\tilde{z}_\ast^{2(d_\theta+\zeta)-3}}{2 d  \, F(\tilde{z}_\ast) \, \rho^{2(d-1)}}
+O(1/\rho^{2(d-2)})\,.
\label{eq:s2}\eea
The homogeneous equation is 
\be
z_1''(\rho)+ 
\frac{d-1}{\rho}
z_1'(\rho) 
+ Q_0  z_1(\rho) = 0\,,
\qquad
Q_0\equiv Q(0)=d_\theta\frac{\tilde{z}_\ast F'(\tilde{z}_\ast)-F(\tilde{z}_\ast)}{\tilde{z}_\ast^2}\,.
\ee 
The independent solutions $j_1,j_2$ of this equation can be expressed in terms of Bessel functions as follows
\be
\label{eq:j1j2}
 j_1(\rho) = \frac{\Gamma(d/2)}{(\sqrt{Q_0} \rho /2)^{\frac{d-2}{2}}} \,J_{\frac{d-2}{2}}(\sqrt{Q_0}\rho)\,,
 \hspace{.3cm}\qquad
 j_2(\rho) =  \left\{ \begin{array}{ll}
 \displaystyle
-\frac{\pi}{2} Y_0 (\sqrt{Q_0} \rho) &d=2\hspace{.2cm}\\
 \displaystyle
 -\frac{\pi}{\Gamma(\frac{d-2}{2})}\left(\frac{\sqrt{Q_0}}{2\rho}\right)^{\frac{d-2}{2}}Y_{\frac{d-2}{2}}(\sqrt{Q_0}\rho) &d>2
 \end{array}\right.
\ee
whose behavior for $\rho \rightarrow  0$ is given respectively by
\be
 \label{eq:j12}
 j_1(\rho) = 1-\frac{Q_0}{2 d }\rho^2+O(\rho^4) \,,
  \hspace{.3cm}\qquad
 j_2(\rho) = \left\{ \begin{array}{ll}
 \log\rho+\log(\sqrt{Q_0}/2)+\gamma_E +\dots &d=2\\
 \rule{0pt}{.5cm}
 \rho^{2-d} +\dots &d>2
 \end{array}\right.
\ee
Considering only the first terms of the expansions (\ref{eq:j12}) and (\ref{eq:s2}) and plugging them into (\ref{eq:inhomo}), one finds
\be 
\label{eq:z1rho}
z_{1,\textrm{\tiny inh}}(\rho)= \left\{ \begin{array}{ll}
\displaystyle 
\frac{E_+^2}{\delta}\,
\frac{\tilde{z}_\ast^{2(1-\theta+\zeta)-1}(\tilde{z}_\ast F'(\tilde{z}_\ast)-(2\zeta-\theta)F(\tilde{z}_\ast))}{4 F(\tilde{z}_\ast)} \, \log^2 \rho  
&\hspace{.5cm} d=2\\
\rule{0pt}{.8cm}
\displaystyle 
\frac{E_+^2}{\delta}\,
\frac{\tilde{z}_\ast^{2(d_\theta+\zeta-1)-1}(d\tilde{z}_\ast F'(\tilde{z}_\ast)-2(d\zeta-\theta)F(\tilde{z}_\ast))}{4d(d-2)^2 F(\tilde{z}_\ast)} \, \rho^{-2(d-2)} 
& \hspace{.5cm} d>2
\end{array} \right.
\ee
In the following $j_1,j_2$ of $z_{1,\textrm{\tiny inh}}$ will indicate only their $\rho$ dependence.

\paragraph{\bf Expansion for $\rho\simeq R$.}
First, let us consider the case $d_\theta\neq1$, when (\ref{eq:limitzx}) can be applied. 
Introducing the variable $\sigma\equiv R-\rho$, from (\ref{Prho def}), (\ref{Qrho def}) and (\ref{Srho def}) we find 
\bea
P(\rho) &=&\frac{d_\theta-3}{2\sigma}+O\big(\sigma^0\big)\,,
\\
Q(\rho) &=&-\frac{d_\theta}{4\sigma^2}+O\big(\sigma^{-1}\big)\,,
\\
S(\rho) &=& \frac{E_+^2}{\delta} 
\left(\frac{2(d_\theta-1)\sigma}{(d-1)R}\right)^{d_\theta+\zeta-5/2}\frac{\zeta(d-1)R^{2(\zeta-\theta-1/2)}}{d_\theta-1}+\dots \label{eq:s1}\,.
\eea
Near the boundary we find the following  homogeneous equation
\be 
\label{eq diff rhoR}
z_1''(\sigma)-\frac{d_\theta-3}{2 \sigma}\, z_1'(\sigma) 
- \frac{d_\theta}{4\sigma^2} \,z_1(\sigma)=0\,,
\ee
whose solutions read
\be
\label{eq:k's}
 k_1(\sigma)\,=\, \sigma^{-1/2}\,,
 \qquad
 k_2(\sigma)\,=\, \sigma^{d_\theta/2} \,.
\ee
Since $z_1(R)=0$ and $k_1(\sigma)$ diverges when $\sigma \rightarrow 0$, the solution of (\ref{eq diff rhoR}) is proportional to $k_2$.
Adapting (\ref{eq:inhomo}) to this case through (\ref{eq:k's}) and (\ref{eq:s1}) we obtain that 
\be
\label{eq:z1bdysigma}
z_{1,\textrm{\tiny inh}}(\sigma) 
=  
\frac{E_+^2}{\delta} \,
\frac{4\zeta R^{3/2-d-\theta+\zeta}}{(d\theta+\zeta)(d\theta+2\zeta-1)}\left(\frac{2(d_\theta-1)}{d-1}\right)^{d_\theta+\zeta-7/2}\sigma^{\zeta-1/2+d_\theta} +\dots \,,
\ee
which vanishes for $\sigma \rightarrow  0$ because $\zeta \geqslant 1$.\\
Note that (\ref{eq:z1bdysigma}) in the limit $\sigma\to 0$, $z_1$ is well behaved and thus in the following calculation the boundary contribution will be ignored being $E_+/\delta \sim \delta\to 0$ when approaching saturation. We have checked that, by employing the parametric reformulation (\ref{0step}), this happens also when $d_\theta =1$.
This is not the case for (\ref{eq:z1rho}) which will play an important role in determine the late time behaviour of the entanglement entropy.

\paragraph{Approaching saturation.}
Let us now try to put things together. First, notice that as the solution approaches its thermal value, we have that
\be
z_c \to \tilde{z}_\ast\,, \qquad z_\ast \to \tilde{z}_\ast\,,
\ee  
where $\tilde{z}_\ast$ is associated to the tip of the static black hole geodesic, and at the same time
\be
\rho_c \to 0\,, \qquad  E_+ \to 0\,, \qquad \delta \to 0\,.
\ee
In the following we will try to relate the above quantities in their approach to equilibrium values. To this purpose it turns out to be useful to relate the three infinitesimal quantities $\rho_c$, $\delta$ and $E_+$ among themselves. \\
First, one introduces a new infinitesimal parameter $\varepsilon$
\be
\rho_c \equiv z_c \varepsilon\,. \label{eq:rhoczc}
\ee
From (\ref{eq:bhexp}) with $F=1$ we have that
\be
z_\ast = z_c \left(1+\frac{d_\theta}{2d}\varepsilon^2+O(\varepsilon^4)\right) \,,
\ee
and also 
\be
z_-'(\rho_c)= - \frac{d_\theta}{d z_\ast} \rho_c +O(\rho_c^3)=   - \frac{d_\theta}{d } \varepsilon+ O(\varepsilon^3) \,,
\label{eq:zminus}
\ee
where we recall that $z_-$ refers to the value of the solution at $\rho=\rho_c$ coming from the hvLif part living in $[0,\rho_c]$.
From (\ref{eq:vprimesphere}) with $E_+=0$ and $F=1$ we find
\be
\label{vc prime app}
v_c'= 
\frac{d_\theta z_c^{\zeta}}{d z_\ast}\, \varepsilon
+ O(\varepsilon^3)=\frac{d_\theta}{d }z_c^{\zeta-1} \varepsilon+ O(\varepsilon^3)\,,
\ee
and finally, plugging (\ref{vc prime app}) and (\ref{eq:rhoczc}) into  (\ref{eq:vprimesphere}), we get that
\be 
E_+= \frac{d_\theta}{2d}(F(z_c)-1) z_c^{\theta-\zeta} \varepsilon^d\,. \label{eq:epluss}
\ee
By employing (\ref{eq:latetimez}), (\ref{eq:j1j2}) and (\ref{eq:j12}) we can write $z(\rho)$ at $\rho=\rho_c$ as follows
\be 
z(\rho_c) =z_0(\rho_c) 
+ \delta \big[ \alpha_1 j_1(\rho_c) + \alpha_2 j_2(\rho_c)+z_{1,\textrm{\tiny inh}}(\rho_c)\big]\,,
\label{eq:zcvareps}
\ee
where the constants $\alpha_1$ and $\alpha_2$ are constrained by the boundary condition $z_1(R)=0$.
Since $z'(\rho)$ has a jump at $\rho=\rho_c$, the matching constraint (\ref{eq:matching1sphere}) allows to relate (\ref{eq:zminus}) and (\ref{eq:zcvareps}) (the latter one gives $z'_+(\rho_c)$), namely we have 
\be
z'_+(\rho_c)-z_-' (\rho_c) = \frac{z_c^{1-\zeta} v_c'}{2} \big(1-F(z_c)\big)  \,,
\ee
which gives 
\be
z_
+'(\rho_c) 
=  z_-'(\rho_c) + \frac{z_c^{1-\zeta} v_c'}{2} \big(1-F(z_c)\big)  \\ 
= z_0'(\rho_c) 
+ \delta \big[\alpha_1 j_1'(\rho_c) + \alpha_2 j_2'(\rho_c)+z_{1,\textrm{\tiny inh}}\,'(\rho_c)\big] \,.
\label{eq:zcvarepsp}
\ee
When $d>2$, (\ref{eq:zcvareps}) and (\ref{eq:zcvarepsp}) become respectively
\bea
 \label{eq:zcastd}
& & 
z_c\;=\; \tilde{z}_\ast- \frac{d_\theta}{2d}F(\tilde{z}_\ast) \frac{z_c^2}{\tilde{z}_\ast} \varepsilon^2+\delta \left(\alpha_1 + \alpha_2 	z_c^{2-d}\varepsilon^{2-d}\right)+O(\varepsilon^4)\,,\\
 \label{eq:zcastd2}
\rule{0pt}{.65cm}
& & 
\frac{d_\theta}{d}\left(\frac{1-F(z_c)}{2} -1 \right) \frac{z_c}{z_\ast}\varepsilon
\;=\;  -\frac{d_\theta}{d}F(\tilde{z}_\ast)  \frac{z_c}{\tilde{z}_\ast} \varepsilon +\delta \alpha_2 (2-d)z_c^{1-d} \varepsilon^{1-d}+O(\varepsilon^3)\,,
\eea
Since $z_c \to \tilde{z}_\ast$ when $\varepsilon\to 0$, at first order we have $z_c/\tilde{z}_\ast=1$ in (\ref{eq:zcastd2}), and therefore
\be
\delta = 
 \frac{d_\theta(1-F(\tilde{z}_\ast))\tilde{z}_\ast^{d-1}}{2d \alpha_2 (d-2)} \,\varepsilon^d  
 +O(\varepsilon^{d+2})\,.
 \label{eq:deltad}
\ee 
Plugging this result into  (\ref{eq:zcastd})	 we obtain
\be
z_c =  \tilde{z}_\ast
\left[1- \frac{d_\theta}{2d}\left(F(\tilde{z}_\ast)  +\frac{1-F(\tilde{z}_\ast)}{2-d}\right)\varepsilon^{2} +O(\varepsilon^4)\right]\,.
\label{eq:zczs}
\ee
Instead, for $d=2$ (\ref{eq:zcvareps}) and (\ref{eq:zcvarepsp}) become 
respectively 
\bea 
&&
z_c= \tilde{z}_\ast- \frac{2-\theta}{4} F(\tilde{z}_\ast)  \frac{z_c^2}{\tilde{z}_\ast} \varepsilon^2+\delta 
\left[\alpha_1 + \alpha_2\left(\log\varepsilon+\gamma_E +\log\frac{\sqrt{Q_0}z_c}{2}\right)\right]
+O(\varepsilon^4 \log^2 \varepsilon) \label{eq:d2zc}\,,\\
&&
\rule{0pt}{.6cm}
\frac{2-\theta}{2}\left(\frac{1-F(z_c)}{2} -1 \right) \frac{z_c}{z_\ast}\varepsilon =  -\frac{2-\theta}{2}F(\tilde{z}_\ast)  \frac{z_c}{\tilde{z}_\ast} \varepsilon + \frac{\alpha_2 \delta}{z_c \varepsilon} +\dots \,.
 \label{eq:d2zc2}
 \eea
Notice that the constant factor multiplying $\alpha_2$ in (\ref{eq:d2zc}) can be reabsorbed by a redefinition of $\alpha_1$
\be 
\tilde{\alpha}_1 \equiv \alpha_1 + \alpha_2 \gamma_E + \alpha_2 \log\frac{\sqrt{Q_0}}{2}\,.
\ee
From (\ref{eq:d2zc2}) we find
\be
\delta = - \frac{(2-\theta)[1-F(\tilde{z}_\ast)]\tilde{z}_\ast}{4\alpha_2} \,\varepsilon^2  +O(\varepsilon^4\log\varepsilon)\,,
\ee
which can be plugged into (\ref{eq:d2zc}), giving 
\be
z_c= \tilde{z}_\ast
\left\{ 1-\frac{(2-\theta)[1-F(\tilde{z}_\ast)] }{4}\,\varepsilon^2 \log\varepsilon
-\frac{2-\theta}{4}
\left[
F(\tilde{z}_\ast)+\frac{1-F(\tilde{z}_\ast)}{\alpha_2}
(\tilde{\alpha}_1+\alpha_2\log \tilde{z}_\ast )
\right]
\varepsilon^2
\right\}\,.
\label{eq:d2zczs}
\ee

\paragraph{Time.}
Now we can proceed by evaluating the boundary time at first nontrivial order in $\varepsilon$. \\
By using (\ref{bdy time sphere}) we get
\be
\label{eq:saturation time}
t =
t_s + 
\frac{z_c-\tilde{z}_\ast}{\tilde{z}_\ast^{1-\zeta}F(\tilde{z}_\ast)}
+ E_+ \int_{\rho_c}^{R} 
\frac{ z_0^{d_\theta+2(\zeta-1)} \sqrt{1+z_0'^2/F(z_0)}}{\rho^{d-1}F(z_0)}\, d\rho+O(E_+^2) \,,
\ee
where we have employed the definition of saturation time given in (\ref{t_s def}) and we have approximated $z$ with $z_0$ in the integral occurring in (\ref{eq:saturation time}) because $E_+\propto \varepsilon^d$.
The integrand in (\ref{eq:saturation time}) can be written as $h(\rho)/\rho^{d-1}$ where $h(0)$ is finite. Thus, when $\rho_c\rightarrow 0$, the divergent part of the integral can be computed as $h(\rho_c)$ times the divergent part of integral of $1/\rho^{d-1}$ between $\rho_c$ and $R$.
This gives for (\ref{eq:saturation time}) the following result
\be
\label{eq:timecorrection} 
t-t_s =
\frac{z_c-\tilde{z}_\ast}{\tilde{z}_\ast^{1-\zeta} F(\tilde{z}_\ast)}
+ \frac{E_+ \tilde{z}_\ast^{d_\theta+2(\zeta-1)}}{F(\tilde{z}_\ast)}
\times  \left\{ \begin{array}{ll}
\displaystyle
 -  \log\varepsilon  - I_0  +\dots \qquad &d=2\\ 
\rule{0pt}{.6cm}
\displaystyle
\frac{(\tilde{z}_\ast\varepsilon)^{2-d}}{d-2}+\dots\qquad &d> 2 
\end{array}\right. 
\ee
where $I_0$ is a numerical constant containing the $O(\varepsilon^0)$ terms of the expansion. Now, using (\ref{eq:epluss}),(\ref{eq:zczs}) and (\ref{eq:d2zczs}) we obtain
\bea
t-t_s &=& \left\{ \begin{array}{ll} 
\displaystyle
-  \frac{(2-\theta)\tilde{z}_\ast^\zeta}{4}
\left[1+\frac{1-F(\tilde{z}_\ast)}{F(\tilde{z}_\ast)} \left(\frac{\tilde{\alpha}_1}{\alpha_2} +  I_0\right)  \right]
\varepsilon^2  \hspace{1cm}&  d=2\\
\rule{0pt}{.6cm}
\displaystyle
-\frac{d_\theta}{2d}\,
\tilde{z}_\ast^{\zeta} \varepsilon^2 &d> 2 \end{array}\right. 
\label{eq:timefinal}
\eea

\paragraph{Area.}

The same strategy can be followed to compute the area. From (\ref{area total sphere}) we find
\bea
\mathcal{A} 
&=&
 \frac{2\pi^{d/2}}{\Gamma(d/2)}
\left[\, \int_{0}^{\rho_c} \frac{\rho^{d-1}\, \sqrt{1+z'^2}}{z^{d_\theta}}\, d\rho +
\int_{\rho_c}^{R}d\rho 
\frac{\rho^{d-1}\, \sqrt{1+ z'^2/F(z)}}{z^{d_\theta}\,\sqrt{1+ A^2 E_+^2/F(z)}}  \, \right]
\\
\label{Ajs def}
&=&
\frac{2\pi^{d/2}}{\Gamma(d/2)}
\left[ \,\mathcal{C}_0+\mathcal{C}_1- \frac{E_+^2}{2}\,\mathcal{C}_2 + O(E_+^4) \right]\,,
\eea
where we have kept only the first non trivial order in $E_+\propto \varepsilon^d$ and the $\mathcal{C}_i$ are defined as follows
\be
\mathcal{C}_0 \equiv  
\int_{0}^{\rho_c} \frac{\rho^{d-1}\, \sqrt{1+z'^2}}{z^{d_\theta}}\, d\rho\,, 
\hspace{.5cm}
\mathcal{C}_1 \equiv
\int_{\rho_c}^{R}
\frac{\rho^{d-1}}{z^{d_\theta}}\, \sqrt{1+ \frac{z'^2}{F(z)}}\,d\rho  \,,
\hspace{.5cm}
\mathcal{C}_2 \equiv
 \int_{\rho_c}^{R}d\rho	 
\rho^{1-d} z^{d_\theta+2(\zeta-1)} \frac{\sqrt{1+ z'^2/F(z)}}{F(z)}\,.
\ee
From (\ref{eq:rhoczc}), (\ref{eq:zminus}), (\ref{eq:zczs}) and (\ref{eq:d2zczs}) we get
\be
\label{eq:a0}
\mathcal{C}_0   
\,\simeq
\left.\frac{\rho^{d-1}\, \sqrt{1+z'^2}}{z^{d_\theta}} \right|_{\rho\to \rho_{c}^{-}} \rho_c
\,=
\frac{z_c^{\theta}  \varepsilon^d}{d} 
-
 \frac{d_\theta^2 z_c^{\theta} \varepsilon^{d+2}}{2d^2(d+2)}
+O(\varepsilon^{d+4}) \,.
\ee
As for the functional $\mathcal{C}_1$, notice that its integrand is the same occurring in (\ref{eq:spherarea}) (which is minimized by $z_0$) but the integration domain is $(\rho_c, R)$ instead of $(0,R)$. 
Since $\rho_c \rightarrow 0$, we have that $\mathcal{C}_1$ is equal to the static black hole area $\mathcal{A}_{\textrm{\tiny bh}}$ plus small corrections, which can be originated both from the fact that now the integration domain is not $(0,R)$ and also from evaluating the integral at $z=z_0+\delta z_1$. The second kind of contribution, obtained by computing the variation of the integrand on $z_0$, gives only a boundary term (computed at $\rho=\rho_c$). Thus we have
\bea
\mathcal{C}_1
& =&\mathcal{A}_{\textrm{\tiny bh}}  
- \int_{0}^{\rho_c}
\frac{\rho^{d-1}}{z_0^{d_\theta}}\,\sqrt{1+ \frac{z_0'^2}{F(z_0)}}\, d\rho 
+ \delta z_1
 \left(\frac{\rho^{d-1}}{z^{d_\theta}} \left. \partial_{z'} \sqrt{1+ z'^2/F(z)}\right|_{z=z_0} \right)
 \bigg|_{\rho=\rho_c}\\
& =& \mathcal{A}_{\textrm{\tiny bh}} 
- \frac{z_c^\theta\varepsilon^d}{d}
- \frac{d_\theta^2 [(d+1)F(z_c)-(d+2)]}{2d^2(d+2)}
 \,z_c^\theta \varepsilon^{d+2}
+\frac{d_\theta }{d}  [z_c-z_0(\rho_c)] \, z^{\theta-1}_c \varepsilon^{d}  
+O(\varepsilon^{d+4})\,,
\hspace{.7cm}
\label{eq:a1}
\eea
where 
\be 
z_c-z_0(\rho_c)= \left\{ 
\begin{array}{ll}
\displaystyle
-\frac{(2-\theta)[1-F(z_c)]}{4} \,z_c \log \varepsilon &d=2\\
\rule{0pt}{.7cm}
\displaystyle
\frac{d_\theta[ 1-F(z_c)]}{2d(d-2)}\, z_c\, \varepsilon^2 &d\neq 2
\end{array} \right.
\ee
As for $\mathcal{C}_2$, since it is already multiplied by $E_+^2$ in (\ref{Ajs def}), 
it is enough to compute it at $z=z_0$ and keep only the most divergent term (at $\rho=\rho_c$).
This turns out to provide the same integral occurring in  (\ref{eq:timecorrection}) and for $\mathcal{C}_2$ we find
\be
\mathcal{C}_2 \big|_{z_0}
\,=\,
-\frac{z_c^{d_\theta+2(\zeta-1)}}{F(z_c)}\times 
\left\{ \begin{array}{ll}
\displaystyle
\log\varepsilon+ I_0  +\dots \qquad &d=2\\ 
\rule{0pt}{.7cm}
\displaystyle
\frac{(z_c \varepsilon)^{2-d}}{2-d}+\dots\qquad &d> 2 \end{array}\right. \label{eq:a2}
\ee
where $I_0$ is  the same quantity as in (\ref{eq:timecorrection}).\\
Finally, putting (\ref{eq:a0}),(\ref{eq:a1}) and (\ref{eq:a2}) together we find
\be
\mathcal{A}_{\textrm{\tiny reg}}^{(2)}
=  
\frac{2\pi^{d/2}}{\Gamma(d/2)} \,
\frac{d_\theta^2[1-F(\tilde{z}_\ast)]\tilde{z}_\ast^{\theta}}{2d^2}
\times 
\left\{ \begin{array}{ll}
\displaystyle
\frac{1-F(\tilde{z}_\ast)}{4F(\tilde{z}_\ast)}\, \varepsilon^4 \log\varepsilon +\dots 
&d=2
\\ 
\rule{0pt}{.7cm}
\displaystyle
\left(\frac{d-2}{d+2}+\frac{1-F(\tilde{z}_\ast)}{4F(\tilde{z}_\ast)}\right)\frac{\varepsilon^{d+2}}{2-d}+\dots \hspace{.7cm} &d> 2 
\end{array}\right. 
\label{eq:areafinal}
\ee
Finally, comparing (\ref{eq:timefinal}) and (\ref{eq:areafinal}), we find (\ref{A2reg saturation sphere}).

\section{Strip in more generic backgrounds}
\label{app generic metrics}

In order to understand the terms of the metric determining the linear regime, let us consider the following static background
\be
\label{generic metric bh}
ds^2 = \frac{1}{z^{2d_\theta/d}}
\left(-\,Q(z)dt^2
-\frac{P(z)^2}{Q(z)}\, dz^2+d\boldsymbol{x}^2\right) \,,
\ee
which reduces to the black hole (\ref{eq:poincareHS}) when $Q(z)=z^{2(1-\zeta)}F(z)$ and $P(z)=z^{1-\zeta}$. By introducing the time coordinate $v$ as
\be
dv=dt-\frac{P(z)}{Q(z)}\,dz\,,
\ee
the metric (\ref{generic metric bh}) can be written as 
\be
ds^2 = \frac{1}{z^{2d_\theta/d}}
\big(-Q(z)dv^2-2P(z)dvdz+d\boldsymbol{x}^2\big) \,.
\ee
Here we consider the Vaidya background obtained by promoting $Q$ to a time dependent function, i.e.
\be
\label{vaidya generic}
ds^2 = \frac{1}{z^{2d_\theta/d}}
\big(-Q(v,z)dv^2-2P(z)dvdz+d\boldsymbol{x}^2\big) \,.
\ee
Considering a strip in the spatial part of the boundary $z=0$, its holographic entanglement entropy is obtained by finding the extremal surface of the following functional area
\be
\label{area func vaidya general}
\mathcal{A}[v(x), z(x)]
=2 \ell_\perp^{d-1}
\int_{0}^{\ell/2}
\frac{\sqrt{\mathcal{B}}}{z^{d_\theta}}
\,dx\,,
\qquad
\mathcal{B} \equiv 1- Q(v,z) v'^{2} -2P(z)z'v'\,,
\ee
and the boundary conditions for $v(x)$ and $z(x)$ are given by (\ref{bc vaidya}).
We only have to adapt the analysis performed in \S\ref{stripvaidyasec} to the background (\ref{vaidya generic}). The equations of motion of (\ref{area func vaidya general}) read
\bea 
\label{eom1 vadya gen}
& & 
\partial_x \big[ Qv'+Pz'\big] = Q_v v'^2/2 \,,
\\
\label{eom2 vadya gen}
& &
\partial_x\big[ P v' \big]= 
d_\theta \mathcal{B}/z + Q_zv'^2/2+P_z v' z' \,.
\eea
Choosing the thin shell profile
\be
Q(v,z) = P(z)^2 + \theta(v) \big[ Q(z) - P(z)^2\big] \,,
\ee
we have that for $v<0$ the backgrounds is
\be
ds^2 = \frac{1}{z^{2d_\theta/d}}
\big(-P(z)^2dt^2+dz^2+d\boldsymbol{x}^2\big) \,,
\ee
while for $v>0$ the metric becomes (\ref{generic metric bh}).
The equation (\ref{eom1 vadya gen}) tells us that $Qv'+Pz'$ is constant for $v\neq 0$ but we recall that it takes two different values $E_-$ (for $v<0$) and $E_+$ (for $v>0$). Since $v'(0) =z'(0)=0$, we have that $E_- = 0$.
Integrating across the shell as in \S\ref{stripvaidyasec}, (\ref{eom2 vadya gen}) implies again that
\be 
v'_+=v'_-\equiv v'_c \,,
\qquad 
\textrm{at $\;x=x_c$}\,.
\label{eq:vpm gen}
\ee
Then (\ref{eom1 vadya gen}) leads to
\be 
z'_+-z'_-=-\frac{1}{2P(z)}(Q(z)-P^2(z))v'_c \,.
\ee
From these equations, we get
\be 
E_+ =\frac{(Q_c-P_c^2)v'_c}{2}
=-\frac{(Q_c-P_c^2) z_-'}{2 P_c} \,,
\ee
where $P_c \equiv P(z_c)$, $Q_c \equiv Q(z_c)$ and again 
$z_-'=-\sqrt{(z_\ast/z_c)^{2d_\theta}-1}$. 
Thus, in the black hole part $x_c < x \leqslant \ell/2$ we have
\bea
v'&=&\frac{E_+ - Q(z) z'}{P(z)} \,,\\
z'^2&=&
\frac{Q(z)}{P(z)^2}
\bigg[ \bigg(\frac{z_\ast}{z}\bigg)^{2d_\theta}-1\bigg]
+ \frac{(Q_c-P_c^2)^2}{4P_c^2 P(z)^2}
\bigg[\bigg(\frac{z_\ast}{z_c}\bigg)^{2d_\theta}-1\bigg]
\equiv H(z)\,.
\label{eq:fogm}
\eea
Repeating the steps explained to get (\ref{t shell tot}) and (\ref{A shell tot}), in this case we find 
\be
t = \int_{0}^{z_c} \frac{P(z)}{Q(z)}\left(\frac{E_+}{P(z)\sqrt{H(z)}}+1\right) dz\,,
\qquad
\mathcal{A} =
2\ell_\perp^{d-1}z_\ast^{d_\theta}\int_{0}^{z_c} \frac{dz}{z^{2d_\theta}\sqrt{H(z)}}\,.
\ee

\subsection{Linear growth}

At this point we take the limit of large $z_\ast$, keeping $z_m$ and $z_c$ finite.
In this limit, (\ref{eq:fogm}) becomes
\be 
z'^2=\left(\frac{Q(z)}{z^{2d_\theta}}
+ \frac{(Q_c-P_c^2)^2}{4z_c^{2d_\theta} P_c^2}\right)
\frac{z_\ast^{2d_\theta}}{P(z)^2 } = H(z) \,.
\ee
The equation $\partial_{z_m}H(z_m)=0$, which defines $z_m$, reads
\be 
( Q'_m P_m - 2 Q_m P'_m) z_m
-2d_\theta P_m Q_m -2P'_m  \gamma_c z_m^{2d_\theta+1}=0\,,
\label{eq:zmgeneric}
\ee
where the subindex $m$ denotes that the corresponding quantity is computed at $z=z_m$ and we defined
\be 
\gamma_c \equiv \frac{(Q_c-P_c^2)^2}{4z_c^{2d_\theta} P_c^2}\,.
\ee
Introducing $\gamma^\ast_c\equiv \gamma_c |_{z_c=z_c^\ast}$, the equation for $z_c^\ast$  reads
\be 
\gamma_c^\ast=-\frac{Q_m}{z_m^{2d_\theta}}\,,\label{eq:zcgeneric}
\ee
which reduces to the second equation of (\ref{eqs zast large}) for the case considered in the Appendix \ref{app computational details}.
Then, plugging (\ref{eq:zcgeneric}) into (\ref{eq:zmgeneric}) we find
\be 
Q'_m z_m -2d_\theta Q_m=0\,,
\qquad
\textrm{at}\;\;z_c =z_c^\ast\,,
\ee
which can also be written as
\be 
\partial_{z_m} \bigg( \frac{Q_m}{z_m^{2d_\theta}}\bigg) = 0\,.
\label{eq:generalzm}
\ee
Repeating the steps done to get (\ref{eq:linearl}), (\ref{eq:lineart}) and (\ref{eq:lineara3}), in this case we obtain 
\bea
\ell /2&=& 
\frac{\sqrt{\pi}\, \Gamma (1/(2 d_\theta)+1/2)}{
\Gamma(1/(2d_\theta))}\,z_\ast
-\frac{\log \varepsilon}{\sqrt{H_2}}\,,
\label{eq:linearlgen}\\
\rule{0pt}{.6cm}
t&=& 
-\,\frac{E_+}{Q_m \sqrt{H_2}} \,\log \varepsilon 
= -\,\frac{z_\ast^{d_\theta}}{z_m^{d_\theta}\sqrt{-H_2 Q_m}} \, \log \epsilon\,,
\label{eq:lineartgen}\\
\rule{0pt}{.6cm}
A^{(3)}_{\textrm{\tiny reg}}&=&
-\,2 \ell_\perp^{d-1} \frac{z_\ast^{d_\theta}}{z_m^{2d_\theta} \sqrt{H_2}} \, \log \varepsilon \,.
\label{eq:lineara3gen}
\eea
Thus, (\ref{eq:lineartgen}) and (\ref{eq:lineara3gen}) allow us to find that
\be
A^{(3)}_{\textrm{\tiny reg}}=2 \ell_\perp^{d-1}\frac{\sqrt{-Q_m}}{z_m^{d_\theta}}\; t  \,.
\label{eq:finalareaappgeneral}
\ee
We conclude that $P(z)$ does not affect the linear growth regime.

\section{Vaidya backgrounds with time dependent exponents}

In this appendix we consider the following generalization of (\ref{vaidya metric})
\be
\label{vaidya metric1}
ds^{2}=z^{2 \theta\left(v\right)/d-2}\left(-z^{2\left(1-\zeta\left(v\right)\right)}F\left(v,z\right)dv^{2}-2z^{1-\zeta\left(v\right)}\, dv\, dz+d\boldsymbol{x}^{2}\right)\,,
\ee
where we have introduced a temporal dependence in the Lifshitz and hyperscaling exponents. 
Let us discuss the energy-momentum tensor when the metric (\ref{vaidya metric1}) is on shell.
For simplicity, we consider only the backgrounds (\ref{vaidya metric1})  with $F(v,z)=1$ identically.\\
The first  case we consider is given by $\theta(v) = \textrm{const}$. The associated energy-momentum tensor reads
\be
\label{Tmunu tot zv}
T_{\mu\nu}=T^{\textrm{\tiny (hs)}}_{\mu\nu} + T^{(\zeta)}_{\mu\nu}\,,
\ee
where $T^{\textrm{\tiny (hs)}}_{\mu\nu} $ is the part containing the hyperscaling exponent, which occurs also when $\zeta(v)$ is constant, namely
\be 
\label{Tmunuhs}
T^{\textrm{\tiny (hs)}}_{\mu\nu}=
\left( \begin{array}{ccc}
-z^{-2\zeta} (d_\theta+1+\theta/d) d_\theta/2 
& -z^{1-\zeta}(d_\theta+1+\theta/d)d_\theta/2   & 0 \\
\rule{0pt}{.4cm}
-z^{1-\zeta}(d_\theta+1+\theta/d)d_\theta/2 & z^{-2} d_\theta (\theta/d-\zeta+1) & 0 \\
\rule{0pt}{.4cm}
0 & 0 &
z^{-2}[d_\theta^2(d-1)/d+2\zeta(\zeta-1+d_\theta)]
\, \mathbb{I}_d/2
\end{array} \right),
\ee
(we have denoted by $\mathbb{I}_d$ the $d$ dimensional identity matrix),
while $T^{(\zeta)}_{\mu\nu}$ is the term due to $\zeta' \neq 0$
\be 
\label{Tmunuzeta}
T^{(\zeta)}_{\mu\nu}=
\left( \begin{array}{ccc}
0 & 0 & 0 \\
0 & 0 & 0 \\
0 & 0 & z^\zeta  \zeta' \,\mathbb{I}_d
\end{array} \right) \,.
\ee 
Similarly, we can consider the situation where $\zeta(v) = \textrm{const}$. It leads to
\be
T_{\mu\nu}=T^{\textrm{\tiny (hs)}}_{\mu\nu} + T^{(\theta)}_{\mu\nu}\,,
\ee
where $T^{\textrm{\tiny (hs)}}_{\mu\nu} $ is (\ref{Tmunuhs}) and
\be 
\label{Tmunutheta}
T^{(\theta)}_{\mu\nu}=
\frac{\theta'}{z}
\left( \begin{array}{ccc}
z^{-1\zeta}\big[2 +\log{z} \big(\zeta-d_\theta -\theta/d+ (\theta'/d)  \log{z}  \big)\big] 
& (1-d_\theta \log z ) & 0 \\ 
\rule{0pt}{.4cm}
(1-d_\theta \log z )
 & 0 & 0 \\
 \rule{0pt}{.4cm}
0 & 0 & z^{\zeta-1}\big[2+ (d-1) (d_\theta/d) \log z\big]\, \mathbb{I}_d
\end{array} \right),
\ee 
which vanishes when $\theta(v)$ is constant, as expected.
When both $\theta'(v)\neq 0$ and $\zeta'(v)\neq 0$, we find that
\be
T_{\mu\nu}=T^{\textrm{\tiny (hs)}}_{\mu\nu} 
+T^{(\zeta)}_{\mu\nu}+T^{(\theta)}_{\mu\nu}
+T^{(\theta\zeta)}_{\mu\nu}\,,
\ee
where $T^{\textrm{\tiny (hs)}}_{\mu\nu} $, $T^{(\zeta)}_{\mu\nu}$ and $T^{(\theta)}_{\mu\nu}$ have been defined respectively in (\ref{Tmunuhs}), (\ref{Tmunuzeta}) and (\ref{Tmunutheta}), while $T^{(\theta\zeta)}_{\mu\nu}$ is given by
\be 
T^{(\theta\zeta)}_{\mu\nu}=
\left( \begin{array}{ccc}
-\zeta'  \theta' \log^2(z) & 0 & 0 \\
0 & 0 & 0 \\
0 & 0 & \boldsymbol{0}_d
\end{array} \right)\,,
\ee
being $\boldsymbol{0}_d$ is the $d\times d$ matrix whose elements are zero.\\
It could be interesting to analyze the Null Energy Condition for these kind of backgrounds.
Unfortunately, since the inequalities turn out to be lengthy and not very illuminating, we will consider here only the case of $\theta(v) = \textrm{const}$.
First, since a null vector with respect to the metric (\ref{vaidya metric1}) is null also with respect to (\ref{vaidya metric}), we can employ the vectors (\ref{null vec vaidya}).
Secondly, given the additive structure of $T_{\mu\nu}$ in (\ref{Tmunu tot zv}), we can consider the results of \S\ref{sec backgrounds} and add to them the contribution of $T^{(\zeta)}_{\mu\nu} N^\mu N^\nu$. The resulting inequalities read
\bea
\label{nec zeta(v)1}
d_\theta \big[\zeta(v)-1-\theta/d\big] & \geqslant & 0\,,
 \\
 \label{nec zeta(v)2}
\big[ \zeta(v)-1\big] \big[d_\theta+\zeta(v)\big]+z^{\zeta(v)} \zeta'(v) & \geqslant & 0\,,
\eea
which reduce respectively to (\ref{eq:hsnec1}) and (\ref{eq:hsnec2}) when $\zeta(v) = \textrm{const}$, as expected. 
When $\theta =0$, the inequality (\ref{nec zeta(v)1}) tells us that $\zeta(v) \geqslant 1$. As for (\ref{nec zeta(v)2}), it allows, for instance, a profile with $\zeta ' (v) \geqslant 0$.
In the critical case $\theta = d-1$, (\ref{nec zeta(v)1}) becomes $\zeta(v) \geqslant 2-1/d \geqslant 1$ while (\ref{nec zeta(v)1}) becomes $[ \zeta(v)^2-1 ] + z^{\zeta(v)} \zeta '(v)  \geqslant 0$. Thus, for instance, profiles having $\zeta ' (v) \geqslant 0$ are again allowed.


\end{document}